\documentclass[prx,amsmath,amssymb,notitlepage,twocolumn,
nofootinbib,
superscriptaddress,
longbibliography
]{revtex4-1}
\usepackage{caption}
\usepackage{subcaption}
\usepackage{MnSymbol}
\usepackage{amsmath}
\usepackage{tabularx,graphicx}
\usepackage{epstopdf}
\usepackage{graphicx}
\usepackage{spectralsequences} 
\usepackage{latexsym}
\usepackage{amssymb}
\usepackage{tikz-cd}
\usepackage{amsmath}
\usepackage{color, colortbl}
\usepackage{psfrag}
\usepackage{bbm}
\usepackage{bm}
\usepackage{titlesec}
\usepackage{dsfont}
\usepackage{feynmp}
\usepackage{slashed}
\usepackage{multirow}
\usepackage{framed}
\newcommand{\beq}{\begin{eqnarray}}
\newcommand{\eeq}{\end{eqnarray}}

\newcommand{\tr}{{\rm tr}}

\newcommand{\bsp}{\begin{split}}
\newcommand{\esp}{\end{split}}

\newcommand{\ie}{{i.e., }}
\newcommand{\eg}{{e.g., }}

\usepackage{color}
\definecolor{darkblue}{rgb}{0.,0.,0.4}
\definecolor{darkred}{rgb}{0.5,0.,0.}
\definecolor{BlueViolet}{RGB}{138,43,226}
\definecolor{SkyBlue}{RGB}{30,144,255}
\definecolor{DarkGreen}{RGB}{0,100,0}
\usepackage[pdftex,colorlinks=true,linkcolor=darkblue,citecolor=blue,urlcolor=darkred]{hyperref}
\usepackage[normalem]{ulem}

\def\H{\mathcal{H}}
\def\Z{\mathbb{Z}}
\usepackage{tabularx}

\usepackage{amsthm}
\theoremstyle{plain}
\newtheorem*{theorem*}{Theorem}
\newtheorem{theorem}{Theorem}
\newtheorem{conjecture}{Conjecture}
\newtheorem{definition}{Definition}

\begin{document}
	\title{Symmetry Protected Topological Phases of Mixed States in the Doubled Space}
	
	\author{Ruochen Ma}
	\affiliation{Kadanoff Center for Theoretical Physics, The University of Chicago, Chicago, Illinois, USA 60637}

	\author{Alex Turzillo}
	\affiliation{Perimeter Institute for Theoretical Physics, Waterloo, Ontario, Canada N2L-2Y5}

\begin{abstract}

The interplay of symmetry and topology in quantum many-body mixed states has recently garnered significant interest. In a phenomenon not seen in pure states, mixed states can exhibit average symmetries – symmetries that act on component states while leaving the ensemble invariant. In this work, we systematically characterize symmetry protected topological (SPT) phases of short-range entangled (SRE) mixed states of spin systems -- protected by both average and exact symmetries -- by studying their pure Choi states in a doubled Hilbert space, where the familiar notions and tools for SRE and SPT pure states apply. This advantage of the doubled space comes with a price: extra symmetries as well as subtleties around how hermiticity and positivity of the original density matrix constrain the possible SPT invariants. Nevertheless, the doubled space perspective allows us to obtain a systematic classification of mixed-state SPT (MSPT) phases. We also investigate the robustness of MSPT invariants under symmetric finite-depth quantum channels, the bulk-boundary correspondence for MSPT phases, and the consequences of the MSPT invariants for the separability of mixed states and the symmetry-protected sign problem. In addition to MSPT phases, we study the patterns of spontaneous symmetry breaking (SSB) of mixed states, including the phenomenon of exact-to-average SSB, and the order parameters that detect them. Mixed state SSB is related to an ingappability constraint on symmetric Lindbladian dynamics.

\end{abstract}

\maketitle

\tableofcontents

\section{Introduction}
\label{sec:intro}

The classification and characterization of quantum phases of matter represent central achievements in the field of many-body physics. In the setting of isolated quantum systems -- where one studies the pure ground states of gapped, local Hamiltonians -- our understanding has reached a mature stage. This category encompasses both conventional symmetry-breaking phases within Landau's paradigm as well as a multitude of topological phases, ranging from Symmetry-Protected Topological (SPT) phases to topological orders \cite{2023McGreevyreview}.

Recently, progress has been made toward generalizing the notion of quantum phases to non-equilibrium open systems, such as noisy intermediate-scale quantum (NISQ) simulators \cite{2017NISQ,2018NISQ}. This program includes studying the impact of local noise on pure SPT or topologically ordered states \cite{2022openspt,2022aspt,2023aspt,2022decoxu,2023decobao,2023decofan,2023decoxu}, exploring the possibility of new phases that arise only in open systems \cite{2022aspt,2023aspt}, and examining state-preparing protocols involving measurement and feedback \cite{2022bimeasure,2023LuMeasure,2023Zhu,2023ZHUPRL}. Considering the growing number of examples, it is desirable to establish a general framework for understanding phases of open quantum systems. In this generalization, the pure states which appeared as ground states of gapped, local Hamiltonians are replaced by mixed states. The appearance of mixed states raises the following questions: can we characterize quantum phases solely based on the properties of its mixed states, without reference to how they are prepared (for example, decoherence of a pure topological state)? Can we leverage well-established concepts and tools developed for pure states, e.g. tensor network methods, to investigate the interplay of symmetry and topology in mixed states?

To gain some intuition, let us briefly revisit the standard concept of pure state quantum phases \cite{2010localunitary}. Two states belong to the same phase if and only if there exists a finite-depth local unitary circuit $U$ that (approximately) connects them. The idea behind this definition is that $U$ can only build correlations over a finite distance, implying that states in the same phase exhibit the same long-distance asymptotics of correlation and entanglement. A pure state is said to be \emph{short-range entangled} (SRE) if it belongs to the trivial phase -- the phase containing product states. Global symmetries can further enrich the phase diagram by imposing constraints on local gates in $U$ \cite{2013chenSPT}. For instance, SPT phases, such as topological insulators \cite{2010TI,2011TI} and the Haldane chain \cite{affleck1988valence}, consist of SRE states that are distinguished from product states by such constraints. They lack topological order and can be connected to product states only by circuits for which the local gates fail to be symmetric. Moreover, adiabaticity guarantees that the ground states of two gapped local Hamiltonians belong to the same phase when the two Hamiltonians are connected by a gapped path \cite{2005adiabatic}.

If we attempt to construct a similar framework for mixed states, several questions arise:

\emph{\textbf{First}, which mixed states should be regarded as SRE, as opposed to long-range entangled (LRE)}? Currently, three proposals have been put forward to define SRE mixed states. Refs.~\cite{2022aspt,2023aspt} define an SRE mixed state as a state that can be purified to an SRE pure state. Equivalently, these states can be prepared from a pure product state via a finite-depth quantum channel. Alternatively, Ref.~\cite{2023separability1,2023separability} defines SRE mixed states in terms of ``separability'' -- a mixed state is SRE if it can be decomposed as a convex sum of SRE pure states. These two proposals (whose relation is not entirely clear) encounter the following problem: Given a generic mixed state, it is challenging to find a purification or decomposition. This difficulty arises because there is no known entanglement measure that is both faithful and easily computable that can verify whether either of these two definitions is satisfied.

There is a third proposal, known as the doubled state formalism, originally employed to examine the stability of a fingerprint of a pure SPT state, namely the strange correlator, under decoherence \cite{2022decoxu,2014strange}. This approach involves mapping the density operator via the state-operator map to a \emph{pure} state in a doubled Hilbert space, referred to as the Choi state. The impact of decoherence becomes an interaction between degrees of freedom in the two Hilbert spaces. One can then leverage the rich insights from pure state topological phases to understand mixed state topology. This idea has been successfully applied to analyze the effect of decoherence on error-correcting codes and quantum critical points \cite{2023decobao,2023decoxu}. Another appealing aspect of this approach lies in avoiding the need to find a convex decomposition or a purification (e.g., the square root of a density operator). After applying a state-operator map, which is easily computable, one can then transform the mixed-state problem into a much better-understood pure state question. Given these advantages, in this work, \emph{we define mixed SRE states as those having an SRE Choi state}. Another motivation for introducing this definition of SRE mixed states is that their density operators can be shown to exhibit an area law for mutual information and operator entanglement entropy. This feature enables the efficient classical simulation of the quantum state. Subsequently, we systematically develop a framework for mixed state SRE phases, subject to certain symmetry conditions, akin to that of pure-state quantum phases.

\emph{\textbf{Second,} what are the invariants of symmetric SRE states, and how do they appear physically?} In the pure state scenario, SRE phases are classified by their SPT invariants (\eg group cohomology), which show up in the presense of order parameters -- for example, in the bulk of a $3d$ topological insulator, a unit magnetic monopole carries half electric charge \cite{2008TImonopole}. How can this understanding of invariants of SRE states in the doubled space be translated back to properties of the original mixed states? Compared to a pure state, a mixed state can exhibit a richer symmetry structure, with exact (strong) as well as average (weak) symmetries \cite{2022openspt,2022aspt}. Further enrichment arises from the fact that $\rho$, being a density operator, must be hermitian, which implies that its Choi state has an additional antiunitary symmetry. These symmetries may protect a diverse family of SPT invariants, many of which have no analogue in pure states. One may wonder whether this signals many-body topology intrinsically defined for mixed states. We demonstrate through a no-go theorem that many of these invariants are inconsistent because that would imply the original density operator $\rho$ has unphysical negative eigenvalues (Theorem~\ref{thm:MSPTclassification},  \ref{thm:MSPTclassificationwithT}, \ref{thm:MSPTclassificationgeneral}). In the study of tensor networks, it is known that determining whether a matrix product operator is non-negative is NP-hard in system size \cite{2014MPOpositivity}. Our result suggests that studying the SPT invariants in the doubled space can provide insights into this problem. We further explore the relation of the doubled space formalism to previous studies by demonstrating that: (1) those allowed, nontrivial mixed state SPT states are all not symmetrically separable in the definition in Ref.~\cite{2023separability}. (2) SPT phases with certain SPT invariants have a symmetry-protected sign problem in the definition of Ref.~\cite{2021signproblem}.

\emph{\textbf{Third}, what is the appropriate equivalence relation on mixed states that categorizes them into phases?} Due to the system-environment coupling, a generic open system is subject to a non-unitary evolution generated by a quantum channel $\mathcal{E}$, which does not need to be reversible, i.e. $\mathcal{E}^{-1}$ may not be a quantum channel. An equivalence relation has been proposed based on two-way connectability by finite-depth quantum channels (with symmetric local gates when the problem has symmetry) \cite{2019channelconnectible,2022aspt,2023channelconnectible}. Similar ideas have also been proposed to define steady-state phases \cite{2023steadystatephase}. However, again, it is generally impractical to determine whether two specific density matrices are two-way connectable. Therefore, a natural question to ask is whether it is possible to define a mixed state phase directly in terms of certain universal long-range properties of $\rho$. Moreover, are these properties robust under channel evolutions (deformations)? In this study, we define mixed state phases based on their SRE nature and their SPT invariants in the doubled Hilbert space. We justify this definition by demonstrating that the SRE property remains perturbatively stable under symmetric finite-depth quantum channels and that SPT invariants are stable when SRE is preserved by the channel (with an additional requirement in $d>1$). More surprisingly, employing tensor network methods, we demonstrate that the SRE property may not be preserved by a symmetric finite-depth channel. This is exemplified by a specific case where the resulting state exhibits spontaneous exact-to-average symmetry breaking, a phenomenon without a counterpart in pure states.

Given the rich symmetry structure of mixed states, exploring possible patterns of spontaneous symmetry breaking (SSB) in mixed states is a natural direction of study. In classical statistical mechanics and pure-state many-body physics, there are two complementary ways to probe SSB -- order parameters and disorder parameters \cite{2017disorderarameter}. We generalize these concepts to mixed states. When the density operator $\rho$ is represented by a tensor network, we discuss the implications of different patterns of SSB for the properties of the local tensors and demonstrate explicitly that, unlike a finite-depth unitary, a symmetric finite-depth channel can induce an exact-to-average SSB by destroying the ``normality" of the local tensor. The stability of this SSB also has interesting implications for open system dynamics: a local Lindbladian with a discrete exact symmetry can have a nonzero dissipative gap in the thermodynamic limit only when the steady state exhibits spontaneous breaking of the exact symmetry. As we will see, a precise relationship between the probes mentioned above and a systematic study of SSB in mixed states are still lacking. We only provide a small step in this direction.

\emph{\textbf{Fourth,} is there a notion of bulk-boundary correspondence in mixed states?} A defining feature of nontrivial pure state SPT phases is that they have nontrivial boundary states, protected by bulk t'Hooft anomaly inflow. For example, at the edge of a Haldane chain, there is a dangling spin-half moment. These nontrivial boundary states, often characterized by localized edge degeneracy or gapless boundary modes, are challenging to define in mixed states due to the absence of a Hamiltonian and energy gap. In this study, utilizing tensor network methods, we clarify the notion of localized boundary modes in mixed states. For $1d$ mixed state SPT phases, we demonstrate that the edge may exhibit distinct projective representation (or ``symmetry-breaking") patterns, depending on the bulk SPT invariant. The mixed state bulk-boundary correspondence in higher dimensions is, however, a more challenging topic due to our limited understanding of LRE mixed states and how to relate long-range entanglement in the doubled space to physical properties of the original mixed state $\rho$. We briefly comment on the higher dimensional case and leave a thorough study to further explorations.

The paper is organized as follows. In Sec. \ref{sec:symmetries}, we introduce the operator-state map and symmetry conditions on mixed states. We then define SRE mixed states and mixed state SPT phases in Sec. \ref{sec:doubledSPT}. The main part of this section is dedicated to the discussion on the relation between SPT invariants in the doubled space and the (violation of) non-negativity of the original density matrix. We highlight the connections between SPT invariants, symmetry-protected sign problems, and separability in Sec. \ref{sec:generalsymmetries}. Section \ref{sec:deformations} examines the condition under which a symmetric finite-depth channel preserves the SRE property and SPT invariants. Additionally, we provide an example where the SRE is violated, leading to a final state that manifests exact-to-average SSB. We discuss possible SSB patterns and their implications for tensor networks in Sec. \ref{sec:SSB}. The boundary properties of mixed-state SPT phases are studied in Sec. \ref{sec:boundary}. We conclude our paper and discuss a few open directions in Sec.~ \ref{sec:summary}.

\section{The operator-state map and symmetries}
\label{sec:symmetries}

\emph{a. The operator-state map.} In this section we shall introduce the operator-state map \cite{preskill1998lecture}, which is a one-to-one correspondence between an operator acting on a Hilbert space $\mathcal{H}$ and a pure quantum state in a doubled Hilbert space $\mathcal{H}\bigotimes \mathcal{H}^*$. This mapping enables us to study a mixed state, represented by a density operator $\rho$, using well developed concepts for pure states.

To start, we consider a lattice Hilbert space with a local tensor product structure $\mathcal{H}=\otimes_i\mathcal{H}_i$ ($i$ labeling lattice sites). The space of quantum operators acting on the Hilbert space is denoted by $\mathcal{B}(\mathcal{H})$. We define the ``relative state" as
\begin{equation}
     | \Omega \rrangle := \frac{1}{\sqrt{\mathcal{N}}}\sum_a | a \rangle | a^* \rangle,
     \label{eq:relativestate}
\end{equation}
where $\mathcal{N}$ is the dimension of $\mathcal{H}$, and $\{ |a\rangle \}$ is an orthonormal basis of $\mathcal{H}$. This is a maximally entangled state in the doubled Hilbert space $\mathcal{H}\bigotimes \mathcal{H}^*$. Here a state $|a^*\rangle \in \mathcal{H}^*$ is defined by taking the complex conjugation of a corresponding state $|a\rangle \in \mathcal{H}$ (written in a certain basis), namely $|a^*\rangle = K |a\rangle$ where $K$ is complex conjugation. We will refer to $\mathcal{H}$ and $\mathcal{H}^*$ as the ket and the bra space, respectively.

Two properties of the relative state $| \Omega \rrangle$ are important for our purpose. Firstly, the definition of $| \Omega \rrangle$ is basis independent. To see this, consider another orthonormal basis $\{ |\Tilde{a} \rangle \}$ of $\mathcal{H}$, which is related to $\{ |a\rangle \}$ by a unitary transformation, $|\Tilde{a}\rangle = \sum_a V_{\Tilde{a} a} | a \rangle$. One can show that the relative state $| \Tilde{\Omega} \rrangle$ defined in terms of this new basis is identical to $| \Omega \rrangle$,
\begin{equation}
    \begin{split}
        & | \Tilde{\Omega} \rrangle =  \frac{1}{\sqrt{\mathcal{N}}} \sum_{\Tilde{a}} | \Tilde{a} \rangle | \Tilde{a}^* \rangle \\
        = & \frac{1}{\sqrt{\mathcal{N}}} \sum_{\Tilde{a}} \sum_a \sum_b V_{\Tilde{a} a} V^*_{\Tilde{a} b} |a\rangle |b\rangle = | \Omega \rrangle.
    \end{split}
\end{equation}
The basis independence can be understood by noting that $| \Omega \rrangle$ is obtained by ``flipping the bras" in the (normalized) identity operator in $\mathcal{B}(\mathcal{H})$, which is basis independent.

Secondly, the relative state $| \Omega \rrangle$ is a short-range entangled (SRE) state in the doubled Hilbert space. In fact, $| \Omega \rrangle$ is a product of maximally entangled states on each site,
\begin{equation}
    | \Omega \rrangle = \otimes_i (\frac{1}{\sqrt{\mathcal{N}_i}} \sum_{a_i} | a_i \rangle | a_i^* \rangle), 
\end{equation}
where $\{ | a_i \rangle \}$ is a set of orthonormal basis for the local Hilbert space $\mathcal{H}_i$. This property will be important when we discuss the notion of short-range entanglement in the doubled space.

\begin{figure}
\begin{center}
  \includegraphics[width=.40\textwidth]{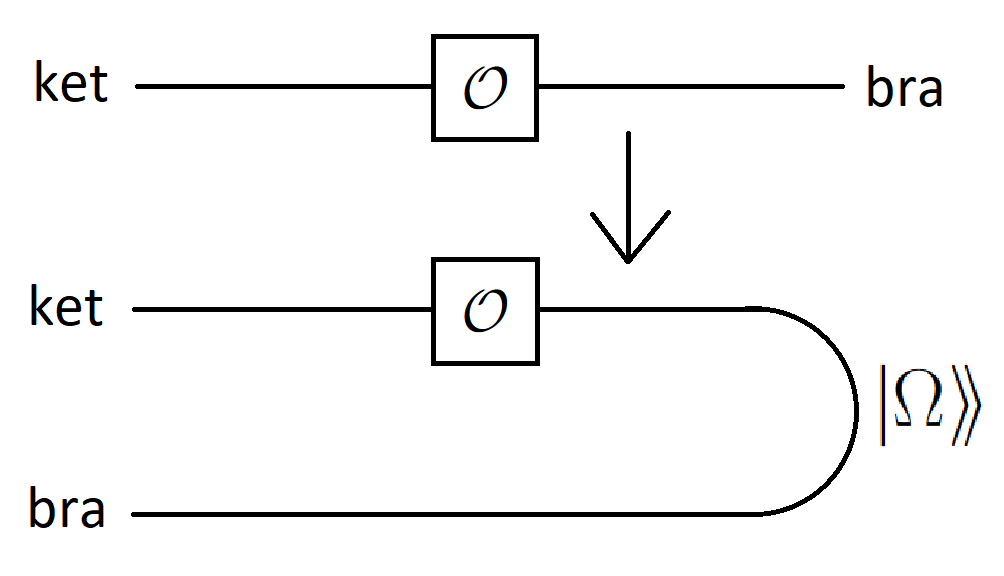} 
\end{center}
\caption{
A diagrammatic representation of the operator-state map. One can use the relative state to freely bend the leg back and forth.
}
\label{Fig:operatorstatemap}
\end{figure}

Now we introduce the operator-state map using the relative state. For each operator $O \in \mathcal{B}(\mathcal{H})$, there is an associated pure state in the doubled space,
\begin{equation}
    | O \rrangle = (O\bigotimes I) | \Omega \rrangle,
    \label{eq:o-smap}
\end{equation}
where $I$ is the identity operator acting on the bra space $\mathcal{H}^*$. We will employ a bigger ``$\bigotimes$" to represent that the operators on its respective sides act on distinct copies of the Hilbert space. $| O \rrangle$ is called the Choi state associated with the operator $O$. In fact, the correspondence between operators in $\mathcal{B}(\mathcal{H})$ and states in $\mathcal{H}\bigotimes \mathcal{H}^*$ is one-to-one, as seen from Fig.~\ref{Fig:operatorstatemap}. Moreover, the doubled space $\mathcal{H}\bigotimes \mathcal{H}^*$ is a Hilbert space endowed with the Hilbert-Schmidt inner product -- for two operators $A,\,B\in \mathcal{B}(\mathcal{H})$, the overlap of the corresponding Choi states is given be 
\begin{equation}
    \llangle A | B \rrangle = \mathrm{tr}(A^\dagger B).
\end{equation}
Note that with the normalization in Eq.~(\ref{eq:relativestate}), the relative state has unit norm, $\llangle \Omega | \Omega \rrangle = 1$. 

Using the relative state $| \Omega \rrangle$, one can also move operators between the two Hilbert spaces $\mathcal{H}$ and $\mathcal{H}^*$, 
\begin{equation}
\begin{split}
    (O\bigotimes I) | \Omega \rrangle & = \frac{1}{\sqrt{\mathcal{N}}} \sum_{a,b} | b \rangle \langle b | O | a \rangle | a^* \rangle \\
    & = \frac{1}{\sqrt{\mathcal{N}}} \sum_{a,b} | b \rangle| a^*\rangle \langle a^* | O^T | b^* \rangle \\
    & = (I\bigotimes O^T) | \Omega \rrangle.
\end{split} 
\end{equation}
One can see that the relative state establishes a one-to-one correspondence between operators acting on $\mathcal{H}$ and those acting on $\mathcal{H}^*$ by taking the transpose.

\emph{b. Symmetries.} Below we will define the topological properties of a density matrix $\rho = \sum_I P_I | \Psi_I \rangle \langle \Psi_I |$, in general representing a mixed state, by studying its Choi state $| \rho \rrangle$ in the doubled space. A crucial ingredient in these nontrivial topologies is symmetry. In contrast to pure states, there are two distinct types of unitary symmetry for mixed states.

The first is the \emph{exact symmetry} (or the \emph{strong symmetry} \cite{Buča_2012}), denoted by $K$. For each element $k\in K$, there is a corresponding unitary operator
$U_k$ acting on $\mathcal{H}$, which forms a linear representation of $K$:
\begin{equation}
    U_k = \otimes_i u_k^i,
\end{equation}
where $u_k^i$ is the (linear) representation of $K$ on a single site $i$. We say a density operator has an exact symmetry $K$ if for any $k \in K$,
\begin{equation}
    U_k\rho = \rho U_k = e^{i\theta_k}\rho.    
\end{equation}
In other words, exact symmetry means that each individual pure state within the convex sum $\rho$ is an eigenstate of $K$ with the same eigenvalue. In the doubled space, this symmetry condition becomes 
\begin{equation}
    [U_k\bigotimes I] |\rho \rrangle = [I \bigotimes U_k^T] |\rho \rrangle = e^{i\theta_k}|\rho \rrangle.
    \label{eq:exactsym}
\end{equation}    
Eq.~(\ref{eq:exactsym}) reveals a feature of the doubled space: the associated Choi state $| \rho \rrangle$ enjoys a ``doubled" exact symmetry, which stems from independent left and right multiplications of $\rho$. 

For a mixed state, there is a second type of unitary symmetry -- the \emph{average symmetry} (or the \emph{weak symmetry} \cite{Buča_2012}), denoted by $G$. The density matrix is invariant only under the adjoint action of an average symmetry:
\begin{equation}
    U_g \rho U_g^\dagger = \rho    
\end{equation}
for any $g\in G$. We again require $U_g = \otimes_i u_g^i$, which is a tensor product of unitary operators, each of which linearly represents $G$ on a single site. In the doubled space, the average symmetry becomes
\begin{equation}
    [U_g\bigotimes U_g^*]|\rho \rrangle = | \rho \rrangle,
    \label{eq:Gtrivialaction}
\end{equation}
which behaves similarly to an ordinary unitary symmetry of a pure state.

For simplicity we often focus on cases where the full unitary symmetry $\mathcal{G}$ of $| \rho \rrangle$ is given by $\mathcal{G} = (K \times K) \times G$. But we note that, similar to the pure state scenario, $K$ and $G$ may form a nontrivial group extension 
\begin{equation}
\begin{tikzcd}
1 \arrow[r] & K \arrow[r] & \Tilde{G} \arrow[r] & G \arrow[r] & 1,
\end{tikzcd}    
\label{eq:internalsymmetry}
\end{equation}
in which the exact symmetry is a normal subgroup.\footnote{The fact that $K$ must be a normal subgroup can be shown as follows: For any $k\in K$ and $g\in G$, $gkg^{-1}\rho = g(kg^{-1}\rho g)g^{-1} = e^{i\theta_k}\rho$. Therefore, $gkg^{-1}\in K$.} The group extension and the $G$ action on the normal subgroup $K$ can be deduced from the unitary operators $U_k$ and $U_g$ that act on the original Hilbert space $\mathcal{H}$.   

Besides the two types of unitary symmetry, a state in the doubled space may also have antiunitary symmetries. In particular, the Choi state $| \rho \rrangle$ of a density operator is always invariant under an antiunitary operator $J$, called the modular conjugation. Physically, the $J$ symmetry is equivalent to hermiticity of the density operator, and can be introduced as 
\begin{equation}
    J (| a \rangle | b^* \rangle) = | b \rangle | a^* \rangle,
\end{equation}
which is a swap that exchanges the ket and the bra spaces, followed by a complex conjugation, $J = K\cdot \mathrm{SWAP}$. It is straightforward to check that the relative state $| \Omega \rrangle$ is invariant under $J$. As a result, we can show that the Choi state $| \rho \rrangle$ is $J$-symmetric, 
\begin{equation}
    J | \rho \rrangle = J (\rho \bigotimes I) J^{-1} J | \Omega \rrangle = (I \bigotimes \rho^*) | \Omega \rrangle = | \rho \rrangle,
\end{equation}
where we used $\rho = \rho^\dagger$.

It is important to note that $J$ does not commute with the exact unitary symmetry. Since $J$ exchanges the quantum states of the bra and the ket, $J$ effectively acts on the exact unitary symmetry $K\times K$ by swapping the two components,\footnote{The specific combination of $K\times K$ and $J$ is also known as the wreath product.}
\begin{equation}
    J (k_1,k_2) J^{-1} = (k_2,k_1),
\end{equation}
for $k_1,\, k_2 \in K$. On the other hand, one can easily check that $J$ commutes with the average unitary symmetry $G$. As an illustrative example, in the case where the unitary symmetry is $\mathcal{G} = (K\times K) \times G$, the full symmetry group in the doubled space is $G_d = (K\times K)\rtimes J \times G$, with the $J$ action on $K\times K$ described above.

Lastly, we direct our attention to another antiunitary symmetry, namely time reversal symmetry $\mathcal{T}$. In the doubled space, we define the $\mathcal{T}$ action as 
\begin{equation}
\begin{split}
    \mathcal{T}: & \, | a \rangle | b^* \rangle \to \sum_{\tilde{a},\tilde{b}} W_{a\tilde{a}} W_{b\tilde{b}}^* | \tilde{a} \rangle | \tilde{b}^* \rangle, \\
    & i \to -i,   
\end{split}
\end{equation}
which is a complex conjugation followed by an adjoint unitary transformation $W$, analogous to its action on pure states. A density operator $\rho$ is time reversal invariant when $\mathcal{T} \rho \mathcal{T}^{-1} = \rho$, which is equivalent to 
\begin{equation}
    \mathcal{T} (| \rho \rrangle) = | \rho \rrangle 
\end{equation}
in the doubled space. 

An important observation is, given the antiunitary nature of $J$ and $\mathcal{T}$, there is no way to factorize them, in particular the $\mathrm{SWAP}$ and complex conjugation, into transformations operating on the ket and the bra space individually. We therefore conclude that, for mixed states, antiunitary symmetries such as time reversal symmetry and the modular conjugation always behave as average symmetries, in the sense that they act on the bra and the ket spaces simultaneously.\footnote{This differs from disordered systems, where we deal with an ensemble of random Hamiltonians and their associated ground states. In such cases, $\mathcal{T}$ may constitute an exact symmetry. Further details can be found in Ref. \cite{2023aspt}.}

As a summary, by the operator-state map, we establish a one-to-one correspondence between density operators and their associated Choi states in the doubled space. The state $| \rho \rrangle$ may be invariant under various unitary or antiunitary symmetries. The full symmetry group in the doubled space, including the unitary and antiunitary ones, is referred to as $G_d$ hereafter. In the upcoming section, we will introduce the concept of mixed state SPT phases by studying topological properties of the Choi state.

\section{Mixed state SPT phases}
\label{sec:doubledSPT}

In this section, we introduce a definition of mixed-state SPT phases. We will also investigate the classification of these phases and their responses to open system evolutions. Before delving into details, let us provide an overall picture of our framework, which generalizes the familiar notion of pure state quantum phases with several important differences:
\begin{enumerate}
    \item We first generalize the concept of SRE states to the context of mixed states. The idea is to map a mixed state $\rho$ to its Choi state in the doubled space, where we can apply the familiar definition of short-range entanglement for pure states. We will then explore the physical properties of these SRE mixed states.
    \item Using the symmetry conditions discussed in Sec.~\ref{sec:symmetries}, we then define the notion of mixed state SPT (MSPT) states in terms of the SPT invariants of the Choi state. 
    \item Naively, one may expect that the mixed state SPT invariants are the same as those of pure SPT states in the doubled space with a symmetry $G_d$. But it turns out that Choi states with certain SPT invariants cannot give rise to physical density operators in the original (single) Hilbert space. In particular, their corresponding density operators would have unphysical negative eigenvalues. We demonstrate this point using a generalized decorated domain wall approach.
    \item We demonstrate that the short-range entangled property, as well as the SPT invariant of the Choi state, are perturbatively stable under natural deformations of open quantum systems, namely under symmetric finite-depth quantum channels. This robustness establishes the notion of \emph{phases} in the context of mixed states.
\end{enumerate}

This work defines phases of mixed states in terms of properties of states -- the SRE property as well as topological invariants of the Choi state -- along with the robustness of these defining properties under symmetric finite-depth channels. This approach, widely employed in the study of phases of matter (even for pure states), stands in contrast to the approach of defining the notion of phase by the state equivalence relation and is favored due to the impracticality of examining all possible adiabatic paths between two given states. Instead, constructing topological invariants proves more useful for distinguishing different phases. An intriguing question for future investigation is whether two arbitrary mixed states with the same topological invariants can be connected through a suitably defined adiabatic deformation (like as been done for some pure states \cite{Schuch_2011}), such as by symmetric finite-depth quantum channels.

\subsection{Short-range entanglement and MSPT phases}
\label{sec:SREmixed}

We now propose the definition of a SRE mixed state as follows.
\begin{definition}
    Let $\rho$ be a density operator of a mixed state, acting on the lattice Hilbert space $\mathcal{H}$. $\rho$ is SRE if its associated Choi state $| \rho \rrangle$ is a SRE state in the doubled space.
    \label{def:SRE}
\end{definition}

Notice that the Choi state of an operator in $\mathcal{B}(\H)$ is always a pure state in the doubled space. The definition of SRE state in the doubled space follows the standard definition for pure states, namely, $| \rho \rrangle$ is SRE if it (more precisely, the normalized state $| \rho \rrangle /\llangle \rho | \rho \rrangle$) can be prepared from a product state in $\H\bigotimes \H^*$ using a finite-depth local unitary circuit acting on the Hilbert space $\mathcal{H}\bigotimes \mathcal{H}^*$. 

Definition \ref{def:SRE} implies that, if a mixed state is SRE, its density operator $\rho$ can be represented by a tensor network with a finite bond dimension, which does not scale with the system size. For example, a $1d$ (translationally invariant) SRE state $| \rho \rrangle$ can be represented by a matrix product state (MPS) \cite{2006mpsrepresentation,2008mpsreview,2011mpsreview,2020mpsreview},
\begin{equation}
    | \rho \rrangle = \sum_{\{ p_i,q_i \} } \mathrm{tr}(A^{p_1q_1}...A^{p_Lq_L})| (p_1q_1)...(p_Lq_L) \rrangle, 
    \label{eq:MPSform}
\end{equation}
Here, $L$ is the length of the system, and the sum is taken over $p_i, q_i = 1,...,d$, with $p$ and $q$ representing the physical indices of $\mathcal{H}$ and $\mathcal{H}^*$ respectively, where $d$ is the local Hilbert space dimension. The symbol $A$ represents a $D\times D$ matrix.  By passing back through the operator-state map, it becomes evident that the operator $\rho$ is a matrix product operator (MPO) with bond dimension $D$, by viewing $p_i,\,q_i$ as the ket and the bra space, respectively. On the other hand, if a density operator $\rho$ can be written as an MPO, its Choi state $| \rho \rrangle = (\rho \bigotimes I)| \Omega \rrangle$ will have an MPS form of Eq.~(\ref{eq:MPSform}), since $| \Omega \rrangle$ is a product state in the doubled space. However, such a Choi state is not necessarily SRE -- an additional normality condition for the local tensor is required. An illustrative example will be discussed in detail in Sec.~\ref{sec:deformations}. The same argument can also be extended to higher dimensions using projected entangled pair states (PEPS). A direct implication of this analysis is that in an SRE mixed state, both the mutual information \cite{2008mpdomutualinfo,2017MPDO} and the operator entanglement entropy \cite{2001opee} (of the density operator) between a region and its complement scale according to the area law. This area law bounds both classical and quantum correlations within an SRE mixed state, thereby enabling efficient classical simulation of the quantum state.

Previous studies \cite{2022aspt,2023aspt} introduced an alternative definition of mixed state SPT phases based on the existence of an SRE purification and two-way connectability via symmetric finite-depth channels. That definition is in fact \emph{inequivalent} to the one presented in this work. It is worth noting that an SRE purification always ensures the existence of a tensor network representation of $\rho$ and can be comprehended within the current formalism. However, as highlighted in Ref.~\cite{2013purification}, in $1d$, there exist matrix product density operators (MPDOs) that are SRE under the current definition but cannot be purified by any SRE state. Therefore, the present proposal, relying on the Choi state and its tensor network representation, encompasses a broader range of states.

We now introduce the following definition of mixed state symmetry-protected topological (MSPT) phase.
\begin{definition}
    A state with density operator $\rho$ is an MSPT state if its Choi state $| \rho \rrangle$ is an SPT state protected by the symmetry $G_d$. Moreover, two mixed states $\rho_1$ and $\rho_2$ are in the same MSPT phase if $| \rho_1 \rrangle$ and $| \rho_2 \rrangle$ belong to the same pure state SPT phase.
    \label{def:MSPT}
\end{definition}

A simple example of MSPT states is given by a standard pure SPT state. To illustrate this, consider a pure SPT state $| \psi \rangle$, which can be prepared by a finite-depth unitary circuit $U$: $| \psi \rangle = U | 0 \rangle$, where $| 0 \rangle$ is a product state. Its Choi state can be expressed as $| \psi \rrangle = U\bigotimes U^* (| 0 \rangle |0^*\rangle)$, which is SRE since $U\bigotimes U^*$ constitutes a finite-depth circuit in the doubled space.

\subsection{Classification of MSPT phases}

We now discuss the classification of MSPT phases as defined above, which is one of our main results. For now, we consider symmetries of the form $G_d=G\times(K\times K)\rtimes J$. Generalizations to states protected by time-reversal symmetry and symmetries that are nontrivial group extensions are considered later in Section~\ref{sec:generalsymmetries}.

We begin by noting that all bosonic SPT phases of pure states with symmetry $G_d$ (described within group cohomology) are classified by $H^{d+1}(G_d,U(1))$ \cite{2013chenSPT}, with $J$ acts on the $U(1)$ coefficient through complex conjugation. This classification encompasses all potential SPT phases of the Choi state $|\rho\rrangle$. Within this classification, a subgroup, as described by the following theorem, represents consistent MSPT phases, in the sense that the corresponding density matrix $\rho$ is non-negative.

\begin{theorem}
    In $d$ spatial dimensions, bosonic MSPT phases protected by a full symmetry $G_d = \mathcal{G}\rtimes J$, where $\mathcal{G} = (K\times K) \times G$, are classified by
    \begin{equation}
        \bigoplus_{p=0}^d H^p[G,H^{d+1-p}(K,U(1))].
        \label{eq:MSPTclassification}
    \end{equation}
    \label{thm:MSPTclassification}
\end{theorem}
More precisely, Theorem \ref{thm:MSPTclassification} means that every class within the subgroup in Eq.(\ref{eq:MSPTclassification}) of $H^{d+1}(G_d,U(1))$ can be realized by some valid density matrix. It also means that, if $|\rho\rrangle$ is a generic SPT state whose invariant lies outside of this subgroup, the corresponding $\rho$ fails to meet the condition $\rho\ge 0$.

Before proving the theorem, let us give some physical interpretations.
\begin{enumerate}
    \item Despite the doubling of the exact symmetry $K$ when considering the Choi state, we will demonstrate that all potential MSPT phases protected jointly by the two copies of $K$ violate the requirement that the density operator in the original Hilbert space is non-negative.
    \item The modular conjugation, or equivalently hermiticity, does not introduce new MSPT phases without a pure state analog. Instead, as $J$ relates the two copies of the exact symmetry (by swapping them), in the classification it effectively reduces the doubled exact symmetry to a single one. 
    \item Eq.~(\ref{eq:MSPTclassification}) has a decorated domain wall interpretation \cite{2014decoreteddomainwall} as follows. A $G$-symmetric state $| \rho \rrangle$ is a superposition of domain wall (or in general, symmetry defect) configurations associated with $G$. On each codimension-$p$ defect we can decorate a $(d-p)$ dimensional SPT of the exact symmetry $K$, labeled by an element in the group cohomology $H^{d-p+1}(K,U(1))$. The decoration pattern is then labeled by an element in $H^p[G,H^{d+1-p}(K,U(1))]$.
    \item Eq.(\ref{eq:MSPTclassification}) represents a subgroup of $H^{d+1}(K\times G,U(1))$, which classifies pure SPT states protected by $K\times G$, both being exact symmetries. The only absent component in Eq.(\ref{eq:MSPTclassification}) describes states protected solely by the average symmetry. We will demonstrate that if such states existed, they would result in negative eigenvalues of $\rho$.
\end{enumerate}
In the following sections, we will establish the proof for Theorem~\ref{thm:MSPTclassification}. Our approach involves decoding the physical meaning of all potential SPT invariants associated with the Choi state, classified by $H^{d+1}(G_d,U(1))$, and examining whether a generic density operator with a specific invariant is consistent with non-negativity. In Theorem~\ref{thm:MSPTclassificationwithT} and Theorem~\ref{thm:MSPTclassificationgeneral} of Sec.~\ref{sec:generalsymmetries}, we will also extend this classification to cases involving time reversal symmetry and situations where the symmetry structure is not a simple direct product.

\subsection{MSPT phases protected by the exact symmetry}\label{subsec:exact}

In this section, we study MSPT phases protected by an exact symmetry $K$, which acts on the Choi state as a doubled symmetry $K_+\times K_-$, where $K_+$ ($K_-$) acts on the ket (bra) Hilbert space. We argue two points: First, in Sec.~\ref{subsec:exactonly} and Sec.~\ref{sec:higherdpositive}, we see that states protected \emph{jointly} by $K_+$ and $K_-$ cannot arise as Choi states of valid density matrices due to non-negativity. Second, in Sec.~\ref{sec:modconjsym}, we see that modular conjugation symmetry constrains the invariants of states protected solely by $K_+$ or $K_-$, so that effectively it is as if there is only one copy of $K$ protecting the state.

\subsubsection{MSPT phases with exact symmetries in $1d$}
\label{subsec:exactonly}

We begin by demonstrating, following the approach in Ref.~\cite{2017positivity}, that non-negativity $\rho \ge 0$ imposes a restriction on the SPT invariant of the Choi state $| \rho \rrangle$ -- intuitively, $| \rho \rrangle$ can not be an SPT protected ``jointly" by the two copies of the exact symmetry.

For the moment, focus on $|\rho \rrangle$ in $1d$. One useful characterization of SPT phases in $1d$ is the string order parameter \cite{20101dSPT,20121dSPT}, which consists of a symmetry operator on a large region, multiplied by two local endpoint operators. For our purpose, let us consider the following string order parameter,
\begin{equation}
    \begin{split}
        & s[(e,k);(\alpha,\alpha')] \\
        & = [(O_\alpha^l\otimes \mathbf{1}^n \otimes O_\alpha^r)\bigotimes(O_{\alpha'}^{l,T}\otimes (u_k^*)^n \otimes O_{\alpha'}^{r,T})],
    \end{split}
    \label{eq:stringorder}
\end{equation}
where $e$ is the identity element, and $k$ is an arbitrary element of $K$. The left and right endpoint operators, $O_\alpha^l$ and $O_\alpha^r$, live in opposite irreducible representations (symmetry charges) of $K$,
\begin{equation}
    \begin{split}
        & U_k^\dagger O_\alpha^l U_k = \chi_\alpha(k) O_\alpha^l, \\
        & U_k^\dagger O_\alpha^{r} U_k = \chi_\alpha(k)^* O_\alpha^r,
    \end{split}
    \label{eq:endpoint}
\end{equation}
and similarly for $O_{\alpha'}^{l,r}$. The length $n$ is taken to be much larger than the correlation length of $| \rho \rrangle$. For SPT states in $1d$, the string order parameter obeys a selection rule \cite{20121dSPTdetection}: If $| \rho \rrangle$ is a \emph{generic} $1d$ SPT state protected by the $K_+\times K_-$ symmetry, for each $k\in K$, one can always find a unique combination of charges $\alpha_k$ and $\alpha_k'$, such that the expectation value $\llangle \rho | s[(e,k);(\alpha,\alpha')] | \rho \rrangle$ is nonzero only when $\alpha=\alpha_k$ and $\alpha'=\alpha_k'$.

For clarity, let us denote the unique string order parameter which has long-range order as $s[(e,k);(\alpha_k,\alpha_k')] = S_1\otimes S_2^T$, where $S_1 = (O_{\alpha_k}^l\otimes \mathbf{1}^n \otimes O_{\alpha_k}^r)$ and $S_2^T=(O_{\alpha_k'}^{l,T}\otimes (u_k^*)^n \otimes O_{\alpha_k'}^{r,T})$ act on the the ket and the bra spaces, respectively. One has
\begin{equation}
    \begin{split}
        0  < & |\llangle \rho | S_1\bigotimes S_2^T | \rho \rrangle|^2 \\
       = & |\mathrm{tr}(\rho S_1 \rho S_2)|^2 \\
        \overset{(*)}{=} & |\mathrm{tr}(\sqrt{\rho} S_1\sqrt{\rho}\cdot \sqrt{\rho} S_2 \sqrt{\rho})|^2 \\
       \overset{(\dagger)}{\le} & \mathrm{tr}[ (\sqrt{\rho} S_1 \sqrt{\rho})(\sqrt{\rho} S_1 \sqrt{\rho})^\dagger ] \\ &
       \times \mathrm{tr} [ (\sqrt{\rho} S_2 \sqrt{\rho})(\sqrt{\rho} S_2 \sqrt{\rho})^\dagger ] \\
       = & \llangle \rho | (O_{\alpha_k}^l\otimes O_{\alpha_k}^r)\bigotimes (O_{\alpha_k}^l \otimes O_{\alpha_k}^r)^* | \rho \rrangle \\
       & \times \llangle \rho | S_2 \bigotimes S_2^* | \rho \rrangle,
    \end{split}
    \label{eq:positivity}
\end{equation}
where we use $\rho\ge 0$ to take the square root in $(*)$ and the Cauchy-Schwarz inequality in $(\dagger)$. In Eq.~(\ref{eq:positivity}), the state $|\rho\rrangle$ has been normalized to ensure a unit Hilbert-Schmidt norm. Since $| \rho \rrangle$ is SRE, the cluster decomposition property implies that $\llangle \rho | O_{\alpha_k}^l \bigotimes O_{\alpha_k}^{l,*} | \rho \rrangle \ne 0$, which can only be satisfied if $\alpha_k$ is the trivial $K$ representation.

The endpoint operator, acting in $\mathcal{H}\otimes \mathcal{H}^*$, does not necessarily have to be factorized as assumed in Eq.(\ref{eq:stringorder}). 
Nevertheless, any generic endpoint operator can be expressed as a sum $O^l_{(\alpha,\alpha')} = \sum_m O_\alpha^{l,m}\otimes O_{\alpha'}^{l,m,T}$. Crucially, this sum has a finite number of terms (due to the locality of the endpoint operator), and there must exist a term with an expectation value of $O(1)$, as required in Eq.(\ref{eq:positivity}), so the conclusion above holds.

This observation imposes a strong constraint on possible SPT invariants of $| \rho \rrangle$. For the moment, let us ignore the other symmetries and focus on the exact symmetry $K_+\times K_-$. All possible bosonic SPT phases of the pure state $|\rho\rrangle$ in $1d$ with $K_+\times K_-$ symmetry are classified by $H^2(K_+\times K_-,U(1))$, which can be decomposed using the Künneth theorem as
\begin{equation}\label{1dexactkuenneth}
\begin{split}
    &H^2(K_+\times K_-,U(1))=H^2(K_+,U(1)) \\
    &\oplus H^1[K_-,H^1(K_+,U(1))] \oplus H^2(K_-,U(1)).
\end{split}
\end{equation}
Here the term $H^1[K_-,H^1(K_+,U(1))]$ has a clear physical interpretation: A domain wall of $K_-$ (associated with a group element $k$) traps a charge of $K_+$ labeled by an element in $H^1(K_+,U(1))$. This charge is exactly the representation $\alpha_k$ of the endpoint operator, such that the string order parameter $s[(e,k);(\alpha_k,\alpha_k')]$ acquires a nonzero expectation value in $| \rho \rrangle$. Indeed, given an element $\omega \in H^2(K_+\times K_-,U(1))$ and a group element $k\in K_-$, the decomposition of $\omega$ in $H^1[K_-,H^1(K_+,U(1))]$ is given by the slant product of $\omega$ with $k$,
\begin{equation}
    \frac{\omega[(\bullet,e),(e,k)]}{\omega[(e,k),(\bullet,e)]} \in H^1(K_+,U(1)),
\end{equation}
which can be shown \cite{2017MPSRyu} to correspond to the $K_+$ charge, $\chi_{\alpha_k}(\bullet)$, associated with the nonzero string order parameter. As a conclusion, non-negativity requires that $| \rho \rrangle$ can not be described by any group cohomology class $\omega$, which has nontrivial decomposition in $H^1[K_-,H^1(K_+,U(1))]$. Below, in Sec.~\ref{sec:modconjsym}, we further argue that modular conjugation symmetry enforces a relation between the $H^2(K_+,U(1))$ and $H^2(K_+,U(1))$ terms of Eq.~\ref{1dexactkuenneth}, implying that the total classification of SPT phases protected by $K_+\times K_-$ symmetry that can arise as Choi states of valid density matrices is given by a single copy of $H^2(K,U(1))$.

\subsubsection{Higher dimensions: the decorated domain wall approach}
\label{sec:higherdpositive}

In order to extend the discussion above to higher space dimensions, let us generalize the decorated domain wall approach \cite{2014decoreteddomainwall}, a systematic construction for pure state SPT phases, to the study of mixed states. We first review the idea of constructing pure SPT states by this approach. To construct a $d$-dimensional $K_+ \times K_-$-symmetric state (here one can simply view $K_+$ and $K_-$ as some symmetry of a pure state), we can start from a state in which $K_-$ is broken spontaneously. The $K_-$ symmetry can be restored by condensation of $K_-$-domain walls. Nontrivial $K_+\times K_-$ SPT states are produced by decorating codimension-$p$ symmetry defects of $K_-$ with $(d-p)$-dimensional SPT states protected by the unbroken symmetry $K_+$. Mathematically, this construction is formulated precisely by the Künneth formula: one may decompose the group cohomology classification of $K_+\times K_-$ bosonic SPT phases as follows,
\begin{equation}
    \begin{split}
      H^{d+1}(K_+\times K_-,U(1)) = \bigoplus_{p=0}^{d+1}H^p[K_-,H^{d+1-p}(K_+,U(1))].  
    \end{split}
    \label{eq:kunneth}
\end{equation}
In particular, for $0<p<d+1$, Eq.~(\ref{eq:kunneth}) implies that possible decoration patterns which are consistent with fusion rules of codimension-$p$ symmetry defects of $K_-$ are labeled by elements in $H^p[K_-,H^{d+1-p}(K_+,U(1))]$. Hereafter we refer to SPT states with such nontrivial decorations as states protected \emph{jointly} by $K_+$ and $K_-$. In contrast, the $p=0$ term [$H^{d+1}(K_+,U(1))$] and the $p=d+1$ term [$H^{d+1}(K_-,U(1))$] label SPT states protected \emph{solely} by $K_+$ or $K_-$, respectively.

A consequence of the decorated domain wall approach is that, string order parameters in $1d$ may be generalized to ``defect order parameters" in higher dimensions. Notice that the decorated domain wall construction can be equivalently formulated as follows: consider an SPT state $| \rho \rrangle$, and act on it with a symmetry element $k\in K_-$, but only in a large but finite region $A$: $U_k^{A}\equiv\otimes_{i\in A}u_k^{i}$. Since the Choi state $|\rho \rrangle$ is simply a symmetric SRE pure state, acting with $U_k^{A}$ has nontrivial effect only near the boundary of $A$ \cite{20101dSPT,2011symlocal}:
\begin{equation}
    I \bigotimes (U_k^A)^*|\rho\rrangle=V_k^{\partial A}|\rho\rrangle,
    \label{fractionalization}
\end{equation}
where $V_k^{\partial A}$ is a finite-depth unitary operator that is nontrivial (non-identity) only near the boundary $\partial A$. For an SPT state $| \rho \rrangle$ protected jointly by $K_+\times K_-$, the decorated domain wall construction implies that $V_k^{\partial A}$ creates a $K_+$ SPT phase in one dimension lower (whose invariant is given by the $k$-slant product of the original SPT invariant), according to the decoration pattern labeled by an element in $H^1[K_-,H^d(K_+,U(1))]$. 

One can introduce a defect order parameter as follows: 
\begin{equation}
    s_k^A = (O^{\partial A}_{+})^\dagger\bigotimes (U_k^A)^*,
    \label{defdefectorder}
\end{equation}
where $O_+^{\partial A}$ is a finite-depth unitary along the boundary of $A$, operating \emph{solely} in the $+$ Hilbert space to create a $K_+$ SPT. Similarly to the $1d$ case, there is a unique choice of the defect order parameter (up to symmetric finite-depth local unitary), where $O^{\partial A}_{+}$ carries the same SPT invariant as $V_k^{\partial A}$, that results in an area law decaying expectation value, \ie $\llangle \rho | s_k^A | \rho \rrangle \sim e^{-| \partial A |}$. In contrast, if $O_+^{\partial A}$ has a different SPT invariant compared to $V_{k,+}^{\partial A}$, the defect order parameter vanishes, \ie $\llangle \rho | s_k^A | \rho \rrangle = 0$.

\begin{figure}
\begin{center}
  \includegraphics[width=.40\textwidth]{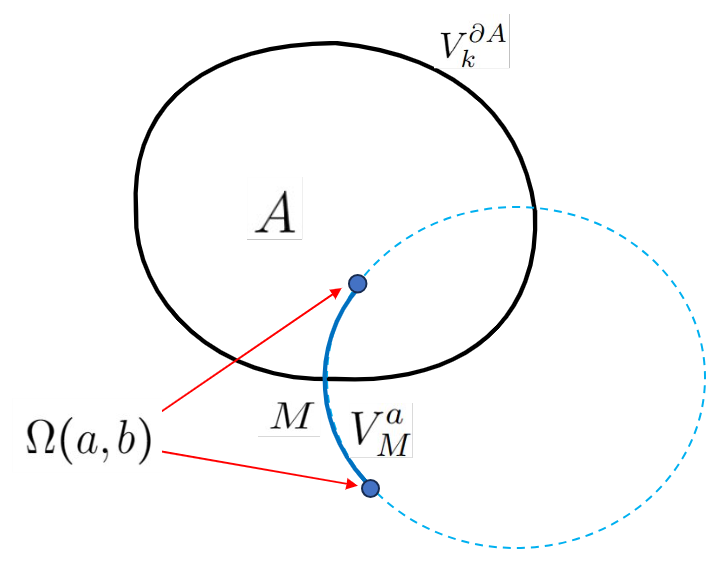} 
\end{center}
\caption{
Due to symmetry localization, as described in Eq. \ref{fractionalization}, one can further restrict the boundary unitary associated with a group element $a$ to an open segment $M$, referred to as $V_M^a$. The set $\{ V_M^a \}$ obeys the group multiplication rule up to a local operator $\Omega(a,b)$ near $\partial M$. 
}
\label{Fig:flux}
\end{figure}

Let us delve into the scaling behavior of the defect order parameter a bit more. For simplicity, we consider the case of $d=2$, though the analysis can be extended to higher dimensions. With symmetry fractionalization as described in Eq. (\ref{fractionalization}), the defect order parameter simplifies to:
\begin{equation}
    \llangle \rho | s_k^A | \rho \rrangle = \llangle \rho | (O_+^{\partial A})^\dagger V_k^{\partial A} | \rho \rrangle.
    \label{SPTentangler}
\end{equation}
Naively, given that the operator $W_{\partial A} =(O_+^{\partial A})^\dagger V_k^{\partial A}$ is localized around $\partial A$, one might expect the defect order described in Eq.(\ref{SPTentangler}) to possess a nonzero value for a generic SRE state $\rho$, decaying in accordance with the boundary area of $\partial A$. However, if the unitary $W_{\partial A}$ generates a \emph{nontrivial} $1d$ $K_+$ SPT along $\partial A$, a nonzero defect order parameter would result in a contradiction. To illustrate, consider that for any arbitrary symmetry group element $a\in \mathcal{G}$, we can further restrict the boundary operator as defined by Eq.(\ref{fractionalization}) to a segment $M$ of $\partial A$ \cite{2014dElse}, producing a unitary operator $V_M^a$, as depicted in Fig \ref{Fig:flux}. Physically, $V_M^a$ introduces an $a$ symmetry flux around each endpoint of $M$.

A crucial property of the operators set $\{ V_M^a \}$ with $a\in \mathcal{G}$, is that they obey the group multiplication rule, up to unitary operators $\Omega(a,b)$ supported in the vicinity of $\partial M$ \cite{2014dElse}. In $2d$, $\Omega(a,b)$ is merely point-like.
\begin{equation}
    V_M^a V_M^b = \Omega(a,b) V_M^{ab},
    \label{eq:elsenayak}
\end{equation}
as illustrated in Fig \ref{Fig:flux}. If we take $a$ and $b$ as two group elements in $K_+$, Eq.(\ref{eq:elsenayak}) would yield the following outcome:
\begin{equation}
    \begin{split}
            \llangle \rho | s_k^A | \rho \rrangle & = \llangle \rho | W_{\partial A} | \rho \rrangle \\
             & =  \llangle \rho | W_{\partial A} V_M^a V_M^b [\Omega(a,b) V_M^{ab}]^\dagger | \rho \rrangle \\
            &  =  e^{i\omega(a,b)}\llangle \rho | V_M^a V_M^b [\Omega(a,b) V_M^{ab}]^\dagger W_{\partial A}  | \rho \rrangle \\
            & = e^{i\omega(a,b)}\llangle \rho | s_k^A  | \rho \rrangle.
            \label{2dselection}
    \end{split}
\end{equation}
Here, one endpoint of $M$ lies deep within region $A$ while the other lies deep outside $A$. In the third line of Eq.(\ref{2dselection}), we use the fact that the commutation of three $K_+$ symmetry fluxes, associated with $a$, $b$, and $(ab)^{-1}$ respectively, with the boundary operator $W_{\partial A}$ results in a phase factor $\omega(a,b)$. This phase factor is precisely determined by the cohomology $H^2(K_+,U(1))$ of the $1d$ $K_+$ SPT created by $W_{\partial A}$. If $W_{\partial A}$ belongs to a nontrivial topological class, $\omega(a,b)$ cannot be trivial for all $a$ and $b$ in $K_+$, which leads to the defect order parameter being zero. Hence, in order for the defect order parameter to not vanish, $O_+^{\partial A}$ must share the same SPT invariant as $V_k^{\partial A}$, which is determined by the decoration pattern $H^1[K_-,H^2(K_+,U(1))]$ of the state $|\rho\rrangle$. 

To extend the selection rule to higher (co-)dimensions, one can use the restriction procedure described in Ref. \cite{2014dElse}. In the $p$-th step of this restriction, one obtains an operator $V_M(a_1,...,a_p)$ indexed by $p$ group elements of $K_+$, acting on a $d-p$ dimensional subregion of the space. The different orders of fusing these operators, in the presence of a defect-creating operator in the $K_-$ Hilbert space that generates a $p$-dimensional symmetry defect (or multi-defect junction) of $K_-$, result in a phase factor determined by the decoration pattern $H^{d-p}[K_-,H^{p+1}(K_+,U(1))]$. The defect order parameter is nonzero only when the boundary unitary $O_+$ in Eq. (\ref{defdefectorder}) (acting in the $+$ Hilbert space) creates an $K_+$ SPT state that matches the decoration pattern of $|\rho\rrangle$. This argument establishes the selection rule in each (co-)dimension \cite{2021decoratedDW}.

Employing the same reasoning as in Eq.~(\ref{eq:positivity}), one can see that a nonzero defect order parameter (taking the linear size of $A$ to be larger than the correlation length of $|\rho\rrangle$ but finite) results in $\llangle \rho | (O_+^{\partial A})^\dagger \bigotimes (O_+^{\partial A})^T | \rho \rrangle \ne 0$. This implies that $O_+^{\partial A}$ hence $V_k^{\partial A}$, can only create a trivial $K_+$ SPT state along $\partial A$. Consequently, non-negativity of $\rho$ requires that the domain wall decoration of $| \rho \rrangle$ is trivial, \ie $| \rho \rrangle$ cannot be an SPT state protected jointly by both $K_+$ and $K_-$ described by a nontrivial element in $H^1[K_-,H^2(K_+,U(1))]$. Extending the analysis to higher (co-)dimensions using the selection rule of defect order parameters, we conclude that a Choi state $| \rho \rrangle$ cannot be an SPT state protected jointly by both $K_+$ and $K_-$.

\subsubsection{Imposing modular conjugation symmetry}\label{sec:modconjsym}

So far we have focused on states jointly protected by $K_+\times K_-$ and found they do not correspond to genuine MSPT phases. On the other hand, SPT states in the double space protected solely by either $K_+$ or $K_-$ have not presented any contradictions. As a result, for generic states, $K_+\times K_-$ MSPT states consistent with non-negativity are classified by pairs
\begin{equation}
    (\alpha,\beta)\in H^{d+1}(K_+,U(1))\oplus H^{d+1}(K_-,U(1)).
\end{equation}

Now, let us incorporate the modular conjugation operator $J$. $J$ serves two distinct roles in the classification of MSPT phases. Firstly, it introduces a new requirement that a state $| \rho \rrangle$ must be $J$-symmetric. Secondly, $J$ may protect new MSPT states that lack a pure state analog. In this section, we will focus on the condition that $| \rho \rrangle$ is $J$-symmetric; we will save the second point for Sec.~\ref{subsec:Jsymmetry}. As detailed in Section \ref{sec:symmetries}, the exact symmetry and the modular conjugation together form a semi-direct (wreath) product denoted as $(K_+\times K_-)\rtimes J$. In the Choi representation, $J$ exchanges the quantum states between the bra and the ket, followed by a conplex conjugation. Consequently, $J$ imposes a constraint on the classes $\alpha$ and $\beta$:
\begin{equation}
(\alpha,\beta) \overset{J}{\to}  (\Bar{\beta},\Bar{\alpha}) 
=(\alpha,\beta).
\label{eq:Kclassofbraandket}
\end{equation}
Here, $\Bar{\alpha}$ and $\Bar{\beta}$ represent the complex conjugated classes due to the antiunitary nature of $J$. We therefore can deduce that $\alpha = \Bar{\beta}$. In conclusion, in $d$ spatial dimensions, the classification of MSPT phases protected by the exact $K\times K$ symmetry is isomorphic to $H^{d+1}(K,U(1))$, which can be represented by a single cohomology class $\alpha$.

From a mathematical perspective, $d$-dimensional SPT phases with $(K\times K)\rtimes J$ symmetry are classified by the group cohomology $H^{d+1}[(K\times K)\rtimes J,U(1)]$, with the $U(1)$ coefficient twisted by $J$. This classification can be decomposed using the Leray-Serre spectral sequence \cite{2019gaiottospectralsequence,2021decoratedDW}, with the $E_2$ page comprising:
\begin{equation}
    \bigoplus_{p+q=d+1} E_2^{p,q} = \bigoplus_{p+q=d+1} H^p[J,H^q(K\times K,U(1))].
    \label{eq:E2page}
\end{equation}
In a physical context, the term with $p=0$ and $q=(d+1)$ corresponds precisely to $J$-symmetric $K\times K$ MSPT phases discussed in the last paragraph. We have demonstrated that a subgroup of it, which is isomorphic to $H^{d+1}(K,U(1))$, complies with the condition $\rho \ge 0$. The other terms with $p>0$ would represent SPT phases protected solely by $J$ or jointly by both $J$ and the exact symmetry. These potential phases, if they exist, would be SPT phases that lack a pure state counterpart. We will explore these possibilities in the following section.

\subsection{SPT phases protected by the modular conjugation}
\label{subsec:Jsymmetry}

In this section, we demonstrate that the modular conjugation $J$, or equivalently hermiticity of $\rho$, does not introduce new SPT phases without a pure state counterpart. More precisely, we will establish that in Eq.~(\ref{eq:E2page}), all terms with $p>0$ will not give rise to nontrivial MSPT states that meet the condition $\rho \ge 0$.

SPT phases protected by an antiunitary symmetry (such as $J$ and $\mathcal{T}$, which will be discussed in Sec.~\ref{sec:generalsymmetries}) can also be understood in terms of the decorated domain wall approach \cite{2017partialtranspose,2018partialtranspose}. Similar to the unitary cases, a $J$-symmetric state $| \rho \rrangle$ is a quantum superposition (condensation) of all possible $J$-domain wall configurations. In the presence of the exact unitary symmetry, a nontrivial SPT phase may be achieved if, on each codimension-$p$ domain wall (or multi-domain wall junction), we decorate a $(d-p)$ dimensional SPT phase protected by the exact symmetry. Crucially, for the superposition of $J$ domain walls to satisfy non-negativity, the following consistency conditions must be met:
\begin{enumerate}
    \item $J$ defects in each codimension must not be decorated with $K\times K$ SPT phases that violate non-negativity on their own.
    \item Within the superposition, the relative phases between $J$ defect configurations must not violate non-negativity.
\end{enumerate}

Let us clarify these consistency conditions and give a physical interpretation of Eq.~(\ref{eq:E2page}). In the doubled space, a representative wave function for an SPT phase protected by $(K\times K)\rtimes J$ symmetry can be expressed as
\begin{equation}
| \rho \rrangle = \sum_D \sqrt{p_D} e^{i\theta_D} | \rho_D \rrangle | a_D^J\rrangle,
\label{eq:wavefunction}
\end{equation}
Here, $D$ runs over networks of $J$ symmetry defects, and $| a_D^J \rrangle$ resides in an ancilla space $\mathcal{A}$, acted upon by $J$ but not $K\times K$, that describes these networks. The state $|\rho_D \rrangle$ resides on a sub-lattice subjected to the action of $K\times K$ (referred to as ``the decoration Hilbert space"), representing the $K\times K$ SPT state decorated on the defect network $D$. Possible $J$-symmetric decoration patterns on codimension-$p$ defects are labeled by elements in $H^p[J,H^{d-p+1}(K\times K,U(1))]$ for $p\le d$ (note that $J$ acts nontrivially on the coefficient group). For each $D$, the defect network $| a_D^J \rrangle$ is $J$-symmetric, as it is itself a superposition of a $J$-domain configuration and its $J$-conjugate, as in Eq.~(\ref{eq:Jwavefunction}). The phase factors $\theta_D$ encode the $H^{d+1}[J,U(1)]$ ($p=d+1$) invariant. Since each $|\rho_D\rrangle$ and $|a_D^J\rrangle$ are $J$-symmetric, each of the phase factors $e^{i\theta_D}$ is a sign $\pm 1$. Upon superposing all $J$ defect configurations, we obtain an SPT state protected by $(K\times K)\rtimes J$ \cite{2021decoratedDW}.

Let us look at a simple example (where the decorations are trivial) to understand a $J$-symmetric state as a superposition of $J$-defect configurations. Suppose the ancilla space $\mathcal{A}$ comprises a qubit on each lattice site, prepared in a product state $| a^J \rrangle \sim \otimes_i | +,+ \rrangle_i$. Here, $| + \rangle = \frac{1}{\sqrt{2}}(| 0 \rangle + |1\rangle)$ is an eigenstate of the Pauli-$X$ operator. This $J$-symmetric state can be alternatively expressed as $| a^J \rrangle \sim  \sum_{\{ A \}} |a_A^J\rrangle$, where $|a_A^J\rrangle =  [\otimes_{i\in A} (|0,0\rrangle_i+|1,0\rrangle_i)]\otimes[\otimes_{j\in \overline{A}} (|1,1\rrangle_j+|0,1\rrangle_j)]$. The set $\{ A \}$ labels different ways to bipartition the space, and corresponds to a configuration $D$ of domain walls. For each bipartition, the state $|a_A^J\rrangle$ breaks the $J$ symmetry (with a local order parameter $\sigma^z_{bra}-\sigma^z_{ket}$) within the region $A$ and its complement $\overline{A}$. After summing over all bipartitions, the $J$ symmetry is restored. A $J$ defect network state $|a_D^J\rrangle$ in Eq.~({\ref{eq:wavefunction}}) corresponds precisely to the superposition of a specific bipartition and its $J$ conjugate, \ie $|a_D^J\rrangle = |a_A^J\rrangle + J|a_A^J\rrangle$. 
Under the operator-state map, the state in Eq.(\ref{eq:wavefunction}) maps to the density matrix in the original Hilbert space given by
\begin{equation}
\rho = \sum_D \sqrt{p_D} e^{i\theta_D} \rho_D\otimes a_D^J.
\label{eq:repdensitymatrix}
\end{equation}
Specifically, the quantum superposition over defect configurations in the doubled space transforms into a convex sum in the original density matrix.\footnote{In the presence of a nontrivial decoration, within a generic $| \rho\rrangle$, the states in the ancilla space and the decoration space may not factorize as a tensor product, as shown in Eq. (\ref{eq:wavefunction}) and Eq. (\ref{eq:repdensitymatrix}). These expressions represent the fixed-point state when nontrivial decorations are present, as illustrated in the explicit construction of group cohomology fixed points in Ref.~\cite{2013chenSPT}.}

We can then extract information about the eigenvalues of $\rho$ in the expression in Eq.~(\ref{eq:repdensitymatrix}). Two observations are important for our subsequent analysis: (1) By definition, distinct domain wall configurations possess orthogonal supports. Each configuration, $a_D^J$, is supported within a sub-Hilbert space labeled by the positions of the domain walls.\footnote{This holds exactly for ``sharp" domain wall configurations -- when $a_D^J$ is a diagonal matrix in a basis describing the position of $J$ domain walls. One can always obtain this property for a generic mixed state by passing it through a maximal dephasing channel (which crucially preserves non-negativity if the initial state is non-negative) to make the domain walls sharp. For instance, if the $J$ order parameter is the Pauli spin $\sigma^z_{bra}-\sigma^z_{ket}$, one may choose the dephasing channel as: 
\begin{equation}
\begin{split}
    \mathcal{E} & = \prod_{<ij>}\mathcal{E}_{ij},\\
    \mathcal{E}_{ij}(\rho)&=\frac{1}{2}[\rho+\sigma^z_i\sigma^z_j\rho\sigma^z_i\sigma^z_j],
\end{split} 
\end{equation}
where $<ij>$ runs over all nearest neighbors.}
Consequently, the eigenvalues of $\rho$ are simply a collection of eigenvalues obtained from each summand in Eq.~(\ref{eq:repdensitymatrix}). 
(2) A trivial $J$ SPT state can always be realized by a valid density matrix consistent with non-negativity. An example is $|a^J\rrangle$ mentioned in the preceding paragraph (which is a product state in the doubled space).

Now one can identify several places where a violation of non-negativity may originate. The decorations on $J$-defects, or equivalently the $E_2$ page as defined in Eq.~(\ref{eq:E2page}) with $p>0$, can be further categorized into the following three groups.

(1) The first is the term with $p=d+1$ in Eq.(\ref{eq:E2page}), given by $H^{d+1}(J,U(1))$, representing states that are protected solely by the modular conjugation $J$. Since the decorations $\rho_D$ are trivial, the topology is entirely a function of the phase factors $e^{i\theta_D}$. From Eq.~(\ref{eq:repdensitymatrix}), it becomes evident that only the trivial phase factor, where $e^{i\theta_D}=1$ for any defect configuration $D$, is consistent with the non-negativity constraint of $\rho$.

Let us make this point more precise. The representative wave function of a $J$ SPT state, characterized by an element in $H^{d+1}(J,U(1))$, has the form
\begin{equation}
\begin{split}
| \rho \rrangle & \sim \sum_{ \{ D_1 \} } | a_{D_1} \rrangle + J | a_{D_1} \rrangle \\ & \pm \sum_{ \{ D_2 \} } ( | a_{D_2} \rrangle + J | a_{D_2} \rrangle ).
\end{split}
\label{eq:Jwavefunction}
\end{equation}
The subsets $\{ D_1 \} $ and $ \{ D_2 \} $ encompass all $J$ domain wall configurations. These domain wall configurations are structured in a way that any configuration within ${ D_2 } $ can be reached from a configuration within ${ D_1 }$ through an odd number of ``F-moves" \cite{2010localunitary}, and vice versa. Conversely, two configurations in the same set can always be transformed into one another through an even number of ``F-moves." The crucial $\pm$ signs are associated with a trivial and a nontrivial $J$ SPT state, respectively. It is worth noting that $H^{d+1}(J,U(1))$ is either trivial (for even values of $d$) or $\Z_2$ (for odd values of $d$), so the relative phase between the two sets of configurations can, at most, be a minus sign. The density matrix associated with the Choi state in Eq.~(\ref{eq:Jwavefunction}) takes the form:
\begin{equation}
\rho \sim \sum_{ \{ D_1 \} } a_{D_1}^J \pm \sum_{ \{ D_2 \} } a_{D_2}^J,
\end{equation}
In a nontrivial $J$ SPT, $a_{D_1}^J$ and $a_{D_2}^J$ cannot have non-negative coefficients simultaneously. If one assumes the associated trivial $J$ SPT is non-negative, $a_{D_1}^J$ ($a_{D_2}^J$) represents a Hermitian, non-negative density matrix describing a $J$ defect configuration $D_1$ ($D_2$). Consequently, non-negativity prohibits a nontrivial $J$ SPT phase, which is described by a nontrivial element in $H^{d+1}(J,U(1))$, from being a valid MSPT state.

In fact, in $1d$, further assuming translation invariance, this first scenario can be understood by investigating the local tensor of the MPDO. Let us ``gauge" the $J$ symmetry, and examining the composition rule of $J$ gauge fluxes \cite{2015gaugingT}. Although $J$ is antiunitary, there is a well-defined procedure to gauge $J$ that applies to a SRE $| \rho \rrangle$ which admits a tensor network representation. By definition, $J$ acts on the local tensor by
\begin{equation}
    J: \, A^{pq} \to (A^{qp})^*.
\end{equation}
If $| \rho \rrangle$ is $J$ symmetric and SRE, one can show that \cite{2020mpsreview}
\begin{equation}
    (A^{qp})^* = V_J^\dagger A^{pq} V_J,
    \label{eq:Jtransformation}
\end{equation}
where $V_J$ is a unitary matrix acting on the entanglement Hilbert space and can be interpreted as a $J$ symmetry flux. In particular, in $1d$ the $J$ SPT invariant can be read out from $V_J$ by
\begin{equation}
    V_J V_J^* = \pm 1,
    \label{eq:Jprojective}
\end{equation}
where $+1$ corresponds to the trivial SPT and $-1$ corresponds to the nontrivial one. If we assume that the tensor $A^{pq}$ generates a density operator with unit trace for all system sizes $L\gg 1$, then we have $\tr(\rho) = \tr [(\sum_p A^{pp})^L] = 1$ for all $L \gg 1$. This straightforwardly implies that the spectrum of the matrix $E^1_{\alpha\beta} = \sum_p A^{pp}_{\alpha\beta}$ contains a one and the rest of the eigenvalues have magnitudes smaller than 1, \ie $E^1 = W \mathrm{diag}(1,\Lambda_2,\Lambda_3...)W^{-1}$, where $\Lambda_{i>1}$ is the Jordan block with diagonal elements being $\lambda_i$ ($|\lambda_{i>1}|<1$), and $W$ is an invertible matrix. Refer to Sec. \ref{sec:SSB} for a physical interpretation of this spectral property as the absence of spontaneous breaking of the average symmetry. Based on Eq.(\ref{eq:Jtransformation}), we have
\begin{equation}
    [\mathrm{diag}(1,\Lambda_2,\Lambda_3...),\Tilde{V}_J] = 0,
    \label{eq:blockdiagonal}
\end{equation}
where $\Tilde{V}_J = W^{-1} V_J W^*$. This implies that $\Tilde{V}_J$ is block diagonal, with all elements in the first column and first row, except $[\Tilde{V}_J]_{11}$, are zero. It follows that $V_JV_J^*=\Tilde{V}_J\Tilde{V}_J^*=1$, ruling out the nontrivial class in Eq.~(\ref{eq:Jprojective}).

(2) Charge decorations: $0d$ point-like $J$ defects in $a_D^J$ may trap $K\times K$ charges, corresponding to the $p=d$, $q=1$ term in Eq.~(\ref{eq:E2page}). The discussion in Sec. \ref{subsec:exactonly} regarding non-negativity imposes no restrictions on charge decorations. According to the Künneth theorem, the complete set of charge decorations is classified by $H^1(K\times K,U(1))= H^1(K,U(1))\oplus H^1(K,U(1))$, which labels charges in the ket and bra spaces. Note that the first two cases (solely $J$ protected phases and charge decorations) encompass all possibilities for $1d$.

(3) Decorations in one or higher dimensions: These scenarios involve decorated states on the condensed $J$ domain walls in one dimension or higher. Mathematically, they correspond to terms in Eq.~(\ref{eq:E2page}) with $0<p<d$ (which only exist for $d>1$).
Here, non-negativity constrains possible decorations on $J$ defects. For clarity, we first illustrate this constraint using fixed-point wavefunctions \cite{2013chenSPT}. In this scenario, for each defect configuration in Eq.(\ref{eq:repdensitymatrix}), the decorated state $\rho_D$ is strictly confined to the defect. Additionally, for each $D$, the decoration and the defect network reside on different sub-lattices and factorize as a tensor product. As discussed in Sec.~\ref{subsec:exactonly}, decorations protected by both $K$ symmetries jointly would lead to $\rho_D$ with negative eigenvalues, and $\rho$ will inherit negative eigenvalues from the decoration $\rho_D$, as the spectrum of the tensor product matrices is simply the product of spectra: $\mathrm{Spec}(\rho_D\otimes a_D^J) = \mathrm{Spec}(\rho_D)\otimes \mathrm{Spec}(a_D^J)$. Consequently, in a fixed-point state, it is only allowed to decorate defects with phases protected solely by one of the two $K$ symmetries.

Now, let us demonstrate that the same conclusion holds for generic wavefunctions. Compared with the fixed-point cases discussed in the previous paragraph, a generic state presents two complications: (1) The state in the ancilla space $\mathcal{A}$ and in the decoration space may not factorize. Unlike Eq.(\ref{eq:repdensitymatrix}), a generic density matrix should be expressed as
\begin{equation}
\rho = \sum_D \widetilde{\rho}_D^J,
\label{highergeneric}
\end{equation}
where $\widetilde{\rho}_D^J$ describes a defect configuration $D$ along with its decoration, operating on both the ancilla and decoration spaces. (2) The decoration may not be strictly confined to the defect but can spread into the bulk by a correlation length $\xi$ of $|\rho \rrangle$. It is not obvious whether one can deduce the spectrum of $\rho$ solely from its decorations on lower dimensional subregions.

\begin{figure}
\begin{center}
  \includegraphics[width=.40\textwidth]{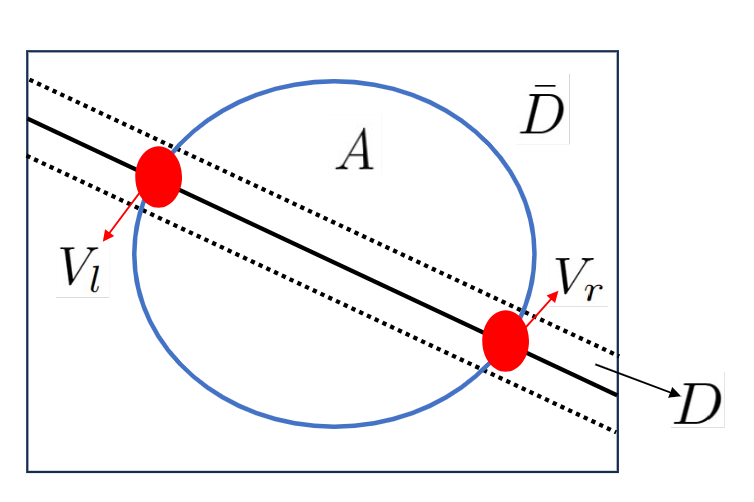} 
\end{center}
\caption{
A specific defect configuration $D$, where we abuse the notation by denoting the region within a distance $\xi$ to the defect as $D$ (inside the dotted lines), while its complement is denoted as $\Bar{D}$. The two local operators $V_l$ and $V_r$ are supported near the left and right endpoints of $D \cap \partial A$, respectively.
}
\label{Fig:defect}
\end{figure}

However, note that the observation that states with distinct defect configurations are supported on orthogonal subspaces still holds. Hence, $\rho$ is non-negative if and only if each $\widetilde{\rho}_D^J$ is non-negative. Let us examine $\widetilde{\rho}_D^J$ associated with a specific defect configuration $D$. We first pass $\widetilde{\rho}_D^J$ through a strongly symmetric measurement channel (see Fig. \ref{Fig:defect}),
\begin{equation}
\mathcal{M}(\widetilde{\rho}_D^J) = \sum_x\tr_{\Bar{D}}(\widetilde{\rho}_D^J | x\rangle \langle x|)| x\rangle \langle x|,
\label{measurementchannel}
\end{equation}
where the orthonormal basis ${ | x \rangle }$, defined in both the decoration space (where the $K \times K$ symmetry acts) and the ancilla, in the spatial region $\bar{D}$ (away from the defect $D$) consists of $K$ symmetric product states. For example, in the case where $K = \mathbb{Z}_2$, ${ | x \rangle }$ represents product states on $\bar{D}$, where each site can independently be either $| + \rangle$ or $| - \rangle$ in the decoration space. The measurement channel is completely positive, so $\mathcal{M}(\widetilde{\rho}_D^J)$ must be positive if $\widetilde{\rho}_D^J$ is. Moreover, since ${ | x \rangle }$ are mutually orthogonal, $\mathcal{M}(\widetilde{\rho}_D^J)$ is positive if and only if each component $\rho_x = \text{tr}_{\bar{D}}(\widetilde{\rho}_D^J | x \rangle \langle x |) | x \rangle \langle x |$ is positive.

Now we demonstrate that $\rho_x$ cannot be positive when the decoration on $D$ is an SPT protected jointly by the two copies of $K$. For clarity, we focus on a $1d$ defect within a $2d$ bulk, see Fig \ref{Fig:defect}. A key observation is that $| \rho_x \rrangle=\llangle x|\tilde\rho_D^J\rrangle|x\rrangle$ effectively represents a $1d$ $K_+\times K_-$ SPT near the defect $D$, with coordinates perpendicular to the defect viewed as internal indices. To substantiate this assertion, consider applying a symmetry operator associated with an element $k$ of, say, $K_-$ to a large but finite region $A$, as depicted in Fig \ref{Fig:defect}:
\begin{equation}
    \begin{split}
        I \bigotimes (U_k^A)^* | \rho_x \rrangle & = \llangle x |(U_k^D)^*| \widetilde{\rho}_D^J \rrangle (U_k^{\Bar{D}})^*|x\rrangle \\
        & = \llangle x |(U_k^{\Bar{D}}\otimes U_k^D)^*| \widetilde{\rho}_D^J \rrangle |x\rrangle \\
        & = \llangle x |V_k^{\partial A} | \widetilde{\rho}_D^J \rrangle |x\rrangle \\
        & = V_l V_r \llangle x | \widetilde{\rho}_D^J \rrangle |x\rrangle = V_l V_r | \rho_x\rrangle,
        \label{defectreduction}
    \end{split}
\end{equation}
where $| x\rrangle$ represents the Choi state of $| x\rangle \langle x |$. In Eq. \ref{defectreduction}, we employed the following observations: (1) In the first line, $U_k^A = U_k^D \otimes U_k^{\Bar{D}}$ is a product of symmetry operators in $D$ and $\Bar{D}$, with the first factor hitting the state $\llangle x|\tilde\rho_D^J\rrangle$ in the $D$ Hilbert space and the second hitting the state $|x\rrangle$ in the $\bar D$ Hilbert space; (2) In the second line, the symmetry of the states $|x\rangle$ is used; (3) In the third line, since each $J$ symmetry-breaking pattern (\ie a configuration of $J$ defect) in the superposition in Eq. \ref{highergeneric} is $K\times K$ symmetric and SRE \cite{2021decoratedDW}, the effect of the symmetry $U_k^A$ on $| \widetilde{\rho}_D^J \rrangle$ is captured by a finite-depth unitary $V_k^{\partial A}$ supported near $\partial A$. (4) In the final line, upon pulling $V_k^{\partial A}$ outside of $|x\rrangle$, it becomes an operator $\tilde V$ supported on $D$. Since $|\widetilde{\rho}_D^J \rrangle$ is a trivial $K\times K$ SPT away from $D$ (which implies that strange correlators between $\llangle x|$ and $|\tilde\rho_D^J\rrangle$ decay exponentially \cite{2014strange}), the fact that this operator is supported near $\partial A$ remains true and is not altered by the measurement channel. As a result, the $\tilde V$ is supported near $D\cap\partial A$ and therefore takes the form $V_lV_r$.

When the decoration on the defect $J$ belongs to an SPT state protected jointly by the two copies of $K$, the local operators $V_l$ and $V_r$ must carry a nontrivial $K_+$ charge, as dictated by the element of $H^1[K_-,H^1(K_+,U(1))]$ describing the SPT. Therefore, a generalized string order parameter
\begin{equation}
  \llangle \rho_x |   s_k^A | \rho_x \rrangle = \llangle \rho_x | (O_l O_r)^\dagger\bigotimes (U_k^A)^* |\rho_x \rrangle
\end{equation}
is nonzero if and only if $O_l$ ($O_r$) carries the same $K_+$ charge as $V_l$ ($V_r$). Employing the Cauchy-Schwarz inequality in a same manner as in Eq.(\ref{eq:positivity}), we arrive at the final conclusion that non-negativity prohibits the decoration from being an SPT state protected jointly by the two copies of $K$.
The same argument also applies to higher (co-)dimensions by considering defect order parameters.

In physical terms, the last two categories described above correspond to MSPT phases that are jointly protected by $J$ and the exact symmetry. In both scenarios, potential decorations of $J$ defects consistent with non-negativity are given by
\begin{equation}
H^p(J,M^p\oplus M^p)\subset E_2^{p,d-p+1}, \, 0<p<d+1.
\end{equation}
Here, $M^p = H^{d-p+1}(K,U(1))$, in which $J$ acts on the coefficient group by swapping the two $M^p$ components, combined with a complex conjugation on the $U(1)$ coefficient. It can be mathematically demonstrated that such cohomology vanishes for any $p>0$ \cite{2023coho,2023groupcoho}. In conclusion, there are no SPT phases that are jointly protected by both $J$ and the exact symmetry while remaining consistent with non-negativity.

In summary, this section has demonstrated that SPT phases in the Choi representation, protected solely by $J$ or jointly by $J$ and the exact symmetry, do not qualify as legitimate MSPT states due to non-negativity of $\rho$. Combined with the discussion in Sec.~\ref{sec:higherdpositive}, we can conclude that consistent MSPT states protected by an exact symmetry $K$ are classified by $H^{d+1}(K,U(1))$. This classification coincides with that of pure state SPT phases with symmetry $K$.


\subsection{MSPT phases with average symmetries}
\label{subsec:Avesym}

In this section, we extend our discussion to include the average symmetry $G$. Based on the discussion in Sec.~\ref{sec:symmetries}, we now consider a Choi state $| \rho \rrangle$, which is invariant under a full symmetry group $G_d = (K \times K) \rtimes J \times G$. Utilizing a decorated domain wall construction, we will demonstrate that possible SPT phases of the Choi state, which result in hermitian, non-negative density matrices in the original Hilbert space, have a classification described by Theorem~\ref{thm:MSPTclassification}. 

In the decorated domain wall representation of the density matrix, as shown in Eq.~(\ref{eq:repdensitymatrix}), defects associated with modular conjugation $J$ trap lower dimensional SPT phases protected by the exact symmetry. This representation allows us to pinpoint potential sources of non-negativity violations. A crucial insight employed in this section is that the decorated domain wall approach is applicable to defects of all average symmetries, including $G$. In particular, a representative fixed point density matrix can be expressed as:
\begin{equation}
\rho = \sum_D \sqrt{p_D} e^{i\theta_D} \rho_D\otimes a_D^G.
\label{eq:repdensitymatrixG}
\end{equation}
Here, $a_D^G$ is in an ancilla space (subject to the action of $G$ but not $K\times K$) describing the quantum state of the defect network associated with the average symmetry $G$. $\rho_D$ represents a decoration, whose physical interpretation will become evident shortly.

We begin with the fact that in a $d$-dimensional space, bosonic SRE pure states with the symmetry group $G_d$ can be classified by $H^{d+1}(G_d,U(1))$. This group cohomology can be decomposed using the Künneth theorem, as follows:
\begin{equation}
H^{d+1}(G_d,U(1)) = \bigoplus_{p+q = d+1} H^p[G,H^q((K\times K)\rtimes J,U(1))].
\label{eq:E2withavesym}
\end{equation}
Once more, the physical interpretation underlying this decomposition is the decorated domain wall construction: $G$ symmetry defects are decorated by lower-dimensional SPT states with $(K\times K)\rtimes J$ symmetry. Similar to what was discussed in Section~\ref{subsec:Jsymmetry}, all the components in this decomposition can be categorized into two groups, based on the value of $p$.

(1) SPT states due to the phase factors: The term with $p=d+1$ in Eq.~(\ref{eq:E2withavesym}) is given by $H^{d+1}(G,\Z_2)$. In this case, the decoration $\rho_D$ is trivial, so the phase factor $e^{i\theta_D}$ in Eq.~(\ref{eq:repdensitymatrixG}), which may assume values of $\pm 1$, completely captures the information of this $\Z_2$-valued cohomology. In other words, $H^{d+1}(G,\Z_2)$ serves as the classification for $G$ SPT phases in the doubled space that are invariant under the $J$ symmetry. By employing a similar argument to that used for $J$ defect networks, we deduce that only the trivial $G$ SPT in the Choi representation, characterized by $e^{i\theta_D}=1$ for all $D$, is consistent non-negativity. Consequently, there are no MSPT phases protected solely by the average symmetry.

As an illustrative example, let us consider the Levin-Gu state as a potential Choi state of a nontrivial $2d$ MSPT state protected by a $\Z_2$ average symmetry. The wavefunction of this state, studied in Ref.~\cite{2012LevinGu}, can be formulated as (after maximal dephasing of the domain wall number operator, i.e., $\sigma^z_i \sigma^z_j$, on each link. Here, $\sigma^z$ represents the $\Z_2$ order parameter):
\begin{equation}
\rho = \sum_D (-1)^{N_{dw}(D)}|D\rangle\langle D|,
\end{equation}
where the summation is taken over all possible $\Z_2$ domain wall configurations, and $N_{dw}(D)$ is the number of closed $\Z_2$ domain walls in the configuration $D$. It is evident that the nontrivial phase factor violates non-negativity. Therefore, the Levin-Gu state is not the Choi state of an MSPT state. The superposition of domain walls of average symmetries in $|\rho\rrangle$ corresponds to a classical convex sum in the original density matrix, where a nontrivial phase factor cannot appear.

\begin{figure}
\begin{center}
  \includegraphics[width=.45\textwidth]{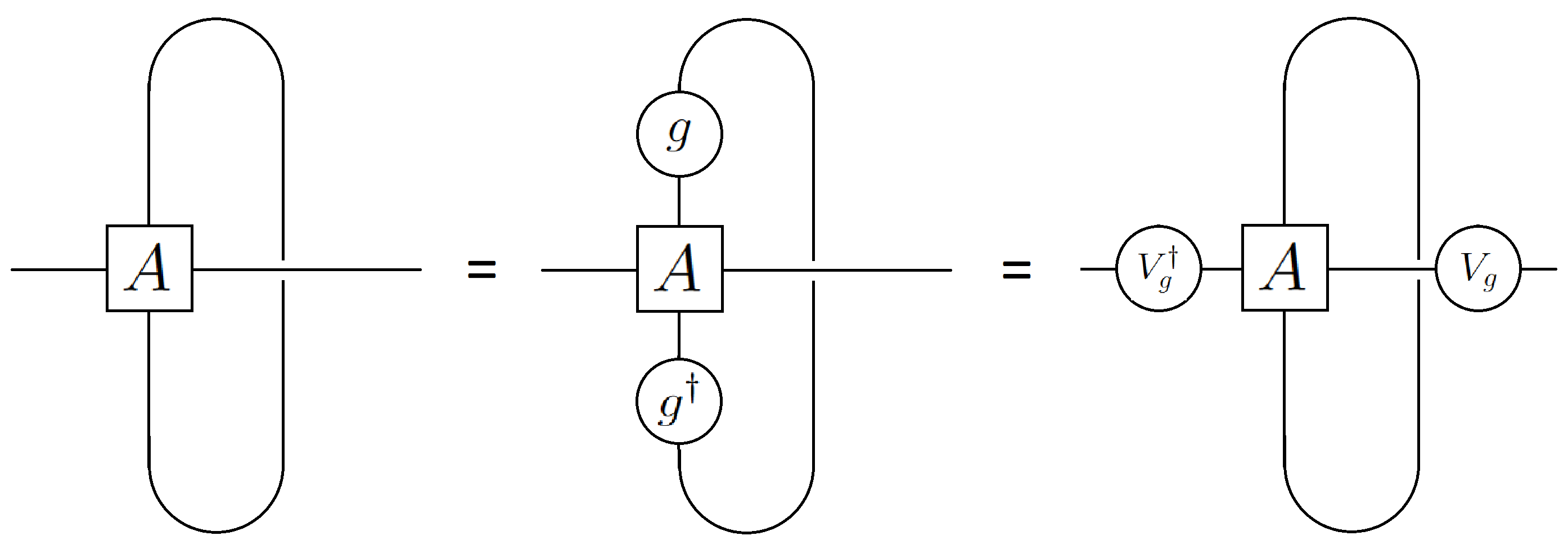} 
\end{center}
\caption{
A diagrammatic derivation of Eq.(\ref{eq:ggaugetransform}).
}
\label{Fig:E1g}
\end{figure}

Once again, in $1d$ systems with translation invariance, the absence of MSPT protected by the average symmetry $G$ (or by $G\times J$) can also be seen by examining the properties of the local tensor $A^{pq}$ of the MPDO. Since $| \rho \rrangle$ is SRE, acting on the local tensor with an average symmetry results in a gauge transformation on the entanglement bond. As illustrated in Fig. \ref{Fig:E1g}, we have the following:
\begin{equation}
    [V_g, E^1] = 0, \quad \forall g\in G. 
    \label{eq:ggaugetransform}
\end{equation}
Similar to the discussion around Eq.~(\ref{eq:Jprojective}), since $E^1 = W \mathrm{diag}(1,\Lambda_2,\Lambda_3...) W^{-1}$ for $|\lambda_{i>1}|<1$ (which physically indicates the absence of spontaneous breaking of $G$, see Sec.~\ref{sec:SSB}), we have $\Tilde{V}_g = W^{-1} V_g W$ is block diagonal with all elements in the first column and first row, except $[\Tilde{V}_g]_{11}$, vanish. This immediately leads to the following results:
\begin{itemize}
    \item $[\Tilde{V}_g]_{11}$ for $g\in G$ forms a one-dimensional representation of $G$, which must be linear. Therefore, $V_g$ for $g\in G$ also forms a linear representation of $G$. Thus, $| \rho \rrangle$ is a trivial $G$ SPT.
    \item By a $U(1)$ gauge transform of $\Tilde{V}_g$ (such that $[\Tilde{V}_g]_{11}$ becomes real), one can always ensure $\Tilde{V}_g \Tilde{V}_J = \Tilde{V}_J \Tilde{V}_g^*$ for all $g\in G$. This, in turn, implies that 
    \begin{equation}
         V_g V_J =V_J V_g^*,\quad \forall g\in G.
    \end{equation}
    Therefore, $| \rho \rrangle$ is a trivial $G\times J$ SPT \cite{2017ShinseiMPS}.
\end{itemize}

(2) SPT states due to decorations: This case involves terms with $0<p<d+1$ in Eq.~(\ref{eq:E2withavesym}), where codimension-$p$ symmetry defects of $G$ are decorated by $(d-p)$-dimensional SPT states protected by the exact symmetry and the modular conjugation. These decorated states are expressed as $\rho_D$ in Eq.~(\ref{eq:repdensitymatrixG}). Specifically,
\begin{itemize}
    \item In $d=1$, the only possibility corresponds to $p=q=1$ in Eq.~(\ref{eq:E2withavesym}). The decoration is a $(K\times K)$ charge invariant under $J$, labeled by an element in $H^1(K,U(1))$.  
    \item In $d>1$, as we concluded in Sec.~\ref{subsec:Jsymmetry}, the $(d-p)$-dimensional decorations $\rho_D$ in Eq.~(\ref{eq:repdensitymatrixG}) consistent with non-negativity are classified by $H^{d-p+1}(K,U(1))$;
\end{itemize}
In both cases stated above, the final decoration patterns fall within the classification of 
\begin{equation}
    \bigoplus_{p=0}^d H^p[G,H^{d+1-p}(K,U(1))],
\end{equation}
which encompass all legitimate MSPT phases, as presented in Theorem~\ref{thm:MSPTclassification}.

To summarize, we have established a no-go theorem, indicating that for any SPT state $|\rho\rrangle$ in the doubled space lying \emph{outside} of the classification in Eq.~(\ref{eq:MSPTclassification}), the corresponding density matrix $\rho$ must violate non-negativity. Conversely, the existence of MSPT states within all classes included in Eq.~(\ref{eq:MSPTclassification}) can be demonstrated through explicit construction: (1) for the $p=0$ class in Eq.~(\ref{eq:MSPTclassification}), an example is a pure $K$ SPT state; (2) for classes protected jointly by the exact and the average symmetry, one can prepare a density matrix within this class using the following procedure. Starting from a pure SPT state $|\psi\rangle$ with a corresponding class in $H^p[G,H^{d+1-p}(K,U(1))]$ (now both $K$ and $G$ are exact symmetries of the pure state), one can apply weak onsite decoherence to reduce $G$ to an average symmetry \cite{2022aspt,2023aspt,2022decoxu}. For instance, the decoherence channel can be chosen as
\begin{equation}
\begin{split}
\mathcal{E} = & \prod_i \mathcal{E}_i, \\
\mathcal{E}_i(\rho) = & (1-p)\rho + p O_i^\dagger \rho O_i,
\end{split}
\end{equation}
with $p\ll 1$, where $O_i$ is a unitary acting on site $i$ that transforms in a faithful representation of $G$. The resulting state $\mathcal{E}(|\psi\rangle\langle\psi|)$ is then an MSPT state within the same class, with $G$ being average. This converse completes the proof of Theorem~\ref{thm:MSPTclassification}.


\subsection{General symmetry groups}
\label{sec:generalsymmetries}

Our analysis can be extended to encompass states protected by more general symmetries. In this section, we will explore two additional scenarios: the case of time-reversal symmetry and the case of a nontrivial extension between the exact and average symmetries.

The first generalization concerns the case where the density matrix exhibits invariance under time-reversal symmetry. As elucidated in Sec.~\ref{sec:symmetries}, time-reversal symmetry operates simultaneously on both the bra and the ket space and should hence be regarded as an average symmetry. For the present discussion, we will assume that $\mathcal{T}$ and the unitary symmetries form a direct product; that is, that the full symmetry group has the form $G_d = [(K \times K)\rtimes J] \times G \times \mathcal{T}$. We have the following theorem:
\begin{theorem}
    In $d$ spatial dimensions, MSPT phases protected by the symmetry group $G_d =  [(K\times K)\rtimes J ] \times G\times \mathcal{T})$ are classified by
    \begin{equation}
        \bigoplus_{p=0}^d H^p[G\times \mathcal{T},H^{d+1-p}(K,U(1))].
        \label{eq:MSPTclassificationT}
    \end{equation}
    Here $\mathcal{T}$ acts on the $U(1)$ coefficient by a complex conjugation.
    \label{thm:MSPTclassificationwithT}
\end{theorem}

The proof of Theorem~\ref{thm:MSPTclassificationwithT} can also be understood through a decorated domain wall construction, akin to the one used in the proof of Theorem~\ref{thm:MSPTclassification}. In a $G\times \mathcal{T}$ symmetric Choi state $| \rho \rrangle$, the wave function is a superposition of all $G\times \mathcal{T}$ defect configurations. On each codimension-$p$ defect, we can decorate a $(d-p)$-dimensional SPT state protected by the $(K\times K) \rtimes J$ symmetry. As demonstrated in Sec.~\ref{subsec:Jsymmetry}, those decorations that are consistent with non-negativity are classified by $H^{d-p+1}(K,U(1))$. Consequently, all MSPT phases which can be constructed using domain wall decorations are described by Eq.~(\ref{eq:MSPTclassificationT}). The $p=d+1$ term is absent in Eq.~(\ref{eq:MSPTclassificationT}), as it encodes the phase factor $e^{i\theta_D}$ in the superposition of $G\times \mathcal{T}$ symmetry defects, which is required to be trivial by non-negativity.

The second generalization involves symmetries that do not form a direct product but rather involve a nontrivial extension, as in Eq.(\ref{eq:internalsymmetry}). We allow the average symmetry group $G$ to contain both unitary and antiunitary elements. Equivalence classes of extensions are characterized by the cohomology group $H^2(G,K)$. Consequently, the full symmetry group $G_d$ is determined by the following morphism of short exact sequences:
\begin{equation}
\begin{tikzcd}
1 \arrow[r] & K^2 \arrow[r]                    & \tilde{G}^2 \arrow[r]                  & G^{2} \arrow[r]          & 1 \\
1 \arrow[r] & K^{2} \arrow[r] \arrow[u, "\cong"] & \mathcal{G} \arrow[r] \arrow[u, ""] & G \arrow[r] \arrow[u, "\Delta"] & 1
\end{tikzcd}
\label{eq:internalsymmetrywithT}
\end{equation}
Here, $\Delta$ represents the diagonal map, and $G_d = \mathcal{G}\rtimes J$. The upper line in Eq.(\ref{eq:internalsymmetrywithT}) shows the group structure of generators (within the original Hilbert space), while the second line represents the symmetry of the density matrix. As clarified in Sec.\ref{sec:symmetries}, the action of $J$ on $\mathcal{G}$ obeys the following rules: (1) $G$ and $J$ commute; (2) $J$ acts on $K\times K$ through $\mathrm{SWAP}$.

Once again, possible MSPT phases can be constructed using the decorated domain wall approach -- although we do not have the Künneth formula and the construction becomes more complicated. Specifically, on each codimension-$p$ defect associated with the average symmetry $G$, we decorate $(d-p)$-dimensional SPT states that are protected by $(K\times K)\rtimes J$. Analogously, the implications of non-negativity can be succinctly summarized as follows: (a) The lower dimensional decorated states must obey the principle of non-negativity individually. (b) The phase factor in the superposition of $G$ defect configurations must be trivial. After taking into account these two requirements, we can efficiently utilize the standard spectral sequence \cite{2019gaiottospectralsequence,2021decoratedDW} for the construction of MSPT phases with general symmetries.

\begin{theorem}
    In $d$ spatial dimensions, MSPT phases can be protected by a general symmetry group, fitting into the extension as given by Eq.~(\ref{eq:internalsymmetrywithT}). MSPT phases, consistent with non-negativity, are classified using a Leray-Serre spectral sequence, where the $E_2$ page consists of:
    \begin{equation}
     \bigoplus_{p+q=d+1} E_2^{p,q}  = \bigoplus_{p+q=d+1} H^p[G,H^q(K,U(1))],
    \label{eq:MSPTclassificationgeneral}
    \end{equation}
    with $q>0$.
\label{thm:MSPTclassificationgeneral}
\end{theorem}
Here, the consistent decorated states are classified by $H^{q}(K,U(1)$), as explained in (a) in the preceding paragraph. The $E_2$ page lacks elements with $q=0$, due to reason (b). We emphasize that not every decoration pattern found on the $E_2$ page can be realized as an MSPT state. Mathematically, all the consistency conditions are organized by the ``differentials" of the spectral sequence, which are exclusively determined by the equivalence class of group extension characterized by $H^2(G,K)$. As a result, the ultimate classification of MSPT, in accordance with non-negativity, aligns precisely with that of decohered average SPT phases exhibiting the corresponding exact and average symmetry, as introduced in Ref.~\cite{2023aspt}.\footnote{In contrast to the usual spectral sequences in pure states, the spectral sequence in Eq.~(\ref{eq:MSPTclassificationgeneral}) does not contain the row with $q=0$. This opens the possibility of intrinsic MSPT states without a pure state analog. See Ref.~\cite{2023aspt} for details.} The mathematical framework is very intricate, and a comprehensive explanation is beyond the scope of this work. For a review of the spectral sequence and its physical interpretation in the context of SPT phases, we direct the interested reader to Ref.~\cite{2023aspt}.

\subsection{Connections to previous studies}
\label{sec:previousstudies}

\subsubsection{Symmetrically separable mixed states}

Next we would like to elaborate on the relationship between our classification of mixed states using their Choi states and the concept of separability as defined in Ref.~\cite{2023separability}. The authors of Ref.~\cite{2023separability} define a symmetrically separable mixed state as one that can be expressed as a convex sum of pure states $\rho = \sum_m p_m | \psi_m \rangle \langle \psi_m |$, each of which is SRE in the sense that it can be prepared from a product state by a finite-depth unitary with symmetric local gates. We present the following statement:

\begin{theorem}
    Assuming $\rho$ is an MSPT state, described by a nontrivial element in Eq.~(\ref{eq:MSPTclassification}), then $\rho$ is not symmetrically separable.
\end{theorem}
For simplicity, we first illustrate the idea in $1d$. Apply a symmetry operator $U^a$ (where $a$ can be an element of either the exact or average symmetry) to a large but finite region $A$. Due to the symmetric SRE nature of $|\rho\rrangle$, $U^a$ acts nontrivially only near the boundary of $A$,
\begin{equation}
U^a |\rho\rrangle = U^a_L U^a_R |\rho\rrangle,
\end{equation}
where $U^a_L$ and $U^a_R$ are localized near the left and right ends of $A$, respectively. By the same argument, we can obtain a set of (say, right) endpoint operators, one for each group element. A crucial observation is that although the symmetry operators form a linear representation of the symmetry group (i.e., $U^a U^b = U^{ab}$), the endpoint operators follow a projective composition rule:
\begin{equation}
U_R^a U_R^b =e^{i\omega(a,b)}U_R^{ab},
\end{equation}
where $\omega(\cdot,\cdot)$ characterizes the second group cohomology class of the nontrivial Choi state $| \rho \rrangle$ -- in fact, this is the idea behind the string (or other nonlocal) order parameters for $1d$ SPT states \cite{20121dSPTdetection,2012nonlocalorder}. On the other hand, suppose $\rho$ is symmetrically separable. Then its Choi state $|\rho\rrangle = \sum_m p_m |\psi_m\rangle|\psi_m^*\rangle$ is a superposition of trivial SPT states (note that each of them can be prepared by a symmetric finite-depth unitary in the doubled space). In particular, the symmetry action $U^a$ fractionalizes as
\begin{equation}
U^a |\rho\rrangle = \sum_m p_m U_{L,m}^a U_{R,m}^a (|\psi_m\rangle |\psi_m^*\rangle ),
\end{equation}
in which all endpoint operators must obey the trivial composition rule. This leads to a contradiction, indicating that $\rho$ is not symmetrically separable. The same consideration can be extended straightforwardly to anti-unitary symmetries \cite{2017partialtranspose,2018partialtranspose} and higher-dimensional cases \cite{2011twodSPT,2014twodtensornetwork}, by considering the composition rule of multiple domain walls. For instance, within a nontrivial MSPT state $|\rho\rrangle$, the point-like operators $\Omega(a,b)$ in Eq.(\ref{eq:elsenayak}) satisfy associativity only up to a nontrivial phase factor dictated by the 3-cocycle of $|\rho\rrangle$ \cite{2014dElse}, thereby excluding the possibility of $|\rho\rrangle$ being a convex sum of trivial SPT states.

\subsubsection{The symmetry-protected sign problem}

We close this section by pointing out a connection between our classification and the concept of the symmetry-protected sign problem \cite{2021signproblem}. A state, denoted by $| \psi \rangle$, exhibits a symmetry-protected sign problem relative to a specific basis $\{ |\alpha \rangle \}$, if it cannot be transformed into a real non-negative wave function in the basis $\{ | \alpha \rangle \}$ using any symmetric finite-depth unitary circuit. In other words, a generic state in the same \emph{phase} as $|\psi\rangle$ has at least one amplitude in the basis $\{ |\alpha\rangle \}$ that falls outside of $\mathbb{R}_{\geq 0}$. This sign problem is recognized as an obstruction to efficiently simulating a quantum many-body system through Monte Carlo methods in certain bases \cite{2016signproblem}. Thus, it can be viewed as a diagnostic of the complexity of topological phases of matter. Building on our previous discussions, a Choi state $ | \rho \rrangle$ will yield a density matrix containing at least one negative eigenvalue if $| \rho \rrangle$ belongs to certain SPT phases, namely those not covered by the classification in Eq.~(\ref{eq:MSPTclassification}). Meanwhile, each eigenvalue of $\rho$ corresponds to the amplitude linked to a $J$-symmetric basis vector in the wave function of $| \rho \rrangle$. Therefore, we can make the following assertion.
\begin{theorem}
    Let $|\psi \rangle$ be a generic state in an SPT phase protected by a symmetry $G_d = \mathcal{G}\rtimes J$, where $\mathcal{G} = (K\times K)\times G$. In a $J$-symmetric basis $\{ |\alpha \rangle \}$, $|\psi \rangle$ exhibits a symmetry-protected sign problem with respect to $\{ |\alpha \rangle \}$ if $|\psi \rangle$ is an SPT state falls outside the scope of Eq.~(\ref{eq:MSPTclassification}).
\end{theorem}

A simple example of an SPT phase with sign problem is the $\Z_2$ SPT in $2d$, where $G = \Z_2 = \otimes_i X_i$ (here $X_i$ is the Pauli-$X$ operator on site $i$). The basis ${ |\alpha\rangle }$ comprises all tensor products of local $Z$-basis vectors on each site. In this basis, the action of $J$ is a complex conjugation of the amplitudes. The amplitude associated with a basis vector $| \alpha \rangle$ is given by
\begin{equation}
    \Psi(| \alpha \rangle) = (-1)^{N_{dw}(\alpha)},
\end{equation}
where $N_{dw}(\alpha)$ is the total number of $\Z_2$ domain walls in $| \alpha \rangle$. The amplitude cannot be non-negative simultaneously for all basis vectors in $\{ | \alpha \rangle \}$.

\subsection{Deformations}
\label{sec:deformations}

So far we have discussed the SPT invariants of mixed states from the perspective of their Choi states in the doubled space. Mixed states in open systems typically undergo evolution through quantum channels. In this section, we address two key questions: (1) whether an SRE mixed state maintains its SRE nature after undergoing a channel, and (2) whether their SPT invariants remain robust when the states are subjected to specific types of quantum channels.

\begin{figure}
\begin{center}
  \includegraphics[width=.40\textwidth]{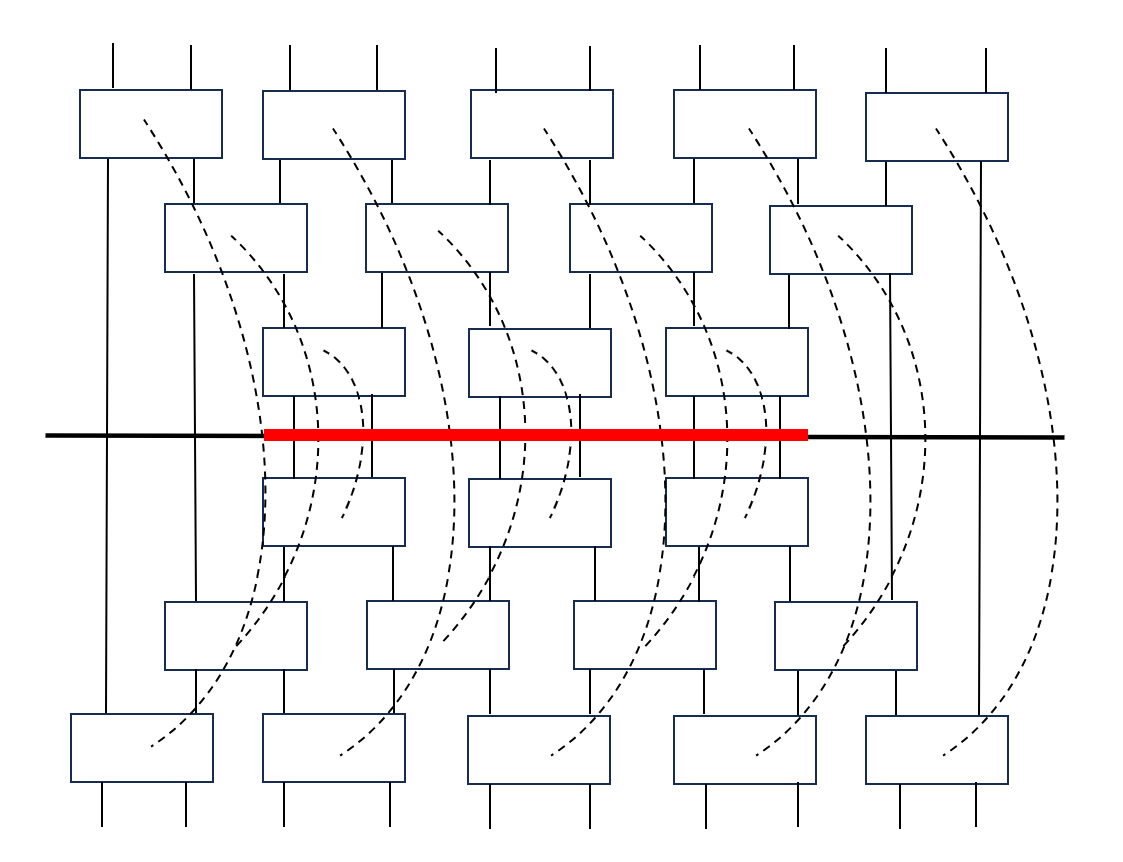} 
\end{center}
\caption{
The truncated channel $\mathcal{E}_A$, wherein we retain only the gates within the ``light cone" of the region $A$ (highlighted in red). The dashed lines indicate the summation over Kraus indices. 
}
\label{Fig:truncatedchannel}
\end{figure}

Similar to symmetric finite-depth unitaries in closed systems, symmetric finite-depth channels will play the role of ``adiabatic deformation" in our framework of MSPT phases. We follow the definitions proposed in Ref.~\cite{2022openspt}: A circuit quantum channel is a channel consisting of layers of disjoint local channels (``gates'') that form a circuit, see Fig.~\ref{Fig:truncatedchannel}. Such a channel is said to be finite-depth if the depth of the circuit is constant in the size of the system. A circuit quantum channel is said to be symmetric if each of the local channels, denoted as $\mathcal{E}_l$, is strongly symmetric with respect to the exact symmetries and weakly symmetric with respect to the average symmetries. Specifically, in terms of the Kraus representation of a local channel, $\mathcal{E}_l(\rho) = \sum_m K_m \rho K_m^\dagger$, (1) each local Kraus operator $K_m$ individually commutes with the exact symmetry. (2) Meanwhile, as a superoperator acting in the doubled space, each $K_m\bigotimes K_m^*$ commutes with the average symmetry $U_g\bigotimes U_g^*$. This symmetry condition ensures that the entire channel, as a superoperator, commutes with the full symmetry group $G_d=(K\times K \times G)\rtimes J$, thereby preserving the exact and average symmetries of the Choi state $| \rho \rrangle$.

Suppose $\rho$ has the property that its Choi state $|\rho\rrangle$ can be represented by an MPS (or, in higher dimensions, a PEPS) of finite bond dimension. This assumption encompasses all SRE mixed states as defined in Section~\ref{sec:SREmixed}. This property is preserved under evolution by a finite-depth channel, by the following argument. First, note that, if the channel can be represented on the doubled space by a projected-entangled-pair operator (PEPO) of finite bond dimension, the result follows because the bond dimensions of the initial PEPS and the PEPO multiply to give a finite bond dimension of PEPS representing the evolved state. It remains to show that a finite-depth channel acts as a finite bond dimension PEPO (as observed in Ref. \cite{2020Piroli}, there are channels that preserve locality yet are not PEPOs, so one must be careful). To construct the PEPO representation of an operator, one performs a Schmidt operator decomposition along a cut separating a region from its complement. If the Schmidt operator rank along this cut scales with the surface area of the region (and not the volume), the bond dimension is finite. A finite-depth channel acts as a circuit of local gates that are not necessarily unitary. In this case, the rank along the cut is bounded from above by the products of the ranks of those gates that cross the cut, each of which is finite. Since the circuit is local, only gates near the cut (and not in the bulk of the region) cross the cut, and, since the circuit is finite depth, only finitely many gates are near the cut per unit surface area. Therefore, the Schmidt rank scales such that the PEPO representing the channel can have finite bond dimension, and we can conclude that finite-depth channels preserve the PEPS-representability of the Choi state $| \rho \rrangle$.

One might then expect that the SRE nature of the Choi state is preserved by any symmetric finite-depth channel. However, this is \emph{not} the case. A crucial observation is that the tensor network form described in Eq.~(\ref{eq:MPSform}) does not always represent an SRE state. In $1d$, for a state $| \rho \rrangle$ to be SRE, the MPS representation must also be \emph{normal}, meaning that the transfer matrix, defined as $E^2_{\alpha\alpha',\beta\beta'} = \sum_{p,q} A^{pq}_{\alpha\beta} A^{pq*}_{\alpha'\beta'}$, has a unique largest eigenvalue \cite{2020mpsreview}. However, there exist symmetric finite-depth channels that, when applied to an SRE state $| \rho \rrangle$, cause it to maintain the MPS form but lose its normality. Let us illustrate this point with a simple example.

\begin{figure}
\begin{center}
  \includegraphics[width=.45\textwidth]{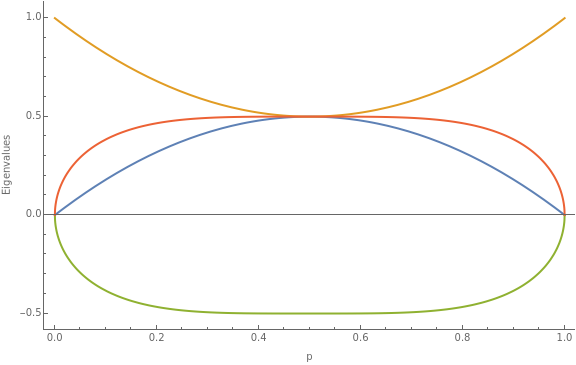} 
\end{center}
\caption{
Non-zero eigenvalues of the transfer matrix $E^2$ as a function of the dephasing strength $p$
}
\label{Fig:dephasing}
\end{figure}

Consider a pure $1d$ cluster state \cite{2001cluster}, whose density matrix can be represented by an MPDO with local tensors $A^{pq} = B^p\otimes B^q$ ($p,q=0,1$), where:
\begin{equation}
    B^0 = \frac{1}{\sqrt{2}}\begin{pmatrix}
    1&1\\&
\end{pmatrix}~, B^1 = \frac{1}{\sqrt{2}}\begin{pmatrix}
    &\\1&-1
\end{pmatrix}. 
\end{equation}
The cluster state has a $\Z_2\times \Z_2$ symmetry, generated by the product of Pauli-$X$ operators on even and odd sites, respectively. Let us pass this state through a symmetric depth-1 dephasing channel defined as: \begin{equation}
\begin{split}
    &\mathcal{E} = \otimes_i \mathcal{E}_i, \\
    &\mathcal{E}_i(\rho)=(1-p)\rho+pX_i \rho X_i,
    \label{eq:dephasingchannel}
\end{split}
\end{equation} 
where $X_i$ represents the Pauli-$X$ operator on site $i$, and $p$ is the dephasing strength. We present a plot of non-zero eigenvalues for the transfer matrix $E^2$ in Fig~\ref{Fig:dephasing}. An immediate observation is that, for a generic quantum channel, the largest eigenvalue is unique, implying that the Choi state $| \rho \rrangle$ remains SRE. However, at the maximal dephasing strength $p=1/2$, the MPS representation of $| \rho \rrangle$ loses its normality.

A straightforward calculation (see Appendix~\ref{cluster}) reveals that at $p=1/2$, the evolved state becomes:
\begin{equation}
\label{eq:dephasedcluster}
\begin{split}
\rho_{1/2}&=\prod_{i\text{ even}}(\frac{1}{4}I_i I_{i+1})+\prod_{i\text{ even}}(\frac{1}{4}I_i X_{i+1})\\
&\qquad+\prod_{i\text{ even}}(\frac{1}{4}X_i I_{i+1})+\prod_{i\text{ even}}(\frac{1}{4}X_iX_{i+1}),
\end{split}
\end{equation}
which is a maximally mixed state (with a definite $\Z_2\times \Z_2$ charge). Its Choi representation corresponds to a GHZ state in the doubled space, which is LRE by our definition. We interpret this state as an instance of spontaneous exact-to-average symmetry breaking, which we will discuss in detail in the following section. This example highlights a fundamental distinction between evolutions in closed and open systems, \ie a symmetric finite-depth channel can potentially change the SRE nature of a mixed state.

In fact, for a $1d$ SRE state $|\rho\rrangle$ subject to a finite-depth local channel, the evolved state remains SRE if the channel is non-degenerate, \ie it is an invertible matrix. Similar observation was initially made in Ref. \cite{2024BaoY}, and we offer here a revised version tailored to our needs (the proof is provided in the Appendix \ref{app:nondeg}). 

\begin{theorem}
    \label{thm:nondeg}
   Let $|\rho\rrangle$ be an SRE state, characterized by having a transfer matrix $E^2$ with a unique largest eigenvalue, and let $\mathcal{E}$ be a finite-depth local channel. If $\mathcal{E}$ is non-degenerate, then the state $\mathcal{E}|\rho\rrangle$ remains SRE.
\end{theorem}
As an example, the dephasing channel described in Eq. \ref{eq:dephasingchannel} becomes degenerate only when $p=1/2$.

In higher dimensions, a PEPS with a finite bond dimension can represent even more LRE states, such as states with algebraic correlation functions \cite{2006criticalPEPS}. Therefore, it is not guaranteed that a SRE Choi state, after passing through a symmetric finite-depth channel, remains SRE. However, we still anticipate that the SRE property is perturbatively stable, meaning the Choi state remains SRE within a finite parameter regime. An example was detailed in Ref. \cite{2023decobao,2023decofan,2023decoxu}, where a $2d$ SRE pure state\footnote{One needs to first ungauge the toric code considered in Ref. \cite{2023decobao,2023decofan,2023decoxu}, which does not alter the threshold.} was subjected to symmetric local decoherence. There, $| \rho \rrangle$ undergoes a phase transition only when $p$ exceeds a finite threshold.

For the remainder of this section, let us focus on scenarios where the evolved states $\mathcal{E}(\rho)$ remain SRE -- this encompasses generic (\ie non-degenerate) quantum channels in $1d$, where the MPS representation of $\mathcal{E}(\rho)$ remains normal. We will now address the second question raised at the beginning of this section, demonstrating that in $1d$, after passing through a symmetric finite-depth channel, as long as the evolved state remains SRE, it must have the same SPT classification as the initial state.

We begin by noting that the topological class of an SPT state (within the group cohomology classification) is ultimately determined by the Berry phase resulting from local rearrangements of symmetry defects \cite{2013chenSPT,2011twodSPT}. We will demonstrate that, fusing symmetry defects in the final state $| \rho_f \rrangle = \mathcal{E}| \rho_i \rrangle$ results in the same Berry phase factor as in the initial state $\rho_i$, as long as $\rho_f$ is a SRE mixed state, and $\mathcal{E}$ is a symmetric finite-depth channel. The first step involves truncating the channel $\mathcal{E}$ to a large yet finite region $A$, where Kraus operators located outside the ``light cone" of region $A$ are set to identity operator, as illustrated in Fig.~\ref{Fig:truncatedchannel}. The truncated channel, referred to as $\mathcal{E}_A$, also operates as a symmetric finite-depth channel. In particular, it evolves a symmetric SRE initial state $\rho_i$ into another symmetric state, denoted as $\rho_f'$, that can be represented by an MPS with a finite bond dimension.

Next, on the state $\rho_f'$, we apply a symmetry operator $U^a$ (where $a$ could be an element of the exact or the average symmetry) to a large but finite region. We assume that one end of $U^a$ (say, the right endpoint) is deep inside $A$, while the left end is deep outside $A$. The key observation is that this symmetry operation fractionalizes on the Choi state $| \rho_f' \rrangle$. In other words, the impact of this symmetry operation $U^a$ is localized only near its endpoints:
\begin{equation}
U^a | \rho_f' \rrangle = U^a_L U_R^{a'}| \rho_f' \rrangle,
\label{eq:truncatedchannel}
\end{equation}
Here, $U_L^a$ and $U_R^{a'}$ are left and right local operators supported in finite regions around their respective endpoints.

Let us further elucidate Eq.~(\ref{eq:truncatedchannel}). Given our assumption that both the initial state $\rho_i$ and the final state of the original channel, $\rho_f$, are SRE, the notion of symmetry fractionalization \cite{2011symlocal,2011symlocalfermion} dictates that when a symmetry operator acts within a large region, it acts nontrivially only near the edges:
\begin{equation}
\begin{split}
& U^a | \rho_i \rrangle = U_L^a U_R^a | \rho_i \rrangle, \\
& U^a | \rho_f \rrangle = U_L^{a'} U_R^{a'} | \rho_f \rrangle.
\end{split}
\label{eq:originalchannel}
\end{equation}
Here, the endpoint operators $U_L^a$ and $U_R^a$ are supported within regions of a length comparable to the correlation length of $| \rho_i \rrangle$, and similarly for $U_L^{a'}$ and $U_R^{a'}$. Furthermore, if $\mathcal{E}$ is a finite-depth channel, the left endpoint operator $U_L^a$ in Eq.(\ref{eq:truncatedchannel}), which resides deep outside region $A$, remains unaffected by the truncated channel $\mathcal{E}_A$. Consequently, $U_L^a$ corresponds to the left endpoint operator resulting from the application of $U^a$ to the initial state $| \rho_i \rrangle$, hence the notation in Eq.(\ref{eq:originalchannel}). Similarly, the right endpoint operator $U_R^{a'}$, located deep within region $A$, remains unaltered by the truncation far away, and thus is identical to the right endpoint operator generated by the application of $U^a$ to the final state $| \rho_f \rrangle$. It is also worth noting that within the bulk of the symmetry action $U^a$, there are no nontrivial effects as indicated in Eq.~(\ref{eq:truncatedchannel}). This is due to the fact that within the bulk, $U^a$ commutes with each local gate of $\mathcal{E}$ (and hence with those of $\mathcal{E}_A$). Therefore, the bulk of $U^a$ effectively acts on the initial state $|\rho_i \rrangle$ and does not induce any non-trivial effects.

Now we consider another symmetry operator $U^b$ acting in a finite but large region, with its left end deep outside $A$ and its right end deep inside. Using a similar argument as before, we conclude that the action of $U^b$ on $| \rho_f'\rrangle$ affects the state only near its endpoints:
\begin{equation}
U^b | \rho_f' \rrangle = U^b_L U_R^{b'}| \rho_f' \rrangle,
\end{equation}
Note that although the symmetry operators form a linear representation of the symmetry group (i.e., $U^a U^b = U^{ab}$), the endpoint operators may follow a projective multiplication rule, namely
\begin{equation}
U_R^a U_R^b =e^{i\omega(a,b)}U_R^{ab},
\end{equation}
where $\omega(\cdot,\cdot)$ precisely characterizes the second group cohomology class of the underlying SRE state \cite{2011symlocal,2013chenSPT}. Since the symmetry operators form a linear representation, it follows that both sets of endpoint operators, $U_R$ and $U_R'$, exhibit an inverse projective multiplication phase $\omega(\cdot,\cdot)$ compared to that of $U_L$. Consequently, $U_R$ and $U_R'$ share the same projective representation. As a result, the two states, $| \rho_i \rrangle$ and $| \rho_f \rrangle$, must belong to the same class in the second group cohomology.

The argument above can be extended to higher dimensions \cite{2020mpsreview,2011twodSPT,2014twodtensornetwork}. However, in $d\ge 2$, in addition to the requirement that the initial and final states be SRE, an additional condition is needed: the state generated by an arbitrary spatially truncated channel (\ie $|\rho_f'\rrangle = \mathcal{E}_A | \rho_i \rrangle$ where $\mathcal{E}_A$ represents the truncation of 
$\mathcal{E}$ within the light-cone of region $A$) must be SRE as well. This ensures that the symmetry fractionalization condition $U^a | \rho_f' \rrangle = V^a_\partial | \rho_f' \rrangle$ is satisfied for an arbitrary symmetry operator $U^a$ acting in a large but finite region, where $V^a_\partial$ is a unitary supported near the boundary of $U^a$. Without this extra condition, the truncated channel $\mathcal{E}_A$ can potentially result in an LRE state at the boundary of the region $A$, leading to the breakdown of symmetry fractionalization, as demonstrated by the example in Eq.(\ref{eq:dephasedcluster}). With this additional requirement, one can then use to characterization of the SPT invariants in Appendix C of Ref.~\cite{2014dElse} to show that the final state $| \rho_f \rrangle$ has the same SPT invariant as the initial state. The idea is once again that $| \rho_f' \rrangle$, as an SRE state, must be characterized by the same SPT invariant in different spatial regions.

In summary, this section has demonstrated the following points: (1) The finite bond dimension tensor network nature of $|\rho\rrangle$ is preserved by finite-depth quantum channels. However, the SRE nature is not necessarily preserved; (2) The SPT invariant of $|\rho\rrangle$ is preserved by a symmetric finite-depth channel, as long as:
\begin{itemize}
    \item In $1d$, the final state remains SRE;
    \item in $d\ge 2$, the final states of spatially truncated channels also remain SRE.
\end{itemize}
Consequently, we can classify any density matrix $\rho$, whose Choi state $| \rho \rrangle$ is a pure SPT state in the doubled space, into an MSPT \emph{phase}, characterized by the SPT invariant of $| \rho \rrangle$.

On the other hand, the example in Eq.(\ref{eq:dephasedcluster}) suggests that, unlike a symmetric finite-depth unitary, a symmetric finite-depth channel can significantly modify the long-distance characteristics of a mixed state, potentially resulting in an LRE state. This prompts the question of how to understand the behaviors of a mixed state after undergoing a symmetric-finite depth channel systematically. How can we quantify these (symmetry-protected) long-distance phenomena from an information-theoretic standpoint? We will delve into these questions in detail in an upcoming work.

\section{Spontaneous symmetry breaking in mixed states}
\label{sec:SSB}

Based on the example above, we can delve further into the concept of spontaneous symmetry breaking (SSB) within the context of mixed states. For simplicity, we mainly focus on $1d$ systems. Given the specific symmetry structure of a mixed state (or equivalently, its Choi representation), such as $(K\times K)\rtimes J \times G$, it is natural to ask: What symmetry breaking patterns can emerge in a mixed state? And how can we diagnose them? In this section, we will provide a detailed examination of SSB in mixed states from three complementary perspectives: (1) order parameters; (2) disorder parameters; (3) MPDO representations.

As a first observation, the modular conjugation $J$ cannot be broken, as long as we require that $\rho$ is hermitian. Regarding the other symmetries, three possible symmetry-breaking patterns may emerge.
\begin{enumerate}
    \item An exact symmetry may be completely broken -- meaning that it remains neither as an unbroken exact symmetry nor as an unbroken average (diagonal) symmetry. This symmetry breaking pattern is detected by an order parameter
    \begin{equation}
        C_1 = \llangle \rho | I \bigotimes O_x O_y^\dagger | \rho \rrangle/\llangle \rho | \rho \rrangle \sim O(1),
        \label{eq:linearor}
    \end{equation}
    where $O$ is charged under the exact symmetry $K$. In the original Hilbert space, $C_1$ can be expressed as
    \begin{equation}
        C_1 = \mathrm{tr}(\rho O_x O_y^\dagger \rho)/\mathrm{tr}(\rho^2).
    \end{equation}
    \item An average symmetry can be broken spontaneously. This pattern is also detected by the correlation $C_1$ in Eq.~(\ref{eq:linearor}), where now $O$ is charged under the average symmetry $G$.
    \item An exact symmetry may be broken spontaneously to its diagonal subgroup,
    \begin{equation}
        K\times K \leadsto K,
    \end{equation}
    which means it becomes an average symmetry. This symmetry breaking pattern can be detected by an order parameter
    \begin{equation}
        C_2 = \llangle \rho | O_x O_y^\dagger \bigotimes O_x^\dagger O_y | \rho \rrangle/\llangle \rho | \rho \rrangle \sim O(1)
        \label{eq:nonlinearor}
    \end{equation}
    for a $K$-charged operator $O$. This correlation function can also be written as
    \begin{equation}
        C_2 = \mathrm{tr}(O_x O_y^\dagger \rho O_x^\dagger O_y \rho)/\mathrm{tr}(\rho^2)
    \end{equation}
    in term of the original density matrix. An example of the third scenario is the cluster chain under decoherence discussed in Sec.~\ref{sec:deformations}.
\end{enumerate}
The expressions of the order parameters, Eq.~(\ref{eq:linearor}) and Eq.~(\ref{eq:nonlinearor}), are the standard order parameters for the (pure) Choi state $|\rho \rrangle$, with $C_1\sim O(1)$ detecting whether the symmetries $(e,k),(k,e),(k,k)\in K\times K$ are broken and $C_2\sim O(1)$ detecting whether just $(e,k)$ and $(k,e)$ but not the diagonal $(k,k)\in K\times K$ (which commutes with $O_x O_y^\dagger \bigotimes O_x^\dagger O_y$) are broken. A state exhibits exact-to-average symmetry breaking if $C_1\sim 0$ yet $C_2\sim O(1)$.

One may ask if a third order parameter, $C_3 = \mathrm{tr}(\rho O_x O_y^\dagger) = \llangle I | O_x O_y^\dagger \bigotimes I | \rho \rrangle$, typically employed for detecting SSB in thermal ensembles, can also detect SSB patterns described above. To see that $C_3$ cannot detect exact-to-average symmetry breaking, notice that the maximally dephased cluster chain described in Eq.(\ref{eq:dephasedcluster}) exhibits an exact-to-average SSB of the $\Z_2$ symmetry, which acts on the even (or odd) sites, as evidenced by a nonzero $C_2$ with $O_x = Z_x$ and $O_{x+2n}=Z_{x+2n}$. However, for any local operator $O$, $C_3$ equals zero. One way to interpret this result is that the state $| I \rrangle$ in the expression for $C_3$ explicitly breaks the exact symmetry down to an average one, making it so that exact-to-average SSB cannot be detected. Thus, we see that detecting exact-to-average symmetry breaking requires using order parameters that are \emph{non-linear} in the density matrix. On the other hand, a long-range order of $C_3$ can still indicate the presence of the first two types of SSB, wherein either an exact or average symmetry is completely broken.

\begin{figure}
\begin{center}
\begin{subfigure}[b]{0.2\textwidth}
  \includegraphics[width=\textwidth]{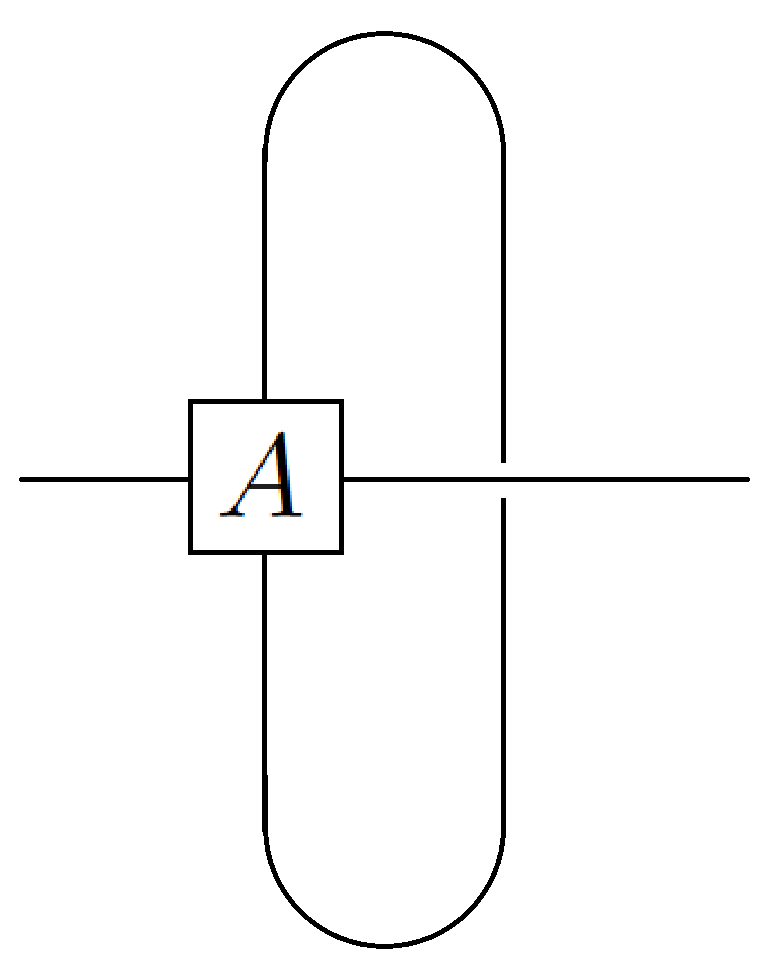} 
  \end{subfigure}
  \begin{subfigure}[b]{0.2\textwidth}
  \includegraphics[width=\textwidth]{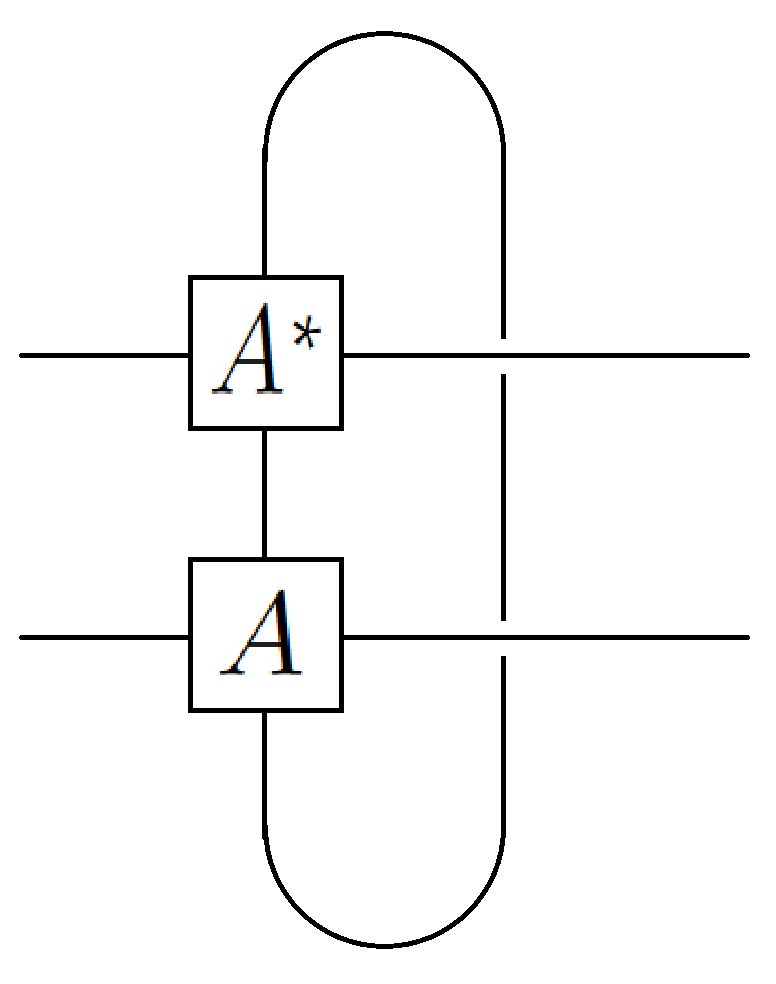} 
  \end{subfigure}
\end{center}
\caption{ A diagrammatic illustration of the transfer matrices, $E^1$ (left) and $E^2$ (right), as defined in Eq.~(\ref{eq:twotransfermatrices}).
}
\label{Fig:twotransfermatrices}
\end{figure}

Now, let us assume that $\rho$ can be represented by a translationally invariant MPDO with a finite bond dimension, and investigate the characteristics of the MPDO representation when the state exhibits any of the three aforementioned types of SSB. We define two types of transfer matrices,
\begin{equation}
    \begin{split}
        & E^1_{\alpha\beta} = \sum_p A^{pp}_{\alpha\beta}, \\
        & E^2_{\alpha\alpha',\beta\beta'} = \sum_{p,q} A^{pq}_{\alpha\beta} A^{pq*}_{\alpha'\beta'}.
    \end{split}
    \label{eq:twotransfermatrices}
\end{equation}
Here, $E^1$ is computed by tracing over the physical index of the local tensor, whereas $E^2$ corresponds to the standard transfer matrix of the Choi state $| \rho \rrangle$, see Fig.~\ref{Fig:twotransfermatrices}. Denote the largest (in magnitude) eigenvalue of $E^1$ as $\lambda_1$. The trace of the density matrix with periodic boundary conditions is given by $\mathrm{tr}\rho = \mathrm{tr}[(E^1)^L]$ and thus scales as $\lambda_1^L$ with $L$ the system size. In the thermodynamic limit, the trace should be finite, and we can rescale the tensors $A^{pq} \to A^{pq}/\lambda_1$ to achieve this. We henthforth assume $\lambda_1 = 1$.

\begin{figure}
\begin{center}
  \includegraphics[width=.35\textwidth]{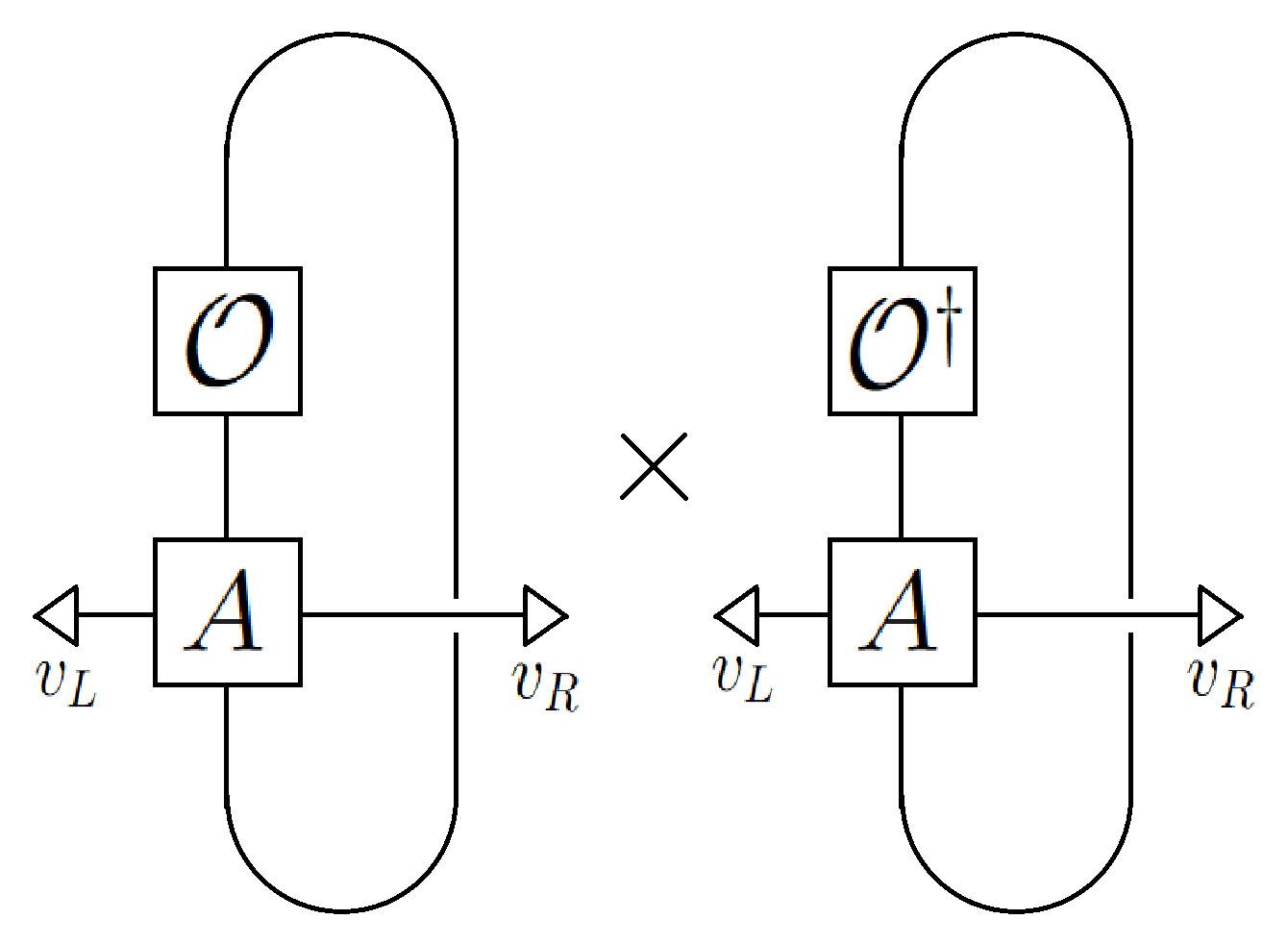} 
\end{center}
\caption{
The asymptotic form of the order parameter $C_3$ when the transfer matrix $E_1$ possesses a non-degenerate largest (in magnitude) eigenvalue.
}
\label{Fig:C3}
\end{figure}

We have the following theorem,
\begin{theorem}
    \label{thm:E1transfer}
    (1) The order parameter $C_3$, with $O$ being in a nontrivial representation of the average symmetry, cannot exhibit long-range order in any symmetric state where $E^1$ possesses a non-degenerate largest eigenvalue. (2) It is impossible to prepare a state with long-range $C_3$ using a symmetric finite-depth channel starting from any state with exponentially decaying $C_3$.
\end{theorem}
The proof of this theorem goes as follows. If $E^1$ has a unique largest eigenvalue with $v_L$ and $v_R$ being the corresponding left and right eigenvectors, the asymptotic form of the order parameter $C_3$, up to corrections that are exponentially small in $|x-y|$, can be represented as shown in Fig.~\ref{Fig:C3}. Namely, $C_3 \xrightarrow{|x-y|\to \infty} \llangle O_x | \rho \rrangle \llangle O_y^\dagger | \rho \rrangle = 0$ factorizes into the product of two one-point functions of charged operators, resulting in zero due to the symmetry of the state $\rho$. Moreover, in any state that can be prepared from a state with exponentially decaying $C_3$ using a symmetric finite-depth channel $\mathcal{E}$, the order parameter exhibits the following behavior:
\begin{equation}
    C_3 = \mathrm{tr}[O_x O_y^\dagger \mathcal{E}(\rho)] = \mathrm{tr}[\mathcal{E}^\dagger(O_x) \mathcal{E}^\dagger(O_y^\dagger) \rho],
\end{equation}
where $\mathcal{E}^\dagger$ represents the adjoint channel of $\mathcal{E}$ in Heisenberg picture. Notably, $\mathcal{E}^\dagger(O_x)$ remains a local operator with the same symmetry charge as $O_x$. Consequently, this correlation function must vanish at large distances.

Theorem \ref{thm:E1transfer} indicates that, starting from a symmetric state, a symmetric finite-depth channel cannot lead to the SSB of an average symmetry (or an average subgroup of an exact symmetry). In Sec.~\ref{sec:deformations}, we have encountered an example, specifically the maximally dephased cluster chain, in which the transfer matrix $E^1$ remains non-degenerate (as can be checked through an explicit calculation), while $E^2$ exhibits a degenerate largest (in magnitude) eigenvalue when $p=1/2$. This phenomenon is precisely an indication of the exact-to-average SSB, as it results in long-range order in $C_2$ (one can choose $O$ to be the Pauli $Z$ operator as the order parameter), but does not manifest in any correlator of the form $C_3$. Physically, order parameters quadratic in the density matrix can detect an additional symmetry-breaking order compared to correlators which are linear in the density matrix, namely the exact-to-average SSB. This observation motivates the following conjecture, which currently lacks a rigorous proof:
\begin{conjecture}
Assuming the tensor $A^{pq}$ generates an MPDO, if the transfer matrix $E^2$ of the Choi state possesses a unique largest eigenvalue, then $E^1$ also exhibits a unique largest (in magnitude) eigenvalue.
\label{conj:transfermatrices}
\end{conjecture}
It is worth noting that the Perron-Frobenius theorem \cite{Wolf2012note} ensures that the largest (in magnitude) eigenvalue of $E^2$ is always a real positive number. The underlying idea of Conjecture \ref{conj:transfermatrices} is that the absence of SSB order in R\'enyi-2 correlators (identified by the non-degeneracy of the largest eigenvalue of $E^2$) implies the absence of SSB order in R\'enyi-1 correlators as well.

In addition to order parameters, another approach for diagnosing (the absence of) SSB is to use disorder parameters \cite{2017disorderarameter}, denoted by $U_A$. These are non-local operators that implement a symmetry transformation within a large but finite region $A$, accompanied by endpoint operators at the boundary of that region, namely $U_A = O_L (\otimes_{i\in A} u^i) O_R$. Whenever a disorder parameter acquires a finite expectation value, it implies that the corresponding symmetry remains unbroken. Like the order parameters discussed earlier, there are three possible disorder parameters in the context of mixed states:
\begin{equation}
    \begin{split}
        D_1 = & \llangle \rho | U_A \bigotimes I | \rho \rrangle / \llangle \rho | \rho \rrangle, \\
        D_2 = & \llangle \rho | U_A \bigotimes U_A^* | \rho \rrangle/\llangle \rho | \rho \rrangle, \\
        D_3 = & \llangle I | U_A | \rho \rrangle. 
    \end{split}
    \label{eq:disorderparameters}
\end{equation}
$D_1$ and $D_2$ are the standard disorder parameters for the pure state $| \rho \rrangle$, where their long-range orders indicate the preservation of an exact symmetry and an average symmetry (which might be an average subgroup of an exact symmetry), respectively. The equivalence between the diagnoses based on order and disorder parameters has been established in $1d$ \cite{2020levinSSB}: A pure state with a finite symmetry group and a finite bond dimension MPS description is guaranteed to exhibit either a nonzero order parameter or a nonzero disorder parameter, but not both, see Table.\ref{tab:SSBpatterns}. Moreover, from a physical perspective, we also expect that a long-range order of $D_3$ similarly indicates the preservation of an exact symmetry. However, it remains unclear whether the two disorder parameters, $D_1$ and $D_3$, are equivalent. In the following, we establish a relation between these two diagnostic tools.  
\begin{table}
\renewcommand\arraystretch{1.4}
\centering
\begin{tabular}{|c|c|c|c|c|}
\hline
Symmetry & $C_1$ & $C_2$ & $D_1$ & $D_2$ \\
\hline
Unbroken & 0 & 0 & $O(1)$ & $O(1)$\\
\hline
$K\times K \leadsto K$  & 0  & $O(1)$ & 0 & $O(1)$\\
\hline
$K\times K \leadsto 0$ & $O(1)$ & $O(1)$ & 0 & 0\\
\hline
\end{tabular}
\caption{The behavior of various diagnostics in a $1d$ Choi state $|\rho\rrangle$ with $K\times K$ exact symmetry and an MPS representation. The operator $O$ in $C_1$ and $C_2$ is chosen as a generic local operator transformed in a nontrivial representation of $K$.}
\label{tab:SSBpatterns}
\end{table}

\begin{theorem}
    In $1d$, if $\rho$ can be represented by an MPDO with a finite bond dimension, the presence of a non-zero disorder parameter $D_1$ implies the existence of a non-zero $D_3$. 
    \label{thm:relationbetweenDs}
\end{theorem}

The theorem is proven in Appendix~\ref{app:proofofD}. The connection between Theorem \ref{thm:relationbetweenDs} and Conjecture \ref{conj:transfermatrices} is as follows. As demonstrated in Appendix~\ref{app:proofofD}, the existence of a nonzero $D_1$ indicates that (after blocking a finite number of sites) the local tensor $A^{pq}$ is symmetric and injective (or a direct sum of symmetric injective blocks). This characteristic, which also ensures a non-zero $D_3$, can be linked to the nondegeneracy of $E^2$ in Conjecture \ref{conj:transfermatrices}. However, the precise connection between the disorder parameter $D_3$ and the spectral properties of $E^1$ remains unclear.

As an illustration, consider the maximally dephased cluster chain with the density matrix given in Eq.~(\ref{eq:dephasedcluster}). This system exhibits an exact-to-average SSB, where both $\Z_2$ exact symmetries are broken down to the average subgroup. This SSB is characterized by a long-range order in $C_2 = \llangle \rho | Z_x Z_y \bigotimes Z_x Z_y | \rho \rrangle/\llangle \rho | \rho \rrangle$, and in this state, the disorder parameters $D_1$ and $D_3$ vanish for any local endpoint operators. In fact, it is possible to generalize this example from $K=\Z_2$ to any finite exact symmetry $K$: the state
\begin{equation}\label{etassb-example}
    \rho = \frac{1}{|K|} \sum_{k\in K} U_k
\end{equation}
exhibits exact-to-average ($K\times K \leadsto K$) SSB. This density matrix can always be expressed as an MPDO with a bond dimension $|K|$ (the order of the group). To understand this, consider an orthonormal local basis labeled by group elements, denoted as $\{| k \rangle \}_{k\in K}$. The tensor of the MPDO representing $\rho = \frac{1}{|K|} \sum_{k\in K} U_k$ takes the form $A^{kk',k'}=\mathrm{diag}(0,...,1,...,0)$, where only the $k$-th element is 1, with all others being 0. 

Finally, we discuss the robustness of the disorder parameter $D_3$ and its implications for open system dynamics.
\begin{theorem}
    If the disorder parameter $D_3$ vanishes in a state $\rho$ for any local endpoint operators, then any symmetric finite-depth channel cannot alter the asymptotic behavior of $D_3$ in $\rho$, in the sense that $D_3$ remains zero for arbitrary local endpoint operators. Note that here we do not require $\rho$ to be represented by an MPDO with a finite bond dimension.
    \label{thm:D3robust}
\end{theorem}
To see this, consider a symmetric finite-depth channel $\mathcal{E}$ and a disorder parameter $U_A$ with an arbitrary local endpoint operator. We have
\begin{equation}
    D_3[\mathcal{E}(\rho)] = \mathrm{tr}[\mathcal{E}^\dagger(U_A)\rho].
\end{equation}
Notably, due to the symmetric finite-depth nature of $\mathcal{E}$, $\mathcal{E}^\dagger(U_A)$ is still a disorder parameter with a local endpoint operator (see Ref. \cite{2022openspt} for a review of this fact). Due to our initial assumption, $\mathcal{E}^\dagger(U_A)$ vanishes within the state $\rho$, thereby establishing the theorem.

Theorem~\ref{thm:D3robust} has a surprising implication for dynamics of open systems. Let us focus on generic open systems governed by a Lindblad master equation,
\begin{equation}
\begin{split}
&\frac{d}{dt} | \rho \rrangle = \mathcal{L} |\rho \rrangle, \\
    \mathcal{L} &= -i(H_+- H_-^*) \\
    & + \sum_m [L_{m,+}L_{m,-}^*-\frac{1}{2}(L_{m,+}^\dagger L_{m,+} + L_{m,-}^T L_{m,-}^*)],   
    \label{eq:Lindbladdynamics}
\end{split}
\end{equation}
where $O_\pm$ represent operators acting on the ket and bra spaces, respectively. The jump operators $L_m$ capture the couplings between the system and its environment. Physically, $\mathcal{L}$ can be viewed as generating a continuous version of a quantum channel. Similar to channels discussed earlier, a Lindbladian $\mathcal{L}$ exhibits an exact symmetry $K$ if both the Hamiltonian and the jump operators commute with the symmetry operator for each $k\in K$:
\begin{equation}
     [U_k,H]=[U_k,L_m]=0.
\label{eq:jumpoperatorexactsym}
\end{equation}
In this study, we concentrate on local Lindbladians, which can be expressed as a summation of local terms, with each term having a norm bounded by a constant independent of the system size.

Due to the trace-preserving and completely-positive properties of Lindblad dynamics, the real part of the eigenvalues of $\mathcal{L}$ must be non-positive, and there must be at least one right eigenvector with an eigenvalue $\lambda = 0$, which corresponds to the steady state of the Lindblad equation. The dissipative gap $\Delta$, is defined as the gap in the real part of the spectrum of $\mathcal{L}$. Similar to the excitation gap in a Hamiltonian system, $\Delta$ plays a crucial role in Lindbladian dynamics. It determines the relaxation time scale to the steady state, as well as the system's stability to local perturbations and the form of correlation functions in the steady state \cite{2004Hastingssteadystate,2015rapidmixing1,2015rapidmixing2}. Regarding the dissipative gap, we can make the following statement:
\begin{theorem}
    For a local Lindbladian $\mathcal{L}$ with a discrete exact symmetry, the dissipative gap $\Delta$ can only be nonzero in the thermodynamic limit when the steady state exhibits spontaneous breaking of the exact symmetry. This is characterized by a non-vanishing R\'enyi-2 order parameter $C_2$ and the vanishing of both $D_1$ and $D_3$ for arbitrary endpoint operators in the steady state.\footnote{Here, we assume that the system possesses translational symmetry, thereby excluding the special cases of non-Hermitian skin effect \cite{2019skineffect}.}
    \label{thm:lindbladingappability}
\end{theorem}

The proof of the theorem proceeds as follows. We consider the initial state $\rho_{\mathrm{in}}$ given by Eq.~(\ref{etassb-example}), which exhibits vanishing $D_3$ for any local endpoint operators. If the gap $\Delta$ is nonvanishing, the system relaxes to a steady state $\rho_s$ within a finite time on the order of $1/\Delta$. The steady state $\rho_s$ must also have vanishing $D_3$. This is because, according to the Lieb-Robinson bound \cite{2010openLiebRobinson}, within a finite time, a disorder parameter evolved by the adjoint Lindbladian remains a disorder parameter with a local endpoint operator, which has a vanishing expectation value with respect to $\rho_{\mathrm{in}}$. Moreover, given that the steady state $\rho_s$ can also be attained from a product state within a finite time, it can be described by an MPDO with a finite bond dimension. By virtue of Theorem~\ref{thm:relationbetweenDs}, it follows that $\rho_s$ exhibits vanishing R\'enyi-2 disorder parameter $D_1$ for arbitrary endpoint operators. Then Table~\ref{tab:SSBpatterns}, shows that $\rho_s$ must also exhibit a nonvanishing $C_2$ with an appropriate order parameter $O$ charged under $K$. This indicates a spontaneous breaking of the exact symmetry. Additionally, as all correlation functions $\mathrm{tr}(\rho O_x O_y^\dagger)$ exponentially decay in $\rho_s$ due to the Lieb-Robinson bound, we anticipate that the exact symmetry breaks down to its average subgroup.

In contrast to the Lieb-Schultz-Mattis (LSM) Theorem \cite{lieb1961two,2000Oshikawa,200HastingsLSM} in closed quantum systems, our ``ingappability" condition of a Lindbladian does not require a projective representation on each unit cell. Exploring the connection between Theorem~\ref{thm:lindbladingappability} and the recently proposed LSM theorem in open systems \cite{2023openLSM} is an interesting question that we leave for future study.

Additionally, one might ask whether a gapped steady state with a non-zero R\'enyi-2 order parameter $C_2$ violates the Lieb-Robinson bound in Lindblad dynamics \cite{2010openLiebRobinson}. It is important to note that the Lieb-Robinson bound restricts the speed of correlation propagation \emph{only} in R\'enyi-1 correlators, for instance, $C_3$. As demonstrated explicitly in the example in Eq.~(\ref{eq:dephasedcluster}), a long-range order in R\'enyi-2 correlators can be generated by a finite-depth local channel. Consequently, we conclude that, \emph{in open systems, there is no Lieb-Robinson bound for R\'enyi-2 correlation functions}.

\section{Mixed state anomalies and boundary properties}
\label{sec:boundary}

One of the defining properties of pure SPT states is that a nontrivial bulk leads to nontrivial boundaries. More precisely, a nontrivial boundary is guaranteed by the t'Hooft anomaly, which is ultimately determined by SPT invariant of the bulk. The anomalous boundary cannot be featureless, in the sense that a unique gapped (SRE) ground state on the boundary is prohibited. A natural question is, how does a bulk nontrivial MSPT phase constrain its boundary properties? In other words, are there any observable quantities that may signal a ``mixed state" anomaly?

The Choi state $|\rho\rrangle$ of a nontrivial MSPT state is a pure SPT state. If we put the state $| \rho \rrangle$ on a space manifold with boundary, the boundary state in the doubled space has a t'Hooft anomaly. As a result, the boundary of $| \rho \rrangle$ must be long-range entangled in the doubled space, which can be further divided into three categories: (1) The boundary state spontaneously breaks the protecting symmetry of $| \rho \rrangle$. (2) Certain correlation functions of local operators may exhibit power-law decay on the boundary. This situation can be understood as arising when a Hamiltonian with $| \rho \rrangle$ as a ground state becomes gapless on a space with boundary. (3) The boundary state of $| \rho \rrangle$ may develop intrinsic topological order. We now illustrate the physical consequences of each of these cases for the MSPT state.

Let us first consider the case where the boundary of $| \rho \rrangle$ breaks the symmetry spontaneously, which means a charged local operator exhibits a long-range (non-decaying) correlation function. We will focus on cases where the bulk SPT is $1d$, where we can offer more rigorous statements by employing the tools of MPS and MPDO. In the context of a familiar $1d$ pure state SPT, boundary SSB simply means the emergence of a localized mode at each boundary. The symmetry acts projectively on these modes, and the two modes on the separate endpoints form a singlet. A natural question to ask is: What is the counterpart of a ``localized projective boundary mode" in the original mixed state? When it comes to MSPT states of various classes in Eq.~(\ref{eq:MSPTclassification}), what are the associated projective transformation (or SSB) patterns at the boundary?

Since the bulk MSPT is SRE, its transfer matrix $E^2$ has a non-degenerate largest eigenvalue. If we focus on the subset of these states for which $E^1$ also has a unique largest (in magnitude) eigenvalue (which, according to Conjecture~\ref{conj:transfermatrices}, is all of them), there is a sense in which they have localized boundaries modes: In such states, any bulk-to-boundary correlation functions, both the R\'enyi-1 or R\'enyi-2 types, exhibit exponential decay.

Next, we may inquire about the potential boundary projective representation (or SSB) patterns for a $1d$ MSPT. We present the following theorem:
\begin{theorem}
    In a $1d$ MSPT solely protected by the exact symmetry, as classified by $H^2(K,U(1))$ in Eq.(\ref{eq:MSPTclassification}), the exact symmetry is broken at the boundary such that no diagonal average subgroup is left unbroken. For an MSPT jointly protected by an exact and an average symmetry, as classified by $H^1[G,H^1(K,U(1))]$ in Eq.(\ref{eq:MSPTclassification}), at the boundary the exact symmetry is broken to its diagonal subgroup, while the average symmetry is broken.\footnote{For either case, there could exist a subgroup of $K$ (or $K\times G$) that remains intact. For example, if $K$ (or $K\times G$) is finite Abelian, this is the projective center.}
\end{theorem}

The proof of this theorem relies on symmetry fractionalization within the Choi state $| \rho \rrangle$. Let us consider a chain with a length $L$ and apply (a single copy of) the exact symmetry, say $K_+$, to $| \rho \rrangle$ within a region denoted as $A$, which comprises all sites ranging from $m$ to $n$, with both $m$ and $L-n$ larger than the bulk correlation length, yet remaining finite. The symmetry operation then fractionalizes, taking the form:
\begin{equation}
\prod_{i=m}^n u_k^i \bigotimes \mathbf{1} | \rho \rrangle = V_lV_r| \rho \rrangle.
\label{eq:boundarySSB1}
\end{equation}
Since the state remains invariant under a global action by $K_+$, Eq.~(\ref{eq:boundarySSB1}) means
\begin{equation}
\begin{split}
    \llangle \rho | [V_l^\dagger (\prod_{i=1}^m u_k^i \bigotimes \mathbf{1}) ] [V_r^\dagger(\prod_{i=n}^L u_k^i \bigotimes \mathbf{1} )] | \rho \rrangle/\llangle \rho | \rho \rrangle & \sim O(1),\\
    \tr [(V_l^\dagger\prod_{i=1}^m u_k^i  ) (V_r^\dagger\prod_{i=n}^L u_k^i)   \rho ] & \sim O(1).
\end{split}
\end{equation}
As $|m-n|\gg 1$, when the support of $V_l^\dagger (\prod_{i=1}^m u_k^i$ and $V_r^\dagger(\prod_{i=n}^L u_k^i \bigotimes \mathbf{1} )$ is small compared to the size $L$ of the bulk, these operators can be viewed as boundary operators. And when $| \rho \rrangle$ represents a $1d$ MSPT protected solely by the exact symmetry $K$, they carry a nontrivial charge under the symmetry $K_+$. Therefore, this correlation function may be interpreted as the exact symmetry being completely broken (there is no average subgroup left) at the boundary. Likewise, when $| \rho \rrangle$ represents a $1d$ MSPT protected jointly by $K$ and $G$, they carry a charge under $G$, which means the average symmetry is broken at the boundary.

Similarly, in the case of an MSPT jointly protected by $K$ and $G$, the symmetry fractionalization pattern
\begin{equation}
    \prod_{i=m}^n u_k^i \bigotimes (u_k^i)^* | \rho \rrangle = V_l V_r | \rho \rrangle
\end{equation}
means there is a correlator
\begin{equation}
\begin{split}
    \llangle \rho | [\prod_{i=1}^m u_g^i \bigotimes (u_g^i)^*] V_l^\dagger [\prod_{i=n}^L u_k^i \bigotimes (u_g^i)^*] & V_r^\dagger | \rho \rrangle/\llangle \rho | \rho \rrangle\\ & \sim O(1),
\end{split}
\end{equation}
of boundary operators. They now carry charges under $K_+$ and $K_-$ but are neutral under $K_+\times K_-$. This demonstrates that the exact symmetry is broken to its average subgroup at the boundary of an MSPT protected jointly by an exact and an average symmetry.

We now turn to the second possibility, which is a power-law correlated boundary state. In particular, the correlation functions, $C_1$ and $C_2$, may exhibit power-law decaying on the boundary. In the case of a $2d$ MSPT state, since it lacks intrinsic topological orders at its boundary, these two possibilities—the power-law correlated boundary state and the boundary state that spontaneously breaks symmetry, as discussed in the preceding paragraph—are all the possibilities at the boundary. As demonstrated in Sec.~\ref{sec:deformations}, this power-law correlated boundary state cannot be prepared by any finite-depth local quantum channel, starting from a product state -- This is due to the fact that a finite-depth channel can only prepare MPDOs with finite bond dimensions. Furthermore, one can argue that a state with power-law correlated R\'enyi-2 correlators is not symmetrically separable. To see this, let us first assume that it is: $\rho = \sum_i p_i |\psi_i \rangle \langle \psi_i |$ where each $|\psi_i\rangle$ can be prepared by a symmetric finite-depth unitary. The R\'enyi-2 correlator can be expressed as
\begin{equation}
    C_2 = \frac{\sum_{ij}p_ip_j|\langle\psi_i|O_x O_y^\dagger| \psi_j\rangle|^2}{\sum_{ij}p_ip_j|\langle\psi_i| \psi_j\rangle|^2}.
\end{equation}
Following Ref.~\cite{2014strange}, a spacetime rotation transforms $\langle\psi_i|O_x O_y^\dagger| \psi_j\rangle$ to a temporal correlation function at the spatial interface between $|\psi_i\rangle$ and $| \psi_j\rangle$. Since all states in the decomposition are trivial SPT states, this correlation should decay exponentially with $|x-y|$, leading to a contradiction.

The last, and the most exotic possibility, is that the boundary state of $| \rho \rrangle$, has intrinsic topological order in the doubled space. This scenario is possible only when the bulk has dimensional three or above. Recently, Abelian topological orders and the phase transitions between them in the doubled space have been proposed as a characterization of the breakdown of topological quantum memory caused by decoherence \cite{2023decobao,2023decofan,2023decoxu}. We leave a detailed study of topological orders in mixed states, and their ability to encode classical and quantum information to future study.  

\section{Summary and outlook}

\label{sec:summary}
In this work, we have established a systematic framework for classifying and characterizing bosonic mixed-state symmetry-protected topological (MSPT) phases. We have also investigated patterns of spontaneous symmetry breaking (SSB) in mixed states. Our approach has been to analyze the properties of mixed states by examining the symmetry and topology of their corresponding Choi states in the doubled space and then exploring the consequences for the original mixed states. Our main results include the following:
\begin{enumerate}
    \item Using the operator-state map, we defined short-range entangled (SRE) mixed states as those that have SRE Choi states in the doubled Hilbert space. SRE mixed states, thus defined, exhibit area law mutual information and operator entanglement entropy. This mapping provides us with the ability to understand the nontrivial topology in mixed states using well-established concepts (e.g., SPT invariants) and tools (e.g., tensor networks) commonly applied to pure states.

    \item Subsequently, we defined MSPT states as those with SPT Choi states and classified their invariants, finding that hermiticity and non-negativity of the density matrix place constraints on the types of SPT invariants that may arise. In total, we demonstrated that the classification is characterized by Theorems~\ref{thm:MSPTclassification}, \ref{thm:MSPTclassificationwithT}, \ref{thm:MSPTclassificationgeneral}. Notably, we proved that the MSPT invariants of states with exact symmetry $K$ and average symmetry $G$ are classified by Theorem ~\ref{thm:MSPTclassification}:
    \begin{equation}
        \bigoplus_{p=0}^d H^p[G,H^{d+1-p}(K,U(1))].
    \end{equation}
    Furthermore, we established that nontrivial MSPT states exhibit boundary properties that depend on the bulk SPT invariants. And we elucidated the connection between the SPT invariants of the Choi state and the separability of the original mixed states, along with the implications for the symmetry-protected sign problem.

    \item We showed that, when the SRE nature of a state is preserved by a symmetric finite-depth quantum channel (and its spatially truncated channels in $d>1$), its MSPT invariant also remains unchanged. We also provided an example where the SRE property is violated under a symmetric finite-depth channel, leading to final states that exhibit an exact-to-average spontaneous symmetry breaking.

    \item We explored potential SSB patterns in mixed states and studied how to detect them with generalized order and disorder parameters. Additionally, we investigated the consequences of these SSB patterns for the properties the local tensors of matrix product density operators. We demonstrated that any local Lindbladian with a discrete exact symmetry can only exhibit a nonzero dissipative gap when the steady state undergoes spontaneous breaking of the exact symmetry.

\end{enumerate}

We end with some open questions:
\begin{enumerate}
    \item \emph{Fermionic MSPTs}: In this work we focused on bosonic MSPT phases. A natural question is how to extend the theory to fermionic systems. Can we classify fermionic mixed states based on the topology of their Choi states, and what are the consequences for the original mixed state, such as separability and the behavior of boundary correlation functions? Since fermionic SPT states can also be analyzed using the decorated domain wall construction \cite{2019gaiottospectralsequence}, we expect our approach can be applied to determine whether a fermionic Choi state is consistent with non-negativity.

    \item \emph{SSB in mixed states}: As previously discussed, various aspects of SSB in mixed states remain unclear. One important direction is establishing the precise relations between the different order and disorder parameters introduced in this study as well as their relations with the spectral properties of tensor networks that exhibit the corresponding SSB patterns. We also wonder whether one can define order parameters for SSB in mixed states using concepts from quantum information theory, such as distinguishability quantified through quantum relative entropy. What is the stability of SSB under environmental decoherence? A comprehensive theory addressing these questions will be presented in an upcoming work.

    \item \emph{Topological orders in mixed states}: In this work, we limited our focus to SRE mixed states and those with LRE due to SSB. A remaining task is to study LRE mixed states, especially mixed state topological orders. The present study motivates the question of whether the doubled state formalism might play a role in defining mixed state topological order. What potential topological orders can exist in the Choi state, taking into account the non-negativity constraint? For example, gauging the symmetry (such as according to the quantum state-gauging procedure of Ref.~\cite{2015stategauging}) seems to produce a family of topological orders in the doubled space -- what is their meaning? It would also be interesting to study connections between topological orders in the doubled state and the capacity to protect quantum information in the original mixed state.
\end{enumerate}

\begin{acknowledgements}

We thank Yimu Bao, Meng Cheng, Jong Yeon Lee, Chong Wang, Matthew Yu and Yijian Zou for the helpful discussions. This work was supported in part by Simons Investigator award number 990660  (R. M.) and by the Simons Collaboration on Ultra-Quantum Matter, which is a grant from the Simons Foundation (No. 651440, R. M.). Research at Perimeter Institute is supported in part by the Government of Canada through the Department of Innovation, Science and Industry Canada and by the Province of Ontario through the Ministry of Colleges and Universities.

\emph{Note}: Upon completion of the manuscript, we became aware of another work \cite{2024BaoY} that also discusses the constraints on the possible SPT phases in the doubled space.

\end{acknowledgements}

\bibliography{Ref.bib} 

\begin{thebibliography}{81}%
\makeatletter
\providecommand \@ifxundefined [1]{%
 \@ifx{#1\undefined}
}%
\providecommand \@ifnum [1]{%
 \ifnum #1\expandafter \@firstoftwo
 \else \expandafter \@secondoftwo
 \fi
}%
\providecommand \@ifx [1]{%
 \ifx #1\expandafter \@firstoftwo
 \else \expandafter \@secondoftwo
 \fi
}%
\providecommand \natexlab [1]{#1}%
\providecommand \enquote  [1]{``#1''}%
\providecommand \bibnamefont  [1]{#1}%
\providecommand \bibfnamefont [1]{#1}%
\providecommand \citenamefont [1]{#1}%
\providecommand \href@noop [0]{\@secondoftwo}%
\providecommand \href [0]{\begingroup \@sanitize@url \@href}%
\providecommand \@href[1]{\@@startlink{#1}\@@href}%
\providecommand \@@href[1]{\endgroup#1\@@endlink}%
\providecommand \@sanitize@url [0]{\catcode `\\12\catcode `\$12\catcode
  `\&12\catcode `\#12\catcode `\^12\catcode `\_12\catcode `\%12\relax}%
\providecommand \@@startlink[1]{}%
\providecommand \@@endlink[0]{}%
\providecommand \url  [0]{\begingroup\@sanitize@url \@url }%
\providecommand \@url [1]{\endgroup\@href {#1}{\urlprefix }}%
\providecommand \urlprefix  [0]{URL }%
\providecommand \Eprint [0]{\href }%
\providecommand \doibase [0]{http://dx.doi.org/}%
\providecommand \selectlanguage [0]{\@gobble}%
\providecommand \bibinfo  [0]{\@secondoftwo}%
\providecommand \bibfield  [0]{\@secondoftwo}%
\providecommand \translation [1]{[#1]}%
\providecommand \BibitemOpen [0]{}%
\providecommand \bibitemStop [0]{}%
\providecommand \bibitemNoStop [0]{.\EOS\space}%
\providecommand \EOS [0]{\spacefactor3000\relax}%
\providecommand \BibitemShut  [1]{\csname bibitem#1\endcsname}%
\let\auto@bib@innerbib\@empty
\bibitem [{\citenamefont {{McGreevy}}(2023)}]{2023McGreevyreview}%
  \BibitemOpen
  \bibfield  {author} {\bibinfo {author} {\bibfnamefont {John}\ \bibnamefont
  {{McGreevy}}},\ }\bibfield  {title} {\enquote {\bibinfo {title} {{Generalized
  Symmetries in Condensed Matter}},}\ }\href {\doibase
  10.1146/annurev-conmatphys-040721-021029} {\bibfield  {journal} {\bibinfo
  {journal} {Annual Review of Condensed Matter Physics}\ }\textbf {\bibinfo
  {volume} {14}},\ \bibinfo {pages} {57--82} (\bibinfo {year} {2023})},\
  \Eprint {http://arxiv.org/abs/2204.03045} {arXiv:2204.03045
  [cond-mat.str-el]} \BibitemShut {NoStop}%
\bibitem [{\citenamefont {{Bernien}}\ \emph {et~al.}(2017)\citenamefont
  {{Bernien}}, \citenamefont {{Schwartz}}, \citenamefont {{Keesling}},
  \citenamefont {{Levine}}, \citenamefont {{Omran}}, \citenamefont {{Pichler}},
  \citenamefont {{Choi}}, \citenamefont {{Zibrov}}, \citenamefont {{Endres}},
  \citenamefont {{Greiner}}, \citenamefont {{Vuleti{\'c}}},\ and\ \citenamefont
  {{Lukin}}}]{2017NISQ}%
  \BibitemOpen
  \bibfield  {author} {\bibinfo {author} {\bibfnamefont {Hannes}\ \bibnamefont
  {{Bernien}}}, \bibinfo {author} {\bibfnamefont {Sylvain}\ \bibnamefont
  {{Schwartz}}}, \bibinfo {author} {\bibfnamefont {Alexander}\ \bibnamefont
  {{Keesling}}}, \bibinfo {author} {\bibfnamefont {Harry}\ \bibnamefont
  {{Levine}}}, \bibinfo {author} {\bibfnamefont {Ahmed}\ \bibnamefont
  {{Omran}}}, \bibinfo {author} {\bibfnamefont {Hannes}\ \bibnamefont
  {{Pichler}}}, \bibinfo {author} {\bibfnamefont {Soonwon}\ \bibnamefont
  {{Choi}}}, \bibinfo {author} {\bibfnamefont {Alexander~S.}\ \bibnamefont
  {{Zibrov}}}, \bibinfo {author} {\bibfnamefont {Manuel}\ \bibnamefont
  {{Endres}}}, \bibinfo {author} {\bibfnamefont {Markus}\ \bibnamefont
  {{Greiner}}}, \bibinfo {author} {\bibfnamefont {Vladan}\ \bibnamefont
  {{Vuleti{\'c}}}}, \ and\ \bibinfo {author} {\bibfnamefont {Mikhail~D.}\
  \bibnamefont {{Lukin}}},\ }\bibfield  {title} {\enquote {\bibinfo {title}
  {{Probing many-body dynamics on a 51-atom quantum simulator}},}\ }\href
  {\doibase 10.1038/nature24622} {\bibfield  {journal} {\bibinfo  {journal}
  {\nat}\ }\textbf {\bibinfo {volume} {551}},\ \bibinfo {pages} {579--584}
  (\bibinfo {year} {2017})},\ \Eprint {http://arxiv.org/abs/1707.04344}
  {arXiv:1707.04344 [quant-ph]} \BibitemShut {NoStop}%
\bibitem [{\citenamefont {{Preskill}}(2018)}]{2018NISQ}%
  \BibitemOpen
  \bibfield  {author} {\bibinfo {author} {\bibfnamefont {John}\ \bibnamefont
  {{Preskill}}},\ }\bibfield  {title} {\enquote {\bibinfo {title} {{Quantum
  Computing in the NISQ era and beyond}},}\ }\href {\doibase
  10.22331/q-2018-08-06-79} {\bibfield  {journal} {\bibinfo  {journal}
  {Quantum}\ }\textbf {\bibinfo {volume} {2}},\ \bibinfo {pages} {79} (\bibinfo
  {year} {2018})},\ \Eprint {http://arxiv.org/abs/1801.00862} {arXiv:1801.00862
  [quant-ph]} \BibitemShut {NoStop}%
\bibitem [{\citenamefont {{de Groot}}\ \emph {et~al.}(2022)\citenamefont {{de
  Groot}}, \citenamefont {{Turzillo}},\ and\ \citenamefont
  {{Schuch}}}]{2022openspt}%
  \BibitemOpen
  \bibfield  {author} {\bibinfo {author} {\bibfnamefont {Caroline}\
  \bibnamefont {{de Groot}}}, \bibinfo {author} {\bibfnamefont {Alex}\
  \bibnamefont {{Turzillo}}}, \ and\ \bibinfo {author} {\bibfnamefont
  {Norbert}\ \bibnamefont {{Schuch}}},\ }\bibfield  {title} {\enquote {\bibinfo
  {title} {{Symmetry Protected Topological Order in Open Quantum Systems}},}\
  }\href {\doibase 10.22331/q-2022-11-10-856} {\bibfield  {journal} {\bibinfo
  {journal} {Quantum}\ }\textbf {\bibinfo {volume} {6}},\ \bibinfo {pages}
  {856} (\bibinfo {year} {2022})},\ \Eprint {http://arxiv.org/abs/2112.04483}
  {arXiv:2112.04483 [quant-ph]} \BibitemShut {NoStop}%
\bibitem [{\citenamefont {{Ma}}\ and\ \citenamefont {{Wang}}(2023)}]{2022aspt}%
  \BibitemOpen
  \bibfield  {author} {\bibinfo {author} {\bibfnamefont {Ruochen}\ \bibnamefont
  {{Ma}}}\ and\ \bibinfo {author} {\bibfnamefont {Chong}\ \bibnamefont
  {{Wang}}},\ }\bibfield  {title} {\enquote {\bibinfo {title} {{Average
  Symmetry-Protected Topological Phases}},}\ }\href {\doibase
  10.1103/PhysRevX.13.031016} {\bibfield  {journal} {\bibinfo  {journal}
  {Physical Review X}\ }\textbf {\bibinfo {volume} {13}},\ \bibinfo {eid}
  {031016} (\bibinfo {year} {2023})},\ \Eprint
  {http://arxiv.org/abs/2209.02723} {arXiv:2209.02723 [cond-mat.str-el]}
  \BibitemShut {NoStop}%
\bibitem [{\citenamefont {{Ma}}\ \emph {et~al.}(2023)\citenamefont {{Ma}},
  \citenamefont {{Zhang}}, \citenamefont {{Bi}}, \citenamefont {{Cheng}},\ and\
  \citenamefont {{Wang}}}]{2023aspt}%
  \BibitemOpen
  \bibfield  {author} {\bibinfo {author} {\bibfnamefont {Ruochen}\ \bibnamefont
  {{Ma}}}, \bibinfo {author} {\bibfnamefont {Jian-Hao}\ \bibnamefont
  {{Zhang}}}, \bibinfo {author} {\bibfnamefont {Zhen}\ \bibnamefont {{Bi}}},
  \bibinfo {author} {\bibfnamefont {Meng}\ \bibnamefont {{Cheng}}}, \ and\
  \bibinfo {author} {\bibfnamefont {Chong}\ \bibnamefont {{Wang}}},\ }\bibfield
   {title} {\enquote {\bibinfo {title} {{Topological Phases with Average
  Symmetries: the Decohered, the Disordered, and the Intrinsic}},}\ }\href
  {\doibase 10.48550/arXiv.2305.16399} {\bibfield  {journal} {\bibinfo
  {journal} {arXiv e-prints}\ ,\ \bibinfo {eid} {arXiv:2305.16399}} (\bibinfo
  {year} {2023})},\ \Eprint {http://arxiv.org/abs/2305.16399} {arXiv:2305.16399
  [cond-mat.str-el]} \BibitemShut {NoStop}%
\bibitem [{\citenamefont {{Lee}}\ \emph
  {et~al.}(2022{\natexlab{a}})\citenamefont {{Lee}}, \citenamefont {{You}},\
  and\ \citenamefont {{Xu}}}]{2022decoxu}%
  \BibitemOpen
  \bibfield  {author} {\bibinfo {author} {\bibfnamefont {Jong~Yeon}\
  \bibnamefont {{Lee}}}, \bibinfo {author} {\bibfnamefont {Yi-Zhuang}\
  \bibnamefont {{You}}}, \ and\ \bibinfo {author} {\bibfnamefont {Cenke}\
  \bibnamefont {{Xu}}},\ }\bibfield  {title} {\enquote {\bibinfo {title}
  {{Symmetry protected topological phases under decoherence}},}\ }\href
  {\doibase 10.48550/arXiv.2210.16323} {\bibfield  {journal} {\bibinfo
  {journal} {arXiv e-prints}\ ,\ \bibinfo {eid} {arXiv:2210.16323}} (\bibinfo
  {year} {2022}{\natexlab{a}})},\ \Eprint {http://arxiv.org/abs/2210.16323}
  {arXiv:2210.16323 [cond-mat.str-el]} \BibitemShut {NoStop}%
\bibitem [{\citenamefont {{Bao}}\ \emph {et~al.}(2023)\citenamefont {{Bao}},
  \citenamefont {{Fan}}, \citenamefont {{Vishwanath}},\ and\ \citenamefont
  {{Altman}}}]{2023decobao}%
  \BibitemOpen
  \bibfield  {author} {\bibinfo {author} {\bibfnamefont {Yimu}\ \bibnamefont
  {{Bao}}}, \bibinfo {author} {\bibfnamefont {Ruihua}\ \bibnamefont {{Fan}}},
  \bibinfo {author} {\bibfnamefont {Ashvin}\ \bibnamefont {{Vishwanath}}}, \
  and\ \bibinfo {author} {\bibfnamefont {Ehud}\ \bibnamefont {{Altman}}},\
  }\bibfield  {title} {\enquote {\bibinfo {title} {{Mixed-state topological
  order and the errorfield double formulation of decoherence-induced
  transitions}},}\ }\href {\doibase 10.48550/arXiv.2301.05687} {\bibfield
  {journal} {\bibinfo  {journal} {arXiv e-prints}\ ,\ \bibinfo {eid}
  {arXiv:2301.05687}} (\bibinfo {year} {2023})},\ \Eprint
  {http://arxiv.org/abs/2301.05687} {arXiv:2301.05687 [quant-ph]} \BibitemShut
  {NoStop}%
\bibitem [{\citenamefont {{Fan}}\ \emph {et~al.}(2023)\citenamefont {{Fan}},
  \citenamefont {{Bao}}, \citenamefont {{Altman}},\ and\ \citenamefont
  {{Vishwanath}}}]{2023decofan}%
  \BibitemOpen
  \bibfield  {author} {\bibinfo {author} {\bibfnamefont {Ruihua}\ \bibnamefont
  {{Fan}}}, \bibinfo {author} {\bibfnamefont {Yimu}\ \bibnamefont {{Bao}}},
  \bibinfo {author} {\bibfnamefont {Ehud}\ \bibnamefont {{Altman}}}, \ and\
  \bibinfo {author} {\bibfnamefont {Ashvin}\ \bibnamefont {{Vishwanath}}},\
  }\bibfield  {title} {\enquote {\bibinfo {title} {{Diagnostics of mixed-state
  topological order and breakdown of quantum memory}},}\ }\href {\doibase
  10.48550/arXiv.2301.05689} {\bibfield  {journal} {\bibinfo  {journal} {arXiv
  e-prints}\ ,\ \bibinfo {eid} {arXiv:2301.05689}} (\bibinfo {year} {2023})},\
  \Eprint {http://arxiv.org/abs/2301.05689} {arXiv:2301.05689 [quant-ph]}
  \BibitemShut {NoStop}%
\bibitem [{\citenamefont {{Lee}}\ \emph {et~al.}(2023)\citenamefont {{Lee}},
  \citenamefont {{Jian}},\ and\ \citenamefont {{Xu}}}]{2023decoxu}%
  \BibitemOpen
  \bibfield  {author} {\bibinfo {author} {\bibfnamefont {Jong~Yeon}\
  \bibnamefont {{Lee}}}, \bibinfo {author} {\bibfnamefont {Chao-Ming}\
  \bibnamefont {{Jian}}}, \ and\ \bibinfo {author} {\bibfnamefont {Cenke}\
  \bibnamefont {{Xu}}},\ }\bibfield  {title} {\enquote {\bibinfo {title}
  {{Quantum Criticality Under Decoherence or Weak Measurement}},}\ }\href
  {\doibase 10.1103/PRXQuantum.4.030317} {\bibfield  {journal} {\bibinfo
  {journal} {PRX Quantum}\ }\textbf {\bibinfo {volume} {4}},\ \bibinfo {eid}
  {030317} (\bibinfo {year} {2023})},\ \Eprint
  {http://arxiv.org/abs/2301.05238} {arXiv:2301.05238 [cond-mat.stat-mech]}
  \BibitemShut {NoStop}%
\bibitem [{\citenamefont {{Lee}}\ \emph
  {et~al.}(2022{\natexlab{b}})\citenamefont {{Lee}}, \citenamefont {{Ji}},
  \citenamefont {{Bi}},\ and\ \citenamefont {{Fisher}}}]{2022bimeasure}%
  \BibitemOpen
  \bibfield  {author} {\bibinfo {author} {\bibfnamefont {Jong~Yeon}\
  \bibnamefont {{Lee}}}, \bibinfo {author} {\bibfnamefont {Wenjie}\
  \bibnamefont {{Ji}}}, \bibinfo {author} {\bibfnamefont {Zhen}\ \bibnamefont
  {{Bi}}}, \ and\ \bibinfo {author} {\bibfnamefont {Matthew P.~A.}\
  \bibnamefont {{Fisher}}},\ }\bibfield  {title} {\enquote {\bibinfo {title}
  {{Decoding Measurement-Prepared Quantum Phases and Transitions: from Ising
  model to gauge theory, and beyond}},}\ }\href {\doibase
  10.48550/arXiv.2208.11699} {\bibfield  {journal} {\bibinfo  {journal} {arXiv
  e-prints}\ ,\ \bibinfo {eid} {arXiv:2208.11699}} (\bibinfo {year}
  {2022}{\natexlab{b}})},\ \Eprint {http://arxiv.org/abs/2208.11699}
  {arXiv:2208.11699 [cond-mat.str-el]} \BibitemShut {NoStop}%
\bibitem [{\citenamefont {{Lu}}\ \emph {et~al.}(2023)\citenamefont {{Lu}},
  \citenamefont {{Zhang}}, \citenamefont {{Vijay}},\ and\ \citenamefont
  {{Hsieh}}}]{2023LuMeasure}%
  \BibitemOpen
  \bibfield  {author} {\bibinfo {author} {\bibfnamefont {Tsung-Cheng}\
  \bibnamefont {{Lu}}}, \bibinfo {author} {\bibfnamefont {Zhehao}\ \bibnamefont
  {{Zhang}}}, \bibinfo {author} {\bibfnamefont {Sagar}\ \bibnamefont
  {{Vijay}}}, \ and\ \bibinfo {author} {\bibfnamefont {Timothy~H.}\
  \bibnamefont {{Hsieh}}},\ }\bibfield  {title} {\enquote {\bibinfo {title}
  {{Mixed-State Long-Range Order and Criticality from Measurement and
  Feedback}},}\ }\href {\doibase 10.1103/PRXQuantum.4.030318} {\bibfield
  {journal} {\bibinfo  {journal} {PRX Quantum}\ }\textbf {\bibinfo {volume}
  {4}},\ \bibinfo {eid} {030318} (\bibinfo {year} {2023})},\ \Eprint
  {http://arxiv.org/abs/2303.15507} {arXiv:2303.15507 [cond-mat.str-el]}
  \BibitemShut {NoStop}%
\bibitem [{\citenamefont {{Chen}}\ \emph {et~al.}(2023)\citenamefont {{Chen}},
  \citenamefont {{Zhu}}, \citenamefont {{Verresen}}, \citenamefont {{Seif}},
  \citenamefont {{B{\"a}umer}}, \citenamefont {{Layden}}, \citenamefont
  {{Tantivasadakarn}}, \citenamefont {{Zhu}}, \citenamefont {{Sheldon}},
  \citenamefont {{Vishwanath}}, \citenamefont {{Trebst}},\ and\ \citenamefont
  {{Kandala}}}]{2023Zhu}%
  \BibitemOpen
  \bibfield  {author} {\bibinfo {author} {\bibfnamefont {Edward~H.}\
  \bibnamefont {{Chen}}}, \bibinfo {author} {\bibfnamefont {Guo-Yi}\
  \bibnamefont {{Zhu}}}, \bibinfo {author} {\bibfnamefont {Ruben}\ \bibnamefont
  {{Verresen}}}, \bibinfo {author} {\bibfnamefont {Alireza}\ \bibnamefont
  {{Seif}}}, \bibinfo {author} {\bibfnamefont {Elisa}\ \bibnamefont
  {{B{\"a}umer}}}, \bibinfo {author} {\bibfnamefont {David}\ \bibnamefont
  {{Layden}}}, \bibinfo {author} {\bibfnamefont {Nathanan}\ \bibnamefont
  {{Tantivasadakarn}}}, \bibinfo {author} {\bibfnamefont {Guanyu}\ \bibnamefont
  {{Zhu}}}, \bibinfo {author} {\bibfnamefont {Sarah}\ \bibnamefont
  {{Sheldon}}}, \bibinfo {author} {\bibfnamefont {Ashvin}\ \bibnamefont
  {{Vishwanath}}}, \bibinfo {author} {\bibfnamefont {Simon}\ \bibnamefont
  {{Trebst}}}, \ and\ \bibinfo {author} {\bibfnamefont {Abhinav}\ \bibnamefont
  {{Kandala}}},\ }\bibfield  {title} {\enquote {\bibinfo {title} {{Realizing
  the Nishimori transition across the error threshold for constant-depth
  quantum circuits}},}\ }\href {\doibase 10.48550/arXiv.2309.02863} {\bibfield
  {journal} {\bibinfo  {journal} {arXiv e-prints}\ ,\ \bibinfo {eid}
  {arXiv:2309.02863}} (\bibinfo {year} {2023})},\ \Eprint
  {http://arxiv.org/abs/2309.02863} {arXiv:2309.02863 [quant-ph]} \BibitemShut
  {NoStop}%
\bibitem [{\citenamefont {{Zhu}}\ \emph {et~al.}(2023)\citenamefont {{Zhu}},
  \citenamefont {{Tantivasadakarn}}, \citenamefont {{Vishwanath}},
  \citenamefont {{Trebst}},\ and\ \citenamefont {{Verresen}}}]{2023ZHUPRL}%
  \BibitemOpen
  \bibfield  {author} {\bibinfo {author} {\bibfnamefont {Guo-Yi}\ \bibnamefont
  {{Zhu}}}, \bibinfo {author} {\bibfnamefont {Nathanan}\ \bibnamefont
  {{Tantivasadakarn}}}, \bibinfo {author} {\bibfnamefont {Ashvin}\ \bibnamefont
  {{Vishwanath}}}, \bibinfo {author} {\bibfnamefont {Simon}\ \bibnamefont
  {{Trebst}}}, \ and\ \bibinfo {author} {\bibfnamefont {Ruben}\ \bibnamefont
  {{Verresen}}},\ }\bibfield  {title} {\enquote {\bibinfo {title} {{Nishimori's
  Cat: Stable Long-Range Entanglement from Finite-Depth Unitaries and Weak
  Measurements}},}\ }\href {\doibase 10.1103/PhysRevLett.131.200201} {\bibfield
   {journal} {\bibinfo  {journal} {\prl}\ }\textbf {\bibinfo {volume} {131}},\
  \bibinfo {eid} {200201} (\bibinfo {year} {2023})},\ \Eprint
  {http://arxiv.org/abs/2208.11136} {arXiv:2208.11136 [quant-ph]} \BibitemShut
  {NoStop}%
\bibitem [{\citenamefont {{Chen}}\ \emph {et~al.}(2010)\citenamefont {{Chen}},
  \citenamefont {{Gu}},\ and\ \citenamefont {{Wen}}}]{2010localunitary}%
  \BibitemOpen
  \bibfield  {author} {\bibinfo {author} {\bibfnamefont {Xie}\ \bibnamefont
  {{Chen}}}, \bibinfo {author} {\bibfnamefont {Zheng-Cheng}\ \bibnamefont
  {{Gu}}}, \ and\ \bibinfo {author} {\bibfnamefont {Xiao-Gang}\ \bibnamefont
  {{Wen}}},\ }\bibfield  {title} {\enquote {\bibinfo {title} {{Local unitary
  transformation, long-range quantum entanglement, wave function
  renormalization, and topological order}},}\ }\href {\doibase
  10.1103/PhysRevB.82.155138} {\bibfield  {journal} {\bibinfo  {journal}
  {\prb}\ }\textbf {\bibinfo {volume} {82}},\ \bibinfo {eid} {155138} (\bibinfo
  {year} {2010})},\ \Eprint {http://arxiv.org/abs/1004.3835} {arXiv:1004.3835
  [cond-mat.str-el]} \BibitemShut {NoStop}%
\bibitem [{\citenamefont {{Chen}}\ \emph {et~al.}(2013)\citenamefont {{Chen}},
  \citenamefont {{Gu}}, \citenamefont {{Liu}},\ and\ \citenamefont
  {{Wen}}}]{2013chenSPT}%
  \BibitemOpen
  \bibfield  {author} {\bibinfo {author} {\bibfnamefont {Xie}\ \bibnamefont
  {{Chen}}}, \bibinfo {author} {\bibfnamefont {Zheng-Cheng}\ \bibnamefont
  {{Gu}}}, \bibinfo {author} {\bibfnamefont {Zheng-Xin}\ \bibnamefont {{Liu}}},
  \ and\ \bibinfo {author} {\bibfnamefont {Xiao-Gang}\ \bibnamefont {{Wen}}},\
  }\bibfield  {title} {\enquote {\bibinfo {title} {{Symmetry protected
  topological orders and the group cohomology of their symmetry group}},}\
  }\href {\doibase 10.1103/PhysRevB.87.155114} {\bibfield  {journal} {\bibinfo
  {journal} {\prb}\ }\textbf {\bibinfo {volume} {87}},\ \bibinfo {eid} {155114}
  (\bibinfo {year} {2013})},\ \Eprint {http://arxiv.org/abs/1106.4772}
  {arXiv:1106.4772 [cond-mat.str-el]} \BibitemShut {NoStop}%
\bibitem [{\citenamefont {{Hasan}}\ and\ \citenamefont
  {{Kane}}(2010)}]{2010TI}%
  \BibitemOpen
  \bibfield  {author} {\bibinfo {author} {\bibfnamefont {M.~Z.}\ \bibnamefont
  {{Hasan}}}\ and\ \bibinfo {author} {\bibfnamefont {C.~L.}\ \bibnamefont
  {{Kane}}},\ }\bibfield  {title} {\enquote {\bibinfo {title} {{Colloquium:
  Topological insulators}},}\ }\href {\doibase 10.1103/RevModPhys.82.3045}
  {\bibfield  {journal} {\bibinfo  {journal} {Reviews of Modern Physics}\
  }\textbf {\bibinfo {volume} {82}},\ \bibinfo {pages} {3045--3067} (\bibinfo
  {year} {2010})},\ \Eprint {http://arxiv.org/abs/1002.3895} {arXiv:1002.3895
  [cond-mat.mes-hall]} \BibitemShut {NoStop}%
\bibitem [{\citenamefont {{Qi}}\ and\ \citenamefont {{Zhang}}(2011)}]{2011TI}%
  \BibitemOpen
  \bibfield  {author} {\bibinfo {author} {\bibfnamefont {Xiao-Liang}\
  \bibnamefont {{Qi}}}\ and\ \bibinfo {author} {\bibfnamefont {Shou-Cheng}\
  \bibnamefont {{Zhang}}},\ }\bibfield  {title} {\enquote {\bibinfo {title}
  {{Topological insulators and superconductors}},}\ }\href {\doibase
  10.1103/RevModPhys.83.1057} {\bibfield  {journal} {\bibinfo  {journal}
  {Reviews of Modern Physics}\ }\textbf {\bibinfo {volume} {83}},\ \bibinfo
  {pages} {1057--1110} (\bibinfo {year} {2011})},\ \Eprint
  {http://arxiv.org/abs/1008.2026} {arXiv:1008.2026 [cond-mat.mes-hall]}
  \BibitemShut {NoStop}%
\bibitem [{\citenamefont {Affleck}\ \emph {et~al.}(1988)\citenamefont
  {Affleck}, \citenamefont {Kennedy}, \citenamefont {Lieb},\ and\ \citenamefont
  {Tasaki}}]{affleck1988valence}%
  \BibitemOpen
  \bibfield  {author} {\bibinfo {author} {\bibfnamefont {Ian}\ \bibnamefont
  {Affleck}}, \bibinfo {author} {\bibfnamefont {Tom}\ \bibnamefont {Kennedy}},
  \bibinfo {author} {\bibfnamefont {Elliott~H}\ \bibnamefont {Lieb}}, \ and\
  \bibinfo {author} {\bibfnamefont {Hal}\ \bibnamefont {Tasaki}},\ }\bibfield
  {title} {\enquote {\bibinfo {title} {Valence bond ground states in isotropic
  quantum antiferromagnets},}\ }\href@noop {} {\bibfield  {journal} {\bibinfo
  {journal} {Communications in Mathematical Physics}\ }\textbf {\bibinfo
  {volume} {115}},\ \bibinfo {pages} {477--528} (\bibinfo {year}
  {1988})}\BibitemShut {NoStop}%
\bibitem [{\citenamefont {{Hastings}}\ and\ \citenamefont
  {{Wen}}(2005)}]{2005adiabatic}%
  \BibitemOpen
  \bibfield  {author} {\bibinfo {author} {\bibfnamefont {M.~B.}\ \bibnamefont
  {{Hastings}}}\ and\ \bibinfo {author} {\bibfnamefont {Xiao-Gang}\
  \bibnamefont {{Wen}}},\ }\bibfield  {title} {\enquote {\bibinfo {title}
  {{Quasiadiabatic continuation of quantum states: The stability of topological
  ground-state degeneracy and emergent gauge invariance}},}\ }\href {\doibase
  10.1103/PhysRevB.72.045141} {\bibfield  {journal} {\bibinfo  {journal}
  {\prb}\ }\textbf {\bibinfo {volume} {72}},\ \bibinfo {eid} {045141} (\bibinfo
  {year} {2005})},\ \Eprint {http://arxiv.org/abs/cond-mat/0503554}
  {arXiv:cond-mat/0503554 [cond-mat.str-el]} \BibitemShut {NoStop}%
\bibitem [{\citenamefont {{Chen}}\ and\ \citenamefont
  {{Grover}}(2023{\natexlab{a}})}]{2023separability1}%
  \BibitemOpen
  \bibfield  {author} {\bibinfo {author} {\bibfnamefont {Yu-Hsueh}\
  \bibnamefont {{Chen}}}\ and\ \bibinfo {author} {\bibfnamefont {Tarun}\
  \bibnamefont {{Grover}}},\ }\bibfield  {title} {\enquote {\bibinfo {title}
  {{Separability transitions in topological states induced by local
  decoherence}},}\ }\href {\doibase 10.48550/arXiv.2309.11879} {\bibfield
  {journal} {\bibinfo  {journal} {arXiv e-prints}\ ,\ \bibinfo {eid}
  {arXiv:2309.11879}} (\bibinfo {year} {2023}{\natexlab{a}})},\ \Eprint
  {http://arxiv.org/abs/2309.11879} {arXiv:2309.11879 [quant-ph]} \BibitemShut
  {NoStop}%
\bibitem [{\citenamefont {{Chen}}\ and\ \citenamefont
  {{Grover}}(2023{\natexlab{b}})}]{2023separability}%
  \BibitemOpen
  \bibfield  {author} {\bibinfo {author} {\bibfnamefont {Yu-Hsueh}\
  \bibnamefont {{Chen}}}\ and\ \bibinfo {author} {\bibfnamefont {Tarun}\
  \bibnamefont {{Grover}}},\ }\bibfield  {title} {\enquote {\bibinfo {title}
  {{Symmetry-enforced many-body separability transitions}},}\ }\href {\doibase
  10.48550/arXiv.2310.07286} {\bibfield  {journal} {\bibinfo  {journal} {arXiv
  e-prints}\ ,\ \bibinfo {eid} {arXiv:2310.07286}} (\bibinfo {year}
  {2023}{\natexlab{b}})},\ \Eprint {http://arxiv.org/abs/2310.07286}
  {arXiv:2310.07286 [quant-ph]} \BibitemShut {NoStop}%
\bibitem [{\citenamefont {{You}}\ \emph {et~al.}(2014)\citenamefont {{You}},
  \citenamefont {{Bi}}, \citenamefont {{Rasmussen}}, \citenamefont {{Slagle}},\
  and\ \citenamefont {{Xu}}}]{2014strange}%
  \BibitemOpen
  \bibfield  {author} {\bibinfo {author} {\bibfnamefont {Yi-Zhuang}\
  \bibnamefont {{You}}}, \bibinfo {author} {\bibfnamefont {Zhen}\ \bibnamefont
  {{Bi}}}, \bibinfo {author} {\bibfnamefont {Alex}\ \bibnamefont
  {{Rasmussen}}}, \bibinfo {author} {\bibfnamefont {Kevin}\ \bibnamefont
  {{Slagle}}}, \ and\ \bibinfo {author} {\bibfnamefont {Cenke}\ \bibnamefont
  {{Xu}}},\ }\bibfield  {title} {\enquote {\bibinfo {title} {{Wave Function and
  Strange Correlator of Short-Range Entangled States}},}\ }\href {\doibase
  10.1103/PhysRevLett.112.247202} {\bibfield  {journal} {\bibinfo  {journal}
  {\prl}\ }\textbf {\bibinfo {volume} {112}},\ \bibinfo {eid} {247202}
  (\bibinfo {year} {2014})},\ \Eprint {http://arxiv.org/abs/1312.0626}
  {arXiv:1312.0626 [cond-mat.str-el]} \BibitemShut {NoStop}%
\bibitem [{\citenamefont {{Qi}}\ \emph {et~al.}(2008)\citenamefont {{Qi}},
  \citenamefont {{Hughes}},\ and\ \citenamefont {{Zhang}}}]{2008TImonopole}%
  \BibitemOpen
  \bibfield  {author} {\bibinfo {author} {\bibfnamefont {Xiao-Liang}\
  \bibnamefont {{Qi}}}, \bibinfo {author} {\bibfnamefont {Taylor~L.}\
  \bibnamefont {{Hughes}}}, \ and\ \bibinfo {author} {\bibfnamefont
  {Shou-Cheng}\ \bibnamefont {{Zhang}}},\ }\bibfield  {title} {\enquote
  {\bibinfo {title} {{Topological field theory of time-reversal invariant
  insulators}},}\ }\href {\doibase 10.1103/PhysRevB.78.195424} {\bibfield
  {journal} {\bibinfo  {journal} {\prb}\ }\textbf {\bibinfo {volume} {78}},\
  \bibinfo {eid} {195424} (\bibinfo {year} {2008})},\ \Eprint
  {http://arxiv.org/abs/0802.3537} {arXiv:0802.3537 [cond-mat.mes-hall]}
  \BibitemShut {NoStop}%
\bibitem [{\citenamefont {{Kliesch}}\ \emph {et~al.}(2014)\citenamefont
  {{Kliesch}}, \citenamefont {{Gross}},\ and\ \citenamefont
  {{Eisert}}}]{2014MPOpositivity}%
  \BibitemOpen
  \bibfield  {author} {\bibinfo {author} {\bibfnamefont {M.}~\bibnamefont
  {{Kliesch}}}, \bibinfo {author} {\bibfnamefont {D.}~\bibnamefont {{Gross}}},
  \ and\ \bibinfo {author} {\bibfnamefont {J.}~\bibnamefont {{Eisert}}},\
  }\bibfield  {title} {\enquote {\bibinfo {title} {{Matrix-Product Operators
  and States: NP-Hardness and Undecidability}},}\ }\href {\doibase
  10.1103/PhysRevLett.113.160503} {\bibfield  {journal} {\bibinfo  {journal}
  {\prl}\ }\textbf {\bibinfo {volume} {113}},\ \bibinfo {eid} {160503}
  (\bibinfo {year} {2014})},\ \Eprint {http://arxiv.org/abs/1404.4466}
  {arXiv:1404.4466 [quant-ph]} \BibitemShut {NoStop}%
\bibitem [{\citenamefont {{Ellison}}\ \emph {et~al.}(2021)\citenamefont
  {{Ellison}}, \citenamefont {{Kato}}, \citenamefont {{Liu}},\ and\
  \citenamefont {{Hsieh}}}]{2021signproblem}%
  \BibitemOpen
  \bibfield  {author} {\bibinfo {author} {\bibfnamefont {Tyler~D.}\
  \bibnamefont {{Ellison}}}, \bibinfo {author} {\bibfnamefont {Kohtaro}\
  \bibnamefont {{Kato}}}, \bibinfo {author} {\bibfnamefont {Zi-Wen}\
  \bibnamefont {{Liu}}}, \ and\ \bibinfo {author} {\bibfnamefont {Timothy~H.}\
  \bibnamefont {{Hsieh}}},\ }\bibfield  {title} {\enquote {\bibinfo {title}
  {{Symmetry-protected sign problem and magic in quantum phases of matter}},}\
  }\href {\doibase 10.22331/q-2021-12-28-612} {\bibfield  {journal} {\bibinfo
  {journal} {Quantum}\ }\textbf {\bibinfo {volume} {5}},\ \bibinfo {pages}
  {612} (\bibinfo {year} {2021})},\ \Eprint {http://arxiv.org/abs/2010.13803}
  {arXiv:2010.13803 [cond-mat.str-el]} \BibitemShut {NoStop}%
\bibitem [{\citenamefont {{Coser}}\ and\ \citenamefont
  {{P{\'e}rez-Garc{\'\i}a}}(2019)}]{2019channelconnectible}%
  \BibitemOpen
  \bibfield  {author} {\bibinfo {author} {\bibfnamefont {Andrea}\ \bibnamefont
  {{Coser}}}\ and\ \bibinfo {author} {\bibfnamefont {David}\ \bibnamefont
  {{P{\'e}rez-Garc{\'\i}a}}},\ }\bibfield  {title} {\enquote {\bibinfo {title}
  {{Classification of phases for mixed states via fast dissipative
  evolution}},}\ }\href {\doibase 10.22331/q-2019-08-12-174} {\bibfield
  {journal} {\bibinfo  {journal} {Quantum}\ }\textbf {\bibinfo {volume} {3}},\
  \bibinfo {pages} {174} (\bibinfo {year} {2019})},\ \Eprint
  {http://arxiv.org/abs/1810.05092} {arXiv:1810.05092 [quant-ph]} \BibitemShut
  {NoStop}%
\bibitem [{\citenamefont {{Sang}}\ \emph {et~al.}(2023)\citenamefont {{Sang}},
  \citenamefont {{Zou}},\ and\ \citenamefont
  {{Hsieh}}}]{2023channelconnectible}%
  \BibitemOpen
  \bibfield  {author} {\bibinfo {author} {\bibfnamefont {Shengqi}\ \bibnamefont
  {{Sang}}}, \bibinfo {author} {\bibfnamefont {Yijian}\ \bibnamefont {{Zou}}},
  \ and\ \bibinfo {author} {\bibfnamefont {Timothy~H.}\ \bibnamefont
  {{Hsieh}}},\ }\bibfield  {title} {\enquote {\bibinfo {title} {{Mixed-state
  Quantum Phases: Renormalization and Quantum Error Correction}},}\ }\href
  {\doibase 10.48550/arXiv.2310.08639} {\bibfield  {journal} {\bibinfo
  {journal} {arXiv e-prints}\ ,\ \bibinfo {eid} {arXiv:2310.08639}} (\bibinfo
  {year} {2023})},\ \Eprint {http://arxiv.org/abs/2310.08639} {arXiv:2310.08639
  [quant-ph]} \BibitemShut {NoStop}%
\bibitem [{\citenamefont {{Rakovszky}}\ \emph {et~al.}(2023)\citenamefont
  {{Rakovszky}}, \citenamefont {{Gopalakrishnan}},\ and\ \citenamefont {{von
  Keyserlingk}}}]{2023steadystatephase}%
  \BibitemOpen
  \bibfield  {author} {\bibinfo {author} {\bibfnamefont {Tibor}\ \bibnamefont
  {{Rakovszky}}}, \bibinfo {author} {\bibfnamefont {Sarang}\ \bibnamefont
  {{Gopalakrishnan}}}, \ and\ \bibinfo {author} {\bibfnamefont {Curt}\
  \bibnamefont {{von Keyserlingk}}},\ }\bibfield  {title} {\enquote {\bibinfo
  {title} {{Defining stable phases of open quantum systems}},}\ }\href
  {\doibase 10.48550/arXiv.2308.15495} {\bibfield  {journal} {\bibinfo
  {journal} {arXiv e-prints}\ ,\ \bibinfo {eid} {arXiv:2308.15495}} (\bibinfo
  {year} {2023})},\ \Eprint {http://arxiv.org/abs/2308.15495} {arXiv:2308.15495
  [quant-ph]} \BibitemShut {NoStop}%
\bibitem [{\citenamefont {{Fradkin}}(2017)}]{2017disorderarameter}%
  \BibitemOpen
  \bibfield  {author} {\bibinfo {author} {\bibfnamefont {Eduardo}\ \bibnamefont
  {{Fradkin}}},\ }\bibfield  {title} {\enquote {\bibinfo {title} {{Disorder
  Operators and Their Descendants}},}\ }\href {\doibase
  10.1007/s10955-017-1737-7} {\bibfield  {journal} {\bibinfo  {journal}
  {Journal of Statistical Physics}\ }\textbf {\bibinfo {volume} {167}},\
  \bibinfo {pages} {427--461} (\bibinfo {year} {2017})},\ \Eprint
  {http://arxiv.org/abs/1610.05780} {arXiv:1610.05780 [cond-mat.stat-mech]}
  \BibitemShut {NoStop}%
\bibitem [{\citenamefont {Preskill}(1998)}]{preskill1998lecture}%
  \BibitemOpen
  \bibfield  {author} {\bibinfo {author} {\bibfnamefont {John}\ \bibnamefont
  {Preskill}},\ }\bibfield  {title} {\enquote {\bibinfo {title} {Lecture notes
  for physics 229: Quantum information and computation},}\ }\href@noop {}
  {\bibfield  {journal} {\bibinfo  {journal} {California Institute of
  Technology}\ }\textbf {\bibinfo {volume} {16}},\ \bibinfo {pages} {1--8}
  (\bibinfo {year} {1998})}\BibitemShut {NoStop}%
\bibitem [{\citenamefont {Buča}\ and\ \citenamefont
  {Prosen}(2012)}]{Buča_2012}%
  \BibitemOpen
  \bibfield  {author} {\bibinfo {author} {\bibfnamefont {Berislav}\
  \bibnamefont {Buča}}\ and\ \bibinfo {author} {\bibfnamefont {Tomaž}\
  \bibnamefont {Prosen}},\ }\bibfield  {title} {\enquote {\bibinfo {title} {A
  note on symmetry reductions of the lindblad equation: transport in
  constrained open spin chains},}\ }\href {\doibase
  10.1088/1367-2630/14/7/073007} {\bibfield  {journal} {\bibinfo  {journal}
  {New Journal of Physics}\ }\textbf {\bibinfo {volume} {14}},\ \bibinfo
  {pages} {073007} (\bibinfo {year} {2012})}\BibitemShut {NoStop}%
\bibitem [{\citenamefont {Schuch}\ \emph {et~al.}(2011)\citenamefont {Schuch},
  \citenamefont {Pérez-García},\ and\ \citenamefont {Cirac}}]{Schuch_2011}%
  \BibitemOpen
  \bibfield  {author} {\bibinfo {author} {\bibfnamefont {Norbert}\ \bibnamefont
  {Schuch}}, \bibinfo {author} {\bibfnamefont {David}\ \bibnamefont
  {Pérez-García}}, \ and\ \bibinfo {author} {\bibfnamefont {Ignacio}\
  \bibnamefont {Cirac}},\ }\bibfield  {title} {\enquote {\bibinfo {title}
  {Classifying quantum phases using matrix product states and projected
  entangled pair states},}\ }\href {\doibase 10.1103/physrevb.84.165139}
  {\bibfield  {journal} {\bibinfo  {journal} {Physical Review B}\ }\textbf
  {\bibinfo {volume} {84}} (\bibinfo {year} {2011}),\
  10.1103/physrevb.84.165139}\BibitemShut {NoStop}%
\bibitem [{\citenamefont {{Verstraete}}\ and\ \citenamefont
  {{Cirac}}(2006)}]{2006mpsrepresentation}%
  \BibitemOpen
  \bibfield  {author} {\bibinfo {author} {\bibfnamefont {F.}~\bibnamefont
  {{Verstraete}}}\ and\ \bibinfo {author} {\bibfnamefont {J.~I.}\ \bibnamefont
  {{Cirac}}},\ }\bibfield  {title} {\enquote {\bibinfo {title} {{Matrix product
  states represent ground states faithfully}},}\ }\href {\doibase
  10.1103/PhysRevB.73.094423} {\bibfield  {journal} {\bibinfo  {journal}
  {\prb}\ }\textbf {\bibinfo {volume} {73}},\ \bibinfo {eid} {094423} (\bibinfo
  {year} {2006})},\ \Eprint {http://arxiv.org/abs/cond-mat/0505140}
  {arXiv:cond-mat/0505140 [cond-mat.str-el]} \BibitemShut {NoStop}%
\bibitem [{\citenamefont {{Verstraete}}\ \emph {et~al.}(2008)\citenamefont
  {{Verstraete}}, \citenamefont {{Murg}},\ and\ \citenamefont
  {{Cirac}}}]{2008mpsreview}%
  \BibitemOpen
  \bibfield  {author} {\bibinfo {author} {\bibfnamefont {F.}~\bibnamefont
  {{Verstraete}}}, \bibinfo {author} {\bibfnamefont {V.}~\bibnamefont
  {{Murg}}}, \ and\ \bibinfo {author} {\bibfnamefont {J.~I.}\ \bibnamefont
  {{Cirac}}},\ }\bibfield  {title} {\enquote {\bibinfo {title} {{Matrix product
  states, projected entangled pair states, and variational renormalization
  group methods for quantum spin systems}},}\ }\href {\doibase
  10.1080/14789940801912366} {\bibfield  {journal} {\bibinfo  {journal}
  {Advances in Physics}\ }\textbf {\bibinfo {volume} {57}},\ \bibinfo {pages}
  {143--224} (\bibinfo {year} {2008})},\ \Eprint
  {http://arxiv.org/abs/0907.2796} {arXiv:0907.2796 [quant-ph]} \BibitemShut
  {NoStop}%
\bibitem [{\citenamefont {{Schollw{\"o}ck}}(2011)}]{2011mpsreview}%
  \BibitemOpen
  \bibfield  {author} {\bibinfo {author} {\bibfnamefont {Ulrich}\ \bibnamefont
  {{Schollw{\"o}ck}}},\ }\bibfield  {title} {\enquote {\bibinfo {title} {{The
  density-matrix renormalization group in the age of matrix product states}},}\
  }\href {\doibase 10.1016/j.aop.2010.09.012} {\bibfield  {journal} {\bibinfo
  {journal} {Annals of Physics}\ }\textbf {\bibinfo {volume} {326}},\ \bibinfo
  {pages} {96--192} (\bibinfo {year} {2011})},\ \Eprint
  {http://arxiv.org/abs/1008.3477} {arXiv:1008.3477 [cond-mat.str-el]}
  \BibitemShut {NoStop}%
\bibitem [{\citenamefont {{Cirac}}\ \emph {et~al.}(2020)\citenamefont
  {{Cirac}}, \citenamefont {{Perez-Garcia}}, \citenamefont {{Schuch}},\ and\
  \citenamefont {{Verstraete}}}]{2020mpsreview}%
  \BibitemOpen
  \bibfield  {author} {\bibinfo {author} {\bibfnamefont {Ignacio}\ \bibnamefont
  {{Cirac}}}, \bibinfo {author} {\bibfnamefont {David}\ \bibnamefont
  {{Perez-Garcia}}}, \bibinfo {author} {\bibfnamefont {Norbert}\ \bibnamefont
  {{Schuch}}}, \ and\ \bibinfo {author} {\bibfnamefont {Frank}\ \bibnamefont
  {{Verstraete}}},\ }\bibfield  {title} {\enquote {\bibinfo {title} {{Matrix
  Product States and Projected Entangled Pair States: Concepts, Symmetries, and
  Theorems}},}\ }\href {\doibase 10.48550/arXiv.2011.12127} {\bibfield
  {journal} {\bibinfo  {journal} {arXiv e-prints}\ ,\ \bibinfo {eid}
  {arXiv:2011.12127}} (\bibinfo {year} {2020})},\ \Eprint
  {http://arxiv.org/abs/2011.12127} {arXiv:2011.12127 [quant-ph]} \BibitemShut
  {NoStop}%
\bibitem [{\citenamefont {{Wolf}}\ \emph {et~al.}(2008)\citenamefont {{Wolf}},
  \citenamefont {{Verstraete}}, \citenamefont {{Hastings}},\ and\ \citenamefont
  {{Cirac}}}]{2008mpdomutualinfo}%
  \BibitemOpen
  \bibfield  {author} {\bibinfo {author} {\bibfnamefont {Michael~M.}\
  \bibnamefont {{Wolf}}}, \bibinfo {author} {\bibfnamefont {Frank}\
  \bibnamefont {{Verstraete}}}, \bibinfo {author} {\bibfnamefont {Matthew~B.}\
  \bibnamefont {{Hastings}}}, \ and\ \bibinfo {author} {\bibfnamefont
  {J.~Ignacio}\ \bibnamefont {{Cirac}}},\ }\bibfield  {title} {\enquote
  {\bibinfo {title} {{Area Laws in Quantum Systems: Mutual Information and
  Correlations}},}\ }\href {\doibase 10.1103/PhysRevLett.100.070502} {\bibfield
   {journal} {\bibinfo  {journal} {\prl}\ }\textbf {\bibinfo {volume} {100}},\
  \bibinfo {eid} {070502} (\bibinfo {year} {2008})},\ \Eprint
  {http://arxiv.org/abs/0704.3906} {arXiv:0704.3906 [quant-ph]} \BibitemShut
  {NoStop}%
\bibitem [{\citenamefont {{Cirac}}\ \emph {et~al.}(2017)\citenamefont
  {{Cirac}}, \citenamefont {{P{\'e}rez-Garc{\'\i}a}}, \citenamefont
  {{Schuch}},\ and\ \citenamefont {{Verstraete}}}]{2017MPDO}%
  \BibitemOpen
  \bibfield  {author} {\bibinfo {author} {\bibfnamefont {J.~I.}\ \bibnamefont
  {{Cirac}}}, \bibinfo {author} {\bibfnamefont {D.}~\bibnamefont
  {{P{\'e}rez-Garc{\'\i}a}}}, \bibinfo {author} {\bibfnamefont
  {N.}~\bibnamefont {{Schuch}}}, \ and\ \bibinfo {author} {\bibfnamefont
  {F.}~\bibnamefont {{Verstraete}}},\ }\bibfield  {title} {\enquote {\bibinfo
  {title} {{Matrix product density operators: Renormalization fixed points and
  boundary theories}},}\ }\href {\doibase 10.1016/j.aop.2016.12.030} {\bibfield
   {journal} {\bibinfo  {journal} {Annals of Physics}\ }\textbf {\bibinfo
  {volume} {378}},\ \bibinfo {pages} {100--149} (\bibinfo {year} {2017})},\
  \Eprint {http://arxiv.org/abs/1606.00608} {arXiv:1606.00608 [quant-ph]}
  \BibitemShut {NoStop}%
\bibitem [{\citenamefont {{Zanardi}}(2001)}]{2001opee}%
  \BibitemOpen
  \bibfield  {author} {\bibinfo {author} {\bibfnamefont {Paolo}\ \bibnamefont
  {{Zanardi}}},\ }\bibfield  {title} {\enquote {\bibinfo {title} {{Entanglement
  of quantum evolutions}},}\ }\href {\doibase 10.1103/PhysRevA.63.040304}
  {\bibfield  {journal} {\bibinfo  {journal} {\pra}\ }\textbf {\bibinfo
  {volume} {63}},\ \bibinfo {eid} {040304} (\bibinfo {year} {2001})},\ \Eprint
  {http://arxiv.org/abs/quant-ph/0010074} {arXiv:quant-ph/0010074 [quant-ph]}
  \BibitemShut {NoStop}%
\bibitem [{\citenamefont {{De las Cuevas}}\ \emph {et~al.}(2013)\citenamefont
  {{De las Cuevas}}, \citenamefont {{Schuch}}, \citenamefont
  {{P{\'e}rez-Garc{\'\i}a}},\ and\ \citenamefont {{Cirac}}}]{2013purification}%
  \BibitemOpen
  \bibfield  {author} {\bibinfo {author} {\bibfnamefont {Gemma}\ \bibnamefont
  {{De las Cuevas}}}, \bibinfo {author} {\bibfnamefont {Norbert}\ \bibnamefont
  {{Schuch}}}, \bibinfo {author} {\bibfnamefont {David}\ \bibnamefont
  {{P{\'e}rez-Garc{\'\i}a}}}, \ and\ \bibinfo {author} {\bibfnamefont
  {J.~Ignacio}\ \bibnamefont {{Cirac}}},\ }\bibfield  {title} {\enquote
  {\bibinfo {title} {{Purifications of multipartite states: limitations and
  constructive methods}},}\ }\href {\doibase 10.1088/1367-2630/15/12/123021}
  {\bibfield  {journal} {\bibinfo  {journal} {New Journal of Physics}\ }\textbf
  {\bibinfo {volume} {15}},\ \bibinfo {eid} {123021} (\bibinfo {year}
  {2013})},\ \Eprint {http://arxiv.org/abs/1308.1914} {arXiv:1308.1914
  [quant-ph]} \BibitemShut {NoStop}%
\bibitem [{\citenamefont {{Chen}}\ \emph {et~al.}(2014)\citenamefont {{Chen}},
  \citenamefont {{Lu}},\ and\ \citenamefont
  {{Vishwanath}}}]{2014decoreteddomainwall}%
  \BibitemOpen
  \bibfield  {author} {\bibinfo {author} {\bibfnamefont {Xie}\ \bibnamefont
  {{Chen}}}, \bibinfo {author} {\bibfnamefont {Yuan-Ming}\ \bibnamefont
  {{Lu}}}, \ and\ \bibinfo {author} {\bibfnamefont {Ashvin}\ \bibnamefont
  {{Vishwanath}}},\ }\bibfield  {title} {\enquote {\bibinfo {title}
  {{Symmetry-protected topological phases from decorated domain walls}},}\
  }\href {\doibase 10.1038/ncomms4507} {\bibfield  {journal} {\bibinfo
  {journal} {Nature Communications}\ }\textbf {\bibinfo {volume} {5}},\
  \bibinfo {eid} {3507} (\bibinfo {year} {2014})},\ \Eprint
  {http://arxiv.org/abs/1303.4301} {arXiv:1303.4301 [cond-mat.str-el]}
  \BibitemShut {NoStop}%
\bibitem [{\citenamefont {{Duivenvoorden}}\ \emph {et~al.}(2017)\citenamefont
  {{Duivenvoorden}}, \citenamefont {{Iqbal}}, \citenamefont {{Haegeman}},
  \citenamefont {{Verstraete}},\ and\ \citenamefont
  {{Schuch}}}]{2017positivity}%
  \BibitemOpen
  \bibfield  {author} {\bibinfo {author} {\bibfnamefont {Kasper}\ \bibnamefont
  {{Duivenvoorden}}}, \bibinfo {author} {\bibfnamefont {Mohsin}\ \bibnamefont
  {{Iqbal}}}, \bibinfo {author} {\bibfnamefont {Jutho}\ \bibnamefont
  {{Haegeman}}}, \bibinfo {author} {\bibfnamefont {Frank}\ \bibnamefont
  {{Verstraete}}}, \ and\ \bibinfo {author} {\bibfnamefont {Norbert}\
  \bibnamefont {{Schuch}}},\ }\bibfield  {title} {\enquote {\bibinfo {title}
  {{Entanglement phases as holographic duals of anyon condensates}},}\ }\href
  {\doibase 10.1103/PhysRevB.95.235119} {\bibfield  {journal} {\bibinfo
  {journal} {\prb}\ }\textbf {\bibinfo {volume} {95}},\ \bibinfo {eid} {235119}
  (\bibinfo {year} {2017})},\ \Eprint {http://arxiv.org/abs/1702.08469}
  {arXiv:1702.08469 [cond-mat.str-el]} \BibitemShut {NoStop}%
\bibitem [{\citenamefont {{Pollmann}}\ \emph {et~al.}(2010)\citenamefont
  {{Pollmann}}, \citenamefont {{Turner}}, \citenamefont {{Berg}},\ and\
  \citenamefont {{Oshikawa}}}]{20101dSPT}%
  \BibitemOpen
  \bibfield  {author} {\bibinfo {author} {\bibfnamefont {Frank}\ \bibnamefont
  {{Pollmann}}}, \bibinfo {author} {\bibfnamefont {Ari~M.}\ \bibnamefont
  {{Turner}}}, \bibinfo {author} {\bibfnamefont {Erez}\ \bibnamefont {{Berg}}},
  \ and\ \bibinfo {author} {\bibfnamefont {Masaki}\ \bibnamefont
  {{Oshikawa}}},\ }\bibfield  {title} {\enquote {\bibinfo {title}
  {{Entanglement spectrum of a topological phase in one dimension}},}\ }\href
  {\doibase 10.1103/PhysRevB.81.064439} {\bibfield  {journal} {\bibinfo
  {journal} {\prb}\ }\textbf {\bibinfo {volume} {81}},\ \bibinfo {eid} {064439}
  (\bibinfo {year} {2010})},\ \Eprint {http://arxiv.org/abs/0910.1811}
  {arXiv:0910.1811 [cond-mat.str-el]} \BibitemShut {NoStop}%
\bibitem [{\citenamefont {{Pollmann}}\ \emph {et~al.}(2012)\citenamefont
  {{Pollmann}}, \citenamefont {{Berg}}, \citenamefont {{Turner}},\ and\
  \citenamefont {{Oshikawa}}}]{20121dSPT}%
  \BibitemOpen
  \bibfield  {author} {\bibinfo {author} {\bibfnamefont {Frank}\ \bibnamefont
  {{Pollmann}}}, \bibinfo {author} {\bibfnamefont {Erez}\ \bibnamefont
  {{Berg}}}, \bibinfo {author} {\bibfnamefont {Ari~M.}\ \bibnamefont
  {{Turner}}}, \ and\ \bibinfo {author} {\bibfnamefont {Masaki}\ \bibnamefont
  {{Oshikawa}}},\ }\bibfield  {title} {\enquote {\bibinfo {title} {{Symmetry
  protection of topological phases in one-dimensional quantum spin systems}},}\
  }\href {\doibase 10.1103/PhysRevB.85.075125} {\bibfield  {journal} {\bibinfo
  {journal} {\prb}\ }\textbf {\bibinfo {volume} {85}},\ \bibinfo {eid} {075125}
  (\bibinfo {year} {2012})},\ \Eprint {http://arxiv.org/abs/0909.4059}
  {arXiv:0909.4059 [cond-mat.str-el]} \BibitemShut {NoStop}%
\bibitem [{\citenamefont {{Pollmann}}\ and\ \citenamefont
  {{Turner}}(2012)}]{20121dSPTdetection}%
  \BibitemOpen
  \bibfield  {author} {\bibinfo {author} {\bibfnamefont {Frank}\ \bibnamefont
  {{Pollmann}}}\ and\ \bibinfo {author} {\bibfnamefont {Ari~M.}\ \bibnamefont
  {{Turner}}},\ }\bibfield  {title} {\enquote {\bibinfo {title} {{Detection of
  symmetry-protected topological phases in one dimension}},}\ }\href {\doibase
  10.1103/PhysRevB.86.125441} {\bibfield  {journal} {\bibinfo  {journal}
  {\prb}\ }\textbf {\bibinfo {volume} {86}},\ \bibinfo {eid} {125441} (\bibinfo
  {year} {2012})},\ \Eprint {http://arxiv.org/abs/1204.0704} {arXiv:1204.0704
  [cond-mat.str-el]} \BibitemShut {NoStop}%
\bibitem [{\citenamefont {{Shiozaki}}\ and\ \citenamefont
  {{Ryu}}(2017{\natexlab{a}})}]{2017MPSRyu}%
  \BibitemOpen
  \bibfield  {author} {\bibinfo {author} {\bibfnamefont {Ken}\ \bibnamefont
  {{Shiozaki}}}\ and\ \bibinfo {author} {\bibfnamefont {Shinsei}\ \bibnamefont
  {{Ryu}}},\ }\bibfield  {title} {\enquote {\bibinfo {title} {{Matrix product
  states and equivariant topological field theories for bosonic
  symmetry-protected topological phases in (1+1) dimensions}},}\ }\href
  {\doibase 10.1007/JHEP04(2017)100} {\bibfield  {journal} {\bibinfo  {journal}
  {Journal of High Energy Physics}\ }\textbf {\bibinfo {volume} {2017}},\
  \bibinfo {eid} {100} (\bibinfo {year} {2017}{\natexlab{a}})},\ \Eprint
  {http://arxiv.org/abs/1607.06504} {arXiv:1607.06504 [cond-mat.str-el]}
  \BibitemShut {NoStop}%
\bibitem [{\citenamefont {{Turner}}\ \emph {et~al.}(2011)\citenamefont
  {{Turner}}, \citenamefont {{Pollmann}},\ and\ \citenamefont
  {{Berg}}}]{2011symlocal}%
  \BibitemOpen
  \bibfield  {author} {\bibinfo {author} {\bibfnamefont {Ari~M.}\ \bibnamefont
  {{Turner}}}, \bibinfo {author} {\bibfnamefont {Frank}\ \bibnamefont
  {{Pollmann}}}, \ and\ \bibinfo {author} {\bibfnamefont {Erez}\ \bibnamefont
  {{Berg}}},\ }\bibfield  {title} {\enquote {\bibinfo {title} {{Topological
  phases of one-dimensional fermions: An entanglement point of view}},}\ }\href
  {\doibase 10.1103/PhysRevB.83.075102} {\bibfield  {journal} {\bibinfo
  {journal} {\prb}\ }\textbf {\bibinfo {volume} {83}},\ \bibinfo {eid} {075102}
  (\bibinfo {year} {2011})},\ \Eprint {http://arxiv.org/abs/1008.4346}
  {arXiv:1008.4346 [cond-mat.str-el]} \BibitemShut {NoStop}%
\bibitem [{\citenamefont {{Else}}\ and\ \citenamefont
  {{Nayak}}(2014)}]{2014dElse}%
  \BibitemOpen
  \bibfield  {author} {\bibinfo {author} {\bibfnamefont {Dominic~V.}\
  \bibnamefont {{Else}}}\ and\ \bibinfo {author} {\bibfnamefont {Chetan}\
  \bibnamefont {{Nayak}}},\ }\bibfield  {title} {\enquote {\bibinfo {title}
  {{Classifying symmetry-protected topological phases through the anomalous
  action of the symmetry on the edge}},}\ }\href {\doibase
  10.1103/PhysRevB.90.235137} {\bibfield  {journal} {\bibinfo  {journal}
  {\prb}\ }\textbf {\bibinfo {volume} {90}},\ \bibinfo {eid} {235137} (\bibinfo
  {year} {2014})},\ \Eprint {http://arxiv.org/abs/1409.5436} {arXiv:1409.5436
  [cond-mat.str-el]} \BibitemShut {NoStop}%
\bibitem [{\citenamefont {{Wang}}\ \emph {et~al.}(2021)\citenamefont {{Wang}},
  \citenamefont {{Ning}},\ and\ \citenamefont {{Cheng}}}]{2021decoratedDW}%
  \BibitemOpen
  \bibfield  {author} {\bibinfo {author} {\bibfnamefont {Qing-Rui}\
  \bibnamefont {{Wang}}}, \bibinfo {author} {\bibfnamefont {Shang-Qiang}\
  \bibnamefont {{Ning}}}, \ and\ \bibinfo {author} {\bibfnamefont {Meng}\
  \bibnamefont {{Cheng}}},\ }\bibfield  {title} {\enquote {\bibinfo {title}
  {{Domain Wall Decorations, Anomalies and Spectral Sequences in Bosonic
  Topological Phases}},}\ }\href {\doibase 10.48550/arXiv.2104.13233}
  {\bibfield  {journal} {\bibinfo  {journal} {arXiv e-prints}\ ,\ \bibinfo
  {eid} {arXiv:2104.13233}} (\bibinfo {year} {2021})},\ \Eprint
  {http://arxiv.org/abs/2104.13233} {arXiv:2104.13233 [cond-mat.str-el]}
  \BibitemShut {NoStop}%
\bibitem [{\citenamefont {{Gaiotto}}\ and\ \citenamefont
  {{Johnson-Freyd}}(2019)}]{2019gaiottospectralsequence}%
  \BibitemOpen
  \bibfield  {author} {\bibinfo {author} {\bibfnamefont {Davide}\ \bibnamefont
  {{Gaiotto}}}\ and\ \bibinfo {author} {\bibfnamefont {Theo}\ \bibnamefont
  {{Johnson-Freyd}}},\ }\bibfield  {title} {\enquote {\bibinfo {title}
  {{Symmetry protected topological phases and generalized cohomology}},}\
  }\href {\doibase 10.1007/JHEP05(2019)007} {\bibfield  {journal} {\bibinfo
  {journal} {Journal of High Energy Physics}\ }\textbf {\bibinfo {volume}
  {2019}},\ \bibinfo {eid} {7} (\bibinfo {year} {2019})},\ \Eprint
  {http://arxiv.org/abs/1712.07950} {arXiv:1712.07950 [hep-th]} \BibitemShut
  {NoStop}%
\bibitem [{\citenamefont {{Shapourian}}\ \emph {et~al.}(2017)\citenamefont
  {{Shapourian}}, \citenamefont {{Shiozaki}},\ and\ \citenamefont
  {{Ryu}}}]{2017partialtranspose}%
  \BibitemOpen
  \bibfield  {author} {\bibinfo {author} {\bibfnamefont {Hassan}\ \bibnamefont
  {{Shapourian}}}, \bibinfo {author} {\bibfnamefont {Ken}\ \bibnamefont
  {{Shiozaki}}}, \ and\ \bibinfo {author} {\bibfnamefont {Shinsei}\
  \bibnamefont {{Ryu}}},\ }\bibfield  {title} {\enquote {\bibinfo {title}
  {{Many-Body Topological Invariants for Fermionic Symmetry-Protected
  Topological Phases}},}\ }\href {\doibase 10.1103/PhysRevLett.118.216402}
  {\bibfield  {journal} {\bibinfo  {journal} {\prl}\ }\textbf {\bibinfo
  {volume} {118}},\ \bibinfo {eid} {216402} (\bibinfo {year} {2017})},\ \Eprint
  {http://arxiv.org/abs/1607.03896} {arXiv:1607.03896 [cond-mat.str-el]}
  \BibitemShut {NoStop}%
\bibitem [{\citenamefont {{Shiozaki}}\ \emph {et~al.}(2018)\citenamefont
  {{Shiozaki}}, \citenamefont {{Shapourian}}, \citenamefont {{Gomi}},\ and\
  \citenamefont {{Ryu}}}]{2018partialtranspose}%
  \BibitemOpen
  \bibfield  {author} {\bibinfo {author} {\bibfnamefont {Ken}\ \bibnamefont
  {{Shiozaki}}}, \bibinfo {author} {\bibfnamefont {Hassan}\ \bibnamefont
  {{Shapourian}}}, \bibinfo {author} {\bibfnamefont {Kiyonori}\ \bibnamefont
  {{Gomi}}}, \ and\ \bibinfo {author} {\bibfnamefont {Shinsei}\ \bibnamefont
  {{Ryu}}},\ }\bibfield  {title} {\enquote {\bibinfo {title} {{Many-body
  topological invariants for fermionic short-range entangled topological phases
  protected by antiunitary symmetries}},}\ }\href {\doibase
  10.1103/PhysRevB.98.035151} {\bibfield  {journal} {\bibinfo  {journal}
  {\prb}\ }\textbf {\bibinfo {volume} {98}},\ \bibinfo {eid} {035151} (\bibinfo
  {year} {2018})},\ \Eprint {http://arxiv.org/abs/1710.01886} {arXiv:1710.01886
  [cond-mat.str-el]} \BibitemShut {NoStop}%
\bibitem [{\citenamefont {{Chen}}\ and\ \citenamefont
  {{Vishwanath}}(2015)}]{2015gaugingT}%
  \BibitemOpen
  \bibfield  {author} {\bibinfo {author} {\bibfnamefont {Xie}\ \bibnamefont
  {{Chen}}}\ and\ \bibinfo {author} {\bibfnamefont {Ashvin}\ \bibnamefont
  {{Vishwanath}}},\ }\bibfield  {title} {\enquote {\bibinfo {title} {{Towards
  Gauging Time-Reversal Symmetry: A Tensor Network Approach}},}\ }\href
  {\doibase 10.1103/PhysRevX.5.041034} {\bibfield  {journal} {\bibinfo
  {journal} {Physical Review X}\ }\textbf {\bibinfo {volume} {5}},\ \bibinfo
  {eid} {041034} (\bibinfo {year} {2015})},\ \Eprint
  {http://arxiv.org/abs/1401.3736} {arXiv:1401.3736 [cond-mat.str-el]}
  \BibitemShut {NoStop}%
\bibitem [{\citenamefont {Seng}()}]{2023coho}%
  \BibitemOpen
  \bibfield  {author} {\bibinfo {author} {\bibfnamefont {Angina}\ \bibnamefont
  {Seng}},\ }\href {https://math.stackexchange.com/q/3051498} {\enquote
  {\bibinfo {title} {Group cohomology of product with swapping (twisting)
  factors},}\ }\bibinfo {howpublished} {Mathematics Stack Exchange},\ \bibinfo
  {note} {uRL:https://math.stackexchange.com/q/3051498 (version: 2018-12-24)},\
  \Eprint {http://arxiv.org/abs/https://math.stackexchange.com/q/3051498}
  {https://math.stackexchange.com/q/3051498} \BibitemShut {NoStop}%
\bibitem [{\citenamefont {{Debray}}(2023)}]{2023groupcoho}%
  \BibitemOpen
  \bibfield  {author} {\bibinfo {author} {\bibfnamefont {Arun}\ \bibnamefont
  {{Debray}}},\ }\bibfield  {title} {\enquote {\bibinfo {title} {{Bordism for
  the 2-group symmetries of the heterotic and CHL strings}},}\ }\href {\doibase
  10.48550/arXiv.2304.14764} {\bibfield  {journal} {\bibinfo  {journal} {arXiv
  e-prints}\ ,\ \bibinfo {eid} {arXiv:2304.14764}} (\bibinfo {year} {2023})},\
  \Eprint {http://arxiv.org/abs/2304.14764} {arXiv:2304.14764 [math.AT]}
  \BibitemShut {NoStop}%
\bibitem [{\citenamefont {{Levin}}\ and\ \citenamefont
  {{Gu}}(2012)}]{2012LevinGu}%
  \BibitemOpen
  \bibfield  {author} {\bibinfo {author} {\bibfnamefont {Michael}\ \bibnamefont
  {{Levin}}}\ and\ \bibinfo {author} {\bibfnamefont {Zheng-Cheng}\ \bibnamefont
  {{Gu}}},\ }\bibfield  {title} {\enquote {\bibinfo {title} {{Braiding
  statistics approach to symmetry-protected topological phases}},}\ }\href
  {\doibase 10.1103/PhysRevB.86.115109} {\bibfield  {journal} {\bibinfo
  {journal} {\prb}\ }\textbf {\bibinfo {volume} {86}},\ \bibinfo {eid} {115109}
  (\bibinfo {year} {2012})},\ \Eprint {http://arxiv.org/abs/1202.3120}
  {arXiv:1202.3120 [cond-mat.str-el]} \BibitemShut {NoStop}%
\bibitem [{\citenamefont {{Shiozaki}}\ and\ \citenamefont
  {{Ryu}}(2017{\natexlab{b}})}]{2017ShinseiMPS}%
  \BibitemOpen
  \bibfield  {author} {\bibinfo {author} {\bibfnamefont {Ken}\ \bibnamefont
  {{Shiozaki}}}\ and\ \bibinfo {author} {\bibfnamefont {Shinsei}\ \bibnamefont
  {{Ryu}}},\ }\bibfield  {title} {\enquote {\bibinfo {title} {{Matrix product
  states and equivariant topological field theories for bosonic
  symmetry-protected topological phases in (1+1) dimensions}},}\ }\href
  {\doibase 10.1007/JHEP04(2017)100} {\bibfield  {journal} {\bibinfo  {journal}
  {Journal of High Energy Physics}\ }\textbf {\bibinfo {volume} {2017}},\
  \bibinfo {eid} {100} (\bibinfo {year} {2017}{\natexlab{b}})},\ \Eprint
  {http://arxiv.org/abs/1607.06504} {arXiv:1607.06504 [cond-mat.str-el]}
  \BibitemShut {NoStop}%
\bibitem [{\citenamefont {{Haegeman}}\ \emph {et~al.}(2012)\citenamefont
  {{Haegeman}}, \citenamefont {{P{\'e}rez-Garc{\'\i}a}}, \citenamefont
  {{Cirac}},\ and\ \citenamefont {{Schuch}}}]{2012nonlocalorder}%
  \BibitemOpen
  \bibfield  {author} {\bibinfo {author} {\bibfnamefont {Jutho}\ \bibnamefont
  {{Haegeman}}}, \bibinfo {author} {\bibfnamefont {David}\ \bibnamefont
  {{P{\'e}rez-Garc{\'\i}a}}}, \bibinfo {author} {\bibfnamefont {Ignacio}\
  \bibnamefont {{Cirac}}}, \ and\ \bibinfo {author} {\bibfnamefont {Norbert}\
  \bibnamefont {{Schuch}}},\ }\bibfield  {title} {\enquote {\bibinfo {title}
  {{Order Parameter for Symmetry-Protected Phases in One Dimension}},}\ }\href
  {\doibase 10.1103/PhysRevLett.109.050402} {\bibfield  {journal} {\bibinfo
  {journal} {\prl}\ }\textbf {\bibinfo {volume} {109}},\ \bibinfo {eid}
  {050402} (\bibinfo {year} {2012})},\ \Eprint {http://arxiv.org/abs/1201.4174}
  {arXiv:1201.4174 [cond-mat.str-el]} \BibitemShut {NoStop}%
\bibitem [{\citenamefont {{Chen}}\ \emph {et~al.}(2011)\citenamefont {{Chen}},
  \citenamefont {{Liu}},\ and\ \citenamefont {{Wen}}}]{2011twodSPT}%
  \BibitemOpen
  \bibfield  {author} {\bibinfo {author} {\bibfnamefont {Xie}\ \bibnamefont
  {{Chen}}}, \bibinfo {author} {\bibfnamefont {Zheng-Xin}\ \bibnamefont
  {{Liu}}}, \ and\ \bibinfo {author} {\bibfnamefont {Xiao-Gang}\ \bibnamefont
  {{Wen}}},\ }\bibfield  {title} {\enquote {\bibinfo {title} {{Two-dimensional
  symmetry-protected topological orders and their protected gapless edge
  excitations}},}\ }\href {\doibase 10.1103/PhysRevB.84.235141} {\bibfield
  {journal} {\bibinfo  {journal} {\prb}\ }\textbf {\bibinfo {volume} {84}},\
  \bibinfo {eid} {235141} (\bibinfo {year} {2011})},\ \Eprint
  {http://arxiv.org/abs/1106.4752} {arXiv:1106.4752 [cond-mat.str-el]}
  \BibitemShut {NoStop}%
\bibitem [{\citenamefont {{Williamson}}\ \emph {et~al.}(2014)\citenamefont
  {{Williamson}}, \citenamefont {{Bultinck}}, \citenamefont {{Mari{\"e}n}},
  \citenamefont {{Sahinoglu}}, \citenamefont {{Haegeman}},\ and\ \citenamefont
  {{Verstraete}}}]{2014twodtensornetwork}%
  \BibitemOpen
  \bibfield  {author} {\bibinfo {author} {\bibfnamefont {Dominic~J.}\
  \bibnamefont {{Williamson}}}, \bibinfo {author} {\bibfnamefont {Nick}\
  \bibnamefont {{Bultinck}}}, \bibinfo {author} {\bibfnamefont {Michael}\
  \bibnamefont {{Mari{\"e}n}}}, \bibinfo {author} {\bibfnamefont {Mehmet~B.}\
  \bibnamefont {{Sahinoglu}}}, \bibinfo {author} {\bibfnamefont {Jutho}\
  \bibnamefont {{Haegeman}}}, \ and\ \bibinfo {author} {\bibfnamefont {Frank}\
  \bibnamefont {{Verstraete}}},\ }\bibfield  {title} {\enquote {\bibinfo
  {title} {{Matrix product operators for symmetry-protected topological phases:
  Gauging and edge theories}},}\ }\href {\doibase 10.48550/arXiv.1412.5604}
  {\bibfield  {journal} {\bibinfo  {journal} {arXiv e-prints}\ ,\ \bibinfo
  {eid} {arXiv:1412.5604}} (\bibinfo {year} {2014})},\ \Eprint
  {http://arxiv.org/abs/1412.5604} {arXiv:1412.5604 [quant-ph]} \BibitemShut
  {NoStop}%
\bibitem [{\citenamefont {{Hastings}}(2016)}]{2016signproblem}%
  \BibitemOpen
  \bibfield  {author} {\bibinfo {author} {\bibfnamefont {M.~B.}\ \bibnamefont
  {{Hastings}}},\ }\bibfield  {title} {\enquote {\bibinfo {title} {{How quantum
  are non-negative wavefunctions?}}}\ }\href {\doibase 10.1063/1.4936216}
  {\bibfield  {journal} {\bibinfo  {journal} {Journal of Mathematical Physics}\
  }\textbf {\bibinfo {volume} {57}},\ \bibinfo {eid} {015210} (\bibinfo {year}
  {2016})},\ \Eprint {http://arxiv.org/abs/1506.08883} {arXiv:1506.08883
  [quant-ph]} \BibitemShut {NoStop}%
\bibitem [{\citenamefont {{Piroli}}\ and\ \citenamefont
  {{Cirac}}(2020)}]{2020Piroli}%
  \BibitemOpen
  \bibfield  {author} {\bibinfo {author} {\bibfnamefont {Lorenzo}\ \bibnamefont
  {{Piroli}}}\ and\ \bibinfo {author} {\bibfnamefont {J.~Ignacio}\ \bibnamefont
  {{Cirac}}},\ }\bibfield  {title} {\enquote {\bibinfo {title} {{Quantum
  Cellular Automata, Tensor Networks, and Area Laws}},}\ }\href {\doibase
  10.1103/PhysRevLett.125.190402} {\bibfield  {journal} {\bibinfo  {journal}
  {\prl}\ }\textbf {\bibinfo {volume} {125}},\ \bibinfo {eid} {190402}
  (\bibinfo {year} {2020})},\ \Eprint {http://arxiv.org/abs/2007.15371}
  {arXiv:2007.15371 [quant-ph]} \BibitemShut {NoStop}%
\bibitem [{\citenamefont {Raussendorf}\ and\ \citenamefont
  {Briegel}(2001)}]{2001cluster}%
  \BibitemOpen
  \bibfield  {author} {\bibinfo {author} {\bibfnamefont {Robert}\ \bibnamefont
  {Raussendorf}}\ and\ \bibinfo {author} {\bibfnamefont {Hans~J}\ \bibnamefont
  {Briegel}},\ }\bibfield  {title} {\enquote {\bibinfo {title} {A one-way
  quantum computer},}\ }\href@noop {} {\bibfield  {journal} {\bibinfo
  {journal} {Physical review letters}\ }\textbf {\bibinfo {volume} {86}},\
  \bibinfo {pages} {5188} (\bibinfo {year} {2001})}\BibitemShut {NoStop}%
\bibitem [{\citenamefont {{Xue}}\ \emph {et~al.}(2024)\citenamefont {{Xue}},
  \citenamefont {{Lee}},\ and\ \citenamefont {{Bao}}}]{2024BaoY}%
  \BibitemOpen
  \bibfield  {author} {\bibinfo {author} {\bibfnamefont {Hanyu}\ \bibnamefont
  {{Xue}}}, \bibinfo {author} {\bibfnamefont {Jong~Yeon}\ \bibnamefont
  {{Lee}}}, \ and\ \bibinfo {author} {\bibfnamefont {Yimu}\ \bibnamefont
  {{Bao}}},\ }\bibfield  {title} {\enquote {\bibinfo {title} {{Tensor network
  formulation of symmetry protected topological phases in mixed states}},}\
  }\href {\doibase 10.48550/arXiv.2403.17069} {\bibfield  {journal} {\bibinfo
  {journal} {arXiv e-prints}\ ,\ \bibinfo {eid} {arXiv:2403.17069}} (\bibinfo
  {year} {2024})},\ \Eprint {http://arxiv.org/abs/2403.17069} {arXiv:2403.17069
  [cond-mat.str-el]} \BibitemShut {NoStop}%
\bibitem [{\citenamefont {{Verstraete}}\ \emph {et~al.}(2006)\citenamefont
  {{Verstraete}}, \citenamefont {{Wolf}}, \citenamefont {{Perez-Garcia}},\ and\
  \citenamefont {{Cirac}}}]{2006criticalPEPS}%
  \BibitemOpen
  \bibfield  {author} {\bibinfo {author} {\bibfnamefont {F.}~\bibnamefont
  {{Verstraete}}}, \bibinfo {author} {\bibfnamefont {M.~M.}\ \bibnamefont
  {{Wolf}}}, \bibinfo {author} {\bibfnamefont {D.}~\bibnamefont
  {{Perez-Garcia}}}, \ and\ \bibinfo {author} {\bibfnamefont {J.~I.}\
  \bibnamefont {{Cirac}}},\ }\bibfield  {title} {\enquote {\bibinfo {title}
  {{Criticality, the Area Law, and the Computational Power of Projected
  Entangled Pair States}},}\ }\href {\doibase 10.1103/PhysRevLett.96.220601}
  {\bibfield  {journal} {\bibinfo  {journal} {\prl}\ }\textbf {\bibinfo
  {volume} {96}},\ \bibinfo {eid} {220601} (\bibinfo {year} {2006})},\ \Eprint
  {http://arxiv.org/abs/quant-ph/0601075} {arXiv:quant-ph/0601075 [quant-ph]}
  \BibitemShut {NoStop}%
\bibitem [{\citenamefont {{Fidkowski}}\ and\ \citenamefont
  {{Kitaev}}(2011)}]{2011symlocalfermion}%
  \BibitemOpen
  \bibfield  {author} {\bibinfo {author} {\bibfnamefont {Lukasz}\ \bibnamefont
  {{Fidkowski}}}\ and\ \bibinfo {author} {\bibfnamefont {Alexei}\ \bibnamefont
  {{Kitaev}}},\ }\bibfield  {title} {\enquote {\bibinfo {title} {{Topological
  phases of fermions in one dimension}},}\ }\href {\doibase
  10.1103/PhysRevB.83.075103} {\bibfield  {journal} {\bibinfo  {journal}
  {\prb}\ }\textbf {\bibinfo {volume} {83}},\ \bibinfo {eid} {075103} (\bibinfo
  {year} {2011})},\ \Eprint {http://arxiv.org/abs/1008.4138} {arXiv:1008.4138
  [cond-mat.str-el]} \BibitemShut {NoStop}%
\bibitem [{\citenamefont {Wolf}(2012)}]{Wolf2012note}%
  \BibitemOpen
  \bibfield  {author} {\bibinfo {author} {\bibfnamefont {M.}~\bibnamefont
  {Wolf}},\ }\bibfield  {title} {\enquote {\bibinfo {title} {Quantum channels
  and operations: Guided tour},}\ }\href
  {https://mediatum.ub.tum.de/node?id=1701036} {\bibfield  {journal} {\bibinfo
  {journal} {Lecture Notes}\ } (\bibinfo {year} {2012})}\BibitemShut {NoStop}%
\bibitem [{\citenamefont {{Levin}}(2020)}]{2020levinSSB}%
  \BibitemOpen
  \bibfield  {author} {\bibinfo {author} {\bibfnamefont {Michael}\ \bibnamefont
  {{Levin}}},\ }\bibfield  {title} {\enquote {\bibinfo {title} {{Constraints on
  Order and Disorder Parameters in Quantum Spin Chains}},}\ }\href {\doibase
  10.1007/s00220-020-03802-4} {\bibfield  {journal} {\bibinfo  {journal}
  {Communications in Mathematical Physics}\ }\textbf {\bibinfo {volume}
  {378}},\ \bibinfo {pages} {1081--1106} (\bibinfo {year} {2020})},\ \Eprint
  {http://arxiv.org/abs/1903.09028} {arXiv:1903.09028 [cond-mat.str-el]}
  \BibitemShut {NoStop}%
\bibitem [{\citenamefont
  {{Hastings}}(2004{\natexlab{a}})}]{2004Hastingssteadystate}%
  \BibitemOpen
  \bibfield  {author} {\bibinfo {author} {\bibfnamefont {M.~B.}\ \bibnamefont
  {{Hastings}}},\ }\bibfield  {title} {\enquote {\bibinfo {title} {{Locality in
  Quantum and Markov Dynamics on Lattices and Networks}},}\ }\href {\doibase
  10.1103/PhysRevLett.93.140402} {\bibfield  {journal} {\bibinfo  {journal}
  {\prl}\ }\textbf {\bibinfo {volume} {93}},\ \bibinfo {eid} {140402} (\bibinfo
  {year} {2004}{\natexlab{a}})},\ \Eprint
  {http://arxiv.org/abs/cond-mat/0405587} {arXiv:cond-mat/0405587
  [cond-mat.stat-mech]} \BibitemShut {NoStop}%
\bibitem [{\citenamefont {{Lucia}}\ \emph {et~al.}(2015)\citenamefont
  {{Lucia}}, \citenamefont {{Cubitt}}, \citenamefont {{Michalakis}},\ and\
  \citenamefont {{P{\'e}rez-Garc{\'\i}a}}}]{2015rapidmixing1}%
  \BibitemOpen
  \bibfield  {author} {\bibinfo {author} {\bibfnamefont {Angelo}\ \bibnamefont
  {{Lucia}}}, \bibinfo {author} {\bibfnamefont {Toby~S.}\ \bibnamefont
  {{Cubitt}}}, \bibinfo {author} {\bibfnamefont {Spyridon}\ \bibnamefont
  {{Michalakis}}}, \ and\ \bibinfo {author} {\bibfnamefont {David}\
  \bibnamefont {{P{\'e}rez-Garc{\'\i}a}}},\ }\bibfield  {title} {\enquote
  {\bibinfo {title} {{Rapid mixing and stability of quantum dissipative
  systems}},}\ }\href {\doibase 10.1103/PhysRevA.91.040302} {\bibfield
  {journal} {\bibinfo  {journal} {\pra}\ }\textbf {\bibinfo {volume} {91}},\
  \bibinfo {eid} {040302} (\bibinfo {year} {2015})},\ \Eprint
  {http://arxiv.org/abs/1409.7809} {arXiv:1409.7809 [quant-ph]} \BibitemShut
  {NoStop}%
\bibitem [{\citenamefont {{Cubitt}}\ \emph {et~al.}(2015)\citenamefont
  {{Cubitt}}, \citenamefont {{Lucia}}, \citenamefont {{Michalakis}},\ and\
  \citenamefont {{Perez-Garcia}}}]{2015rapidmixing2}%
  \BibitemOpen
  \bibfield  {author} {\bibinfo {author} {\bibfnamefont {Toby~S.}\ \bibnamefont
  {{Cubitt}}}, \bibinfo {author} {\bibfnamefont {Angelo}\ \bibnamefont
  {{Lucia}}}, \bibinfo {author} {\bibfnamefont {Spyridon}\ \bibnamefont
  {{Michalakis}}}, \ and\ \bibinfo {author} {\bibfnamefont {David}\
  \bibnamefont {{Perez-Garcia}}},\ }\bibfield  {title} {\enquote {\bibinfo
  {title} {{Stability of Local Quantum Dissipative Systems}},}\ }\href
  {\doibase 10.1007/s00220-015-2355-3} {\bibfield  {journal} {\bibinfo
  {journal} {Communications in Mathematical Physics}\ }\textbf {\bibinfo
  {volume} {337}},\ \bibinfo {pages} {1275--1315} (\bibinfo {year} {2015})},\
  \Eprint {http://arxiv.org/abs/1303.4744} {arXiv:1303.4744 [quant-ph]}
  \BibitemShut {NoStop}%
\bibitem [{\citenamefont {{Song}}\ \emph {et~al.}(2019)\citenamefont {{Song}},
  \citenamefont {{Yao}},\ and\ \citenamefont {{Wang}}}]{2019skineffect}%
  \BibitemOpen
  \bibfield  {author} {\bibinfo {author} {\bibfnamefont {Fei}\ \bibnamefont
  {{Song}}}, \bibinfo {author} {\bibfnamefont {Shunyu}\ \bibnamefont {{Yao}}},
  \ and\ \bibinfo {author} {\bibfnamefont {Zhong}\ \bibnamefont {{Wang}}},\
  }\bibfield  {title} {\enquote {\bibinfo {title} {{Non-Hermitian skin effect
  and chiral damping in open quantum systems}},}\ }\href {\doibase
  10.48550/arXiv.1904.08432} {\bibfield  {journal} {\bibinfo  {journal} {arXiv
  e-prints}\ ,\ \bibinfo {eid} {arXiv:1904.08432}} (\bibinfo {year} {2019})},\
  \Eprint {http://arxiv.org/abs/1904.08432} {arXiv:1904.08432
  [cond-mat.quant-gas]} \BibitemShut {NoStop}%
\bibitem [{\citenamefont {{Poulin}}(2010)}]{2010openLiebRobinson}%
  \BibitemOpen
  \bibfield  {author} {\bibinfo {author} {\bibfnamefont {David}\ \bibnamefont
  {{Poulin}}},\ }\bibfield  {title} {\enquote {\bibinfo {title} {{Lieb-Robinson
  Bound and Locality for General Markovian Quantum Dynamics}},}\ }\href
  {\doibase 10.1103/PhysRevLett.104.190401} {\bibfield  {journal} {\bibinfo
  {journal} {\prl}\ }\textbf {\bibinfo {volume} {104}},\ \bibinfo {eid}
  {190401} (\bibinfo {year} {2010})},\ \Eprint {http://arxiv.org/abs/1003.3675}
  {arXiv:1003.3675 [quant-ph]} \BibitemShut {NoStop}%
\bibitem [{\citenamefont {Lieb}\ \emph {et~al.}(1961)\citenamefont {Lieb},
  \citenamefont {Schultz},\ and\ \citenamefont {Mattis}}]{lieb1961two}%
  \BibitemOpen
  \bibfield  {author} {\bibinfo {author} {\bibfnamefont {Elliott}\ \bibnamefont
  {Lieb}}, \bibinfo {author} {\bibfnamefont {Theodore}\ \bibnamefont
  {Schultz}}, \ and\ \bibinfo {author} {\bibfnamefont {Daniel}\ \bibnamefont
  {Mattis}},\ }\bibfield  {title} {\enquote {\bibinfo {title} {Two soluble
  models of an antiferromagnetic chain},}\ }\href@noop {} {\bibfield  {journal}
  {\bibinfo  {journal} {Annals of Physics}\ }\textbf {\bibinfo {volume} {16}},\
  \bibinfo {pages} {407--466} (\bibinfo {year} {1961})}\BibitemShut {NoStop}%
\bibitem [{\citenamefont {{Oshikawa}}(2000)}]{2000Oshikawa}%
  \BibitemOpen
  \bibfield  {author} {\bibinfo {author} {\bibfnamefont {Masaki}\ \bibnamefont
  {{Oshikawa}}},\ }\bibfield  {title} {\enquote {\bibinfo {title}
  {{Commensurability, Excitation Gap, and Topology in Quantum Many-Particle
  Systems on a Periodic Lattice}},}\ }\href {\doibase
  10.1103/PhysRevLett.84.1535} {\bibfield  {journal} {\bibinfo  {journal}
  {\prl}\ }\textbf {\bibinfo {volume} {84}},\ \bibinfo {pages} {1535--1538}
  (\bibinfo {year} {2000})},\ \Eprint {http://arxiv.org/abs/cond-mat/9911137}
  {arXiv:cond-mat/9911137 [cond-mat.str-el]} \BibitemShut {NoStop}%
\bibitem [{\citenamefont {{Hastings}}(2004{\natexlab{b}})}]{200HastingsLSM}%
  \BibitemOpen
  \bibfield  {author} {\bibinfo {author} {\bibfnamefont {M.~B.}\ \bibnamefont
  {{Hastings}}},\ }\bibfield  {title} {\enquote {\bibinfo {title}
  {{Lieb-Schultz-Mattis in higher dimensions}},}\ }\href {\doibase
  10.1103/PhysRevB.69.104431} {\bibfield  {journal} {\bibinfo  {journal}
  {\prb}\ }\textbf {\bibinfo {volume} {69}},\ \bibinfo {eid} {104431} (\bibinfo
  {year} {2004}{\natexlab{b}})},\ \Eprint
  {http://arxiv.org/abs/cond-mat/0305505} {arXiv:cond-mat/0305505
  [cond-mat.str-el]} \BibitemShut {NoStop}%
\bibitem [{\citenamefont {{Kawabata}}\ \emph {et~al.}(2023)\citenamefont
  {{Kawabata}}, \citenamefont {{Sohal}},\ and\ \citenamefont
  {{Ryu}}}]{2023openLSM}%
  \BibitemOpen
  \bibfield  {author} {\bibinfo {author} {\bibfnamefont {Kohei}\ \bibnamefont
  {{Kawabata}}}, \bibinfo {author} {\bibfnamefont {Ramanjit}\ \bibnamefont
  {{Sohal}}}, \ and\ \bibinfo {author} {\bibfnamefont {Shinsei}\ \bibnamefont
  {{Ryu}}},\ }\bibfield  {title} {\enquote {\bibinfo {title}
  {{Lieb-Schultz-Mattis Theorem in Open Quantum Systems}},}\ }\href {\doibase
  10.48550/arXiv.2305.16496} {\bibfield  {journal} {\bibinfo  {journal} {arXiv
  e-prints}\ ,\ \bibinfo {eid} {arXiv:2305.16496}} (\bibinfo {year} {2023})},\
  \Eprint {http://arxiv.org/abs/2305.16496} {arXiv:2305.16496
  [cond-mat.stat-mech]} \BibitemShut {NoStop}%
\bibitem [{\citenamefont {{Haegeman}}\ \emph {et~al.}(2015)\citenamefont
  {{Haegeman}}, \citenamefont {{Van Acoleyen}}, \citenamefont {{Schuch}},
  \citenamefont {{Cirac}},\ and\ \citenamefont
  {{Verstraete}}}]{2015stategauging}%
  \BibitemOpen
  \bibfield  {author} {\bibinfo {author} {\bibfnamefont {Jutho}\ \bibnamefont
  {{Haegeman}}}, \bibinfo {author} {\bibfnamefont {Karel}\ \bibnamefont {{Van
  Acoleyen}}}, \bibinfo {author} {\bibfnamefont {Norbert}\ \bibnamefont
  {{Schuch}}}, \bibinfo {author} {\bibfnamefont {J.~Ignacio}\ \bibnamefont
  {{Cirac}}}, \ and\ \bibinfo {author} {\bibfnamefont {Frank}\ \bibnamefont
  {{Verstraete}}},\ }\bibfield  {title} {\enquote {\bibinfo {title} {{Gauging
  Quantum States: From Global to Local Symmetries in Many-Body Systems}},}\
  }\href {\doibase 10.1103/PhysRevX.5.011024} {\bibfield  {journal} {\bibinfo
  {journal} {Physical Review X}\ }\textbf {\bibinfo {volume} {5}},\ \bibinfo
  {eid} {011024} (\bibinfo {year} {2015})},\ \Eprint
  {http://arxiv.org/abs/1407.1025} {arXiv:1407.1025 [quant-ph]} \BibitemShut
  {NoStop}%
\bibitem [{\citenamefont {{Micha{\l}ek}}\ and\ \citenamefont
  {{Shitov}}(2018)}]{2018blocking}%
  \BibitemOpen
  \bibfield  {author} {\bibinfo {author} {\bibfnamefont {Mateusz}\ \bibnamefont
  {{Micha{\l}ek}}}\ and\ \bibinfo {author} {\bibfnamefont {Yaroslav}\
  \bibnamefont {{Shitov}}},\ }\bibfield  {title} {\enquote {\bibinfo {title}
  {{Quantum version of Wielandt's Inequality revisited}},}\ }\href {\doibase
  10.48550/arXiv.1809.04387} {\bibfield  {journal} {\bibinfo  {journal} {arXiv
  e-prints}\ ,\ \bibinfo {eid} {arXiv:1809.04387}} (\bibinfo {year} {2018})},\
  \Eprint {http://arxiv.org/abs/1809.04387} {arXiv:1809.04387 [math.AC]}
  \BibitemShut {NoStop}%
\bibitem [{\citenamefont {{Bridgeman}}\ and\ \citenamefont
  {{Chubb}}(2017)}]{2017jacobmps}%
  \BibitemOpen
  \bibfield  {author} {\bibinfo {author} {\bibfnamefont {Jacob~C.}\
  \bibnamefont {{Bridgeman}}}\ and\ \bibinfo {author} {\bibfnamefont
  {Christopher~T.}\ \bibnamefont {{Chubb}}},\ }\bibfield  {title} {\enquote
  {\bibinfo {title} {{Hand-waving and interpretive dance: an introductory
  course on tensor networks}},}\ }\href {\doibase 10.1088/1751-8121/aa6dc3}
  {\bibfield  {journal} {\bibinfo  {journal} {Journal of Physics A Mathematical
  General}\ }\textbf {\bibinfo {volume} {50}},\ \bibinfo {eid} {223001}
  (\bibinfo {year} {2017})},\ \Eprint {http://arxiv.org/abs/1603.03039}
  {arXiv:1603.03039 [quant-ph]} \BibitemShut {NoStop}%
\end{thebibliography}%

\appendix

\section{Details of the cluster state example}
\label{cluster}

Consider the example of the cluster state. In this appendix, we illustrate how a symmetric finite-depth channel can result in spontaneous exact-to-average symmetry breaking. Specifically, we demonstrate that passing the cluster state through the depth-one channel
    \begin{equation}
        \mathcal{E}=\bigotimes_i((1-p)(\cdot)+pX_i(\cdot)X_i)
    \end{equation}
with $p=1/2$ yields a mixed state
\begin{align}
    \label{eq:dephasedclusterchain}\begin{split}\rho_{1/2}&=\prod_{i\text{ even}}(\frac{1}{4}I_i I_{i+1})+\prod_{i\text{ even}}(\frac{1}{4}I_i X_{i+1})\\&\qquad+\prod_{i\text{ even}}(\frac{1}{4}X_i I_{i+1})+\prod_{i\text{ even}}(\frac{1}{4}X_iX_{i+1})~.
    \end{split}
\end{align}

The cluster state is defined by the MPS tensors
$$A^0=|0)(+|=\frac{1}{\sqrt{2}}\begin{pmatrix}
    1&1\\&
\end{pmatrix}~,\qquad A^1=|1)(-|=\frac{1}{\sqrt{2}}\begin{pmatrix}
    &\\1&-1
\end{pmatrix}~.$$
Upon acting with the channel, the cluster state becomes the MPDO with tensor
$$B^{00}=B^{11}=\frac{1}{2}\left(A^0\otimes A^{0*}+A^1\otimes A^{1*}\right)=\frac{1}{4}\begin{pmatrix}1&1&1&1\\&&&\\&&&\\1&-1&-1&1\end{pmatrix}~,$$
$$B^{01}=B^{10}=\frac{1}{2}\left(A^0\otimes A^{1*}+A^1\otimes A^{0*}\right)=\frac{1}{4}\begin{pmatrix}&&&\\1&-1&1&-1\\1&1&-1&-1\\&&&\end{pmatrix}~.$$
Now block pairs of sites by multiplying tensors $B^{ij|kl}:=B^{ij}B^{kl}$ and follow this with a swap of the third and fourth basis vectors and a conjugation by $\begin{pmatrix}1&1\\1&-1\end{pmatrix}\otimes\mathds{1}$ to obtain
$$B^{00|00}=B^{00|11}=B^{11|00}=B^{11|11}=\frac{1}{4}\begin{pmatrix}1&\\&\end{pmatrix}\otimes\begin{pmatrix}1&\\&\end{pmatrix}~,$$
$$B^{00|01}=B^{00|10}=B^{11|01}=B^{11|10}=\frac{1}{4}\begin{pmatrix}&\\&1\end{pmatrix}\otimes\begin{pmatrix}1&\\&\end{pmatrix}~,$$
$$B^{01|00}=B^{01|11}=B^{10|00}=B^{10|11}=\frac{1}{4}\begin{pmatrix}1&\\&\end{pmatrix}\otimes\begin{pmatrix}&\\&1\end{pmatrix}~,$$
$$B^{01|01}=B^{01|10}=B^{10|01}=B^{10|10}=\frac{1}{4}\begin{pmatrix}&\\&1\end{pmatrix}\otimes\begin{pmatrix}&\\&1\end{pmatrix}~.$$
This form tells us that the state is a superposition of four injective (on $2$-site blocks) MPDOs. The first line represents the component with bra and ket indices equal, i.e. the state $\rho\sim I$. The second line has indices equal on even sites and opposite on odds. In total, we can read off the state $\rho_{1/2}$, written out above. $\qquad\square$

\section{A proof of Theorem \ref{thm:nondeg}}
\label{app:nondeg}

\begin{figure}
     \centering
     \begin{subfigure}[b]{0.4\textwidth}
         \centering
         \includegraphics[width=\textwidth]{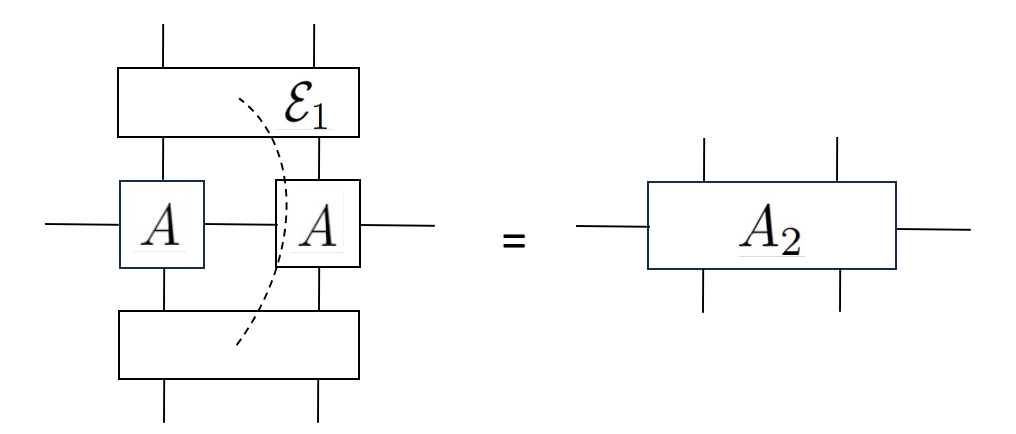}
         \caption{First layer of the channel $\mathcal{E}$}
         \label{firstlayer}
     \end{subfigure}
     \hfill
     \begin{subfigure}[b]{0.5\textwidth}
         \centering
         \includegraphics[width=0.48\textwidth]{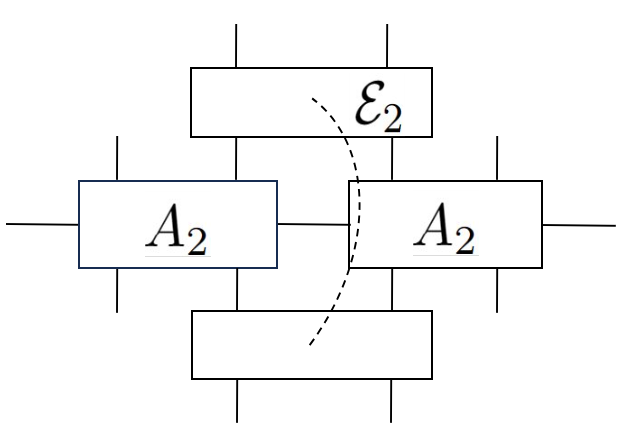}
         \caption{Each local gate $\mathcal{E}_2$ of the second layer acts on two local tensors.}
         \label{fig:secondlayer}
     \end{subfigure}
     \hfill
     \begin{subfigure}[b]{0.4\textwidth}
         \centering
         \includegraphics[width=\textwidth]{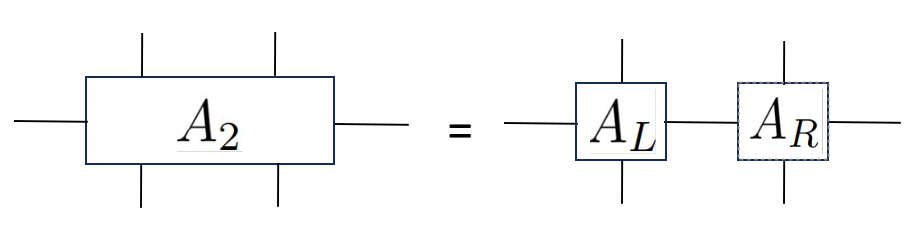}
         \caption{The Schmidt decomposition of $A_2$ gives rise to a left component $A_L$ and a right component $A_R$.}
         \label{Schmidt}
     \end{subfigure}
     \hfill
     \begin{subfigure}[b]{0.4\textwidth}
         \centering
         \includegraphics[width=\textwidth]{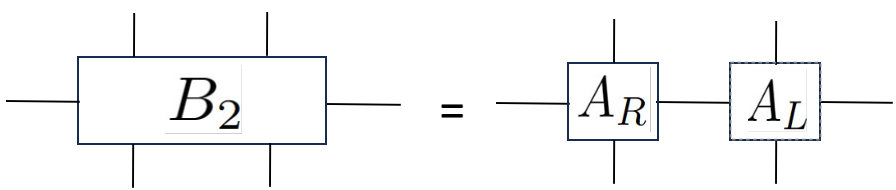}
         \caption{We recombine $A_R$ and $A_L$ in a different order to form $B_2$.}
         \label{recombine}
     \end{subfigure}
     \hfill
     \begin{subfigure}[b]{0.5\textwidth}
         \centering
         \includegraphics[width=0.35\textwidth]{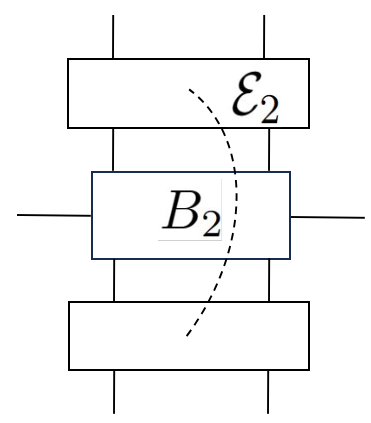}
         \caption{The local tensor remains injective after the second layer of local gates if $\mathcal{E}_2$ is non-degenerate.}
         \label{newevol}
     \end{subfigure}
      \caption{Proof of Theorem \ref{thm:nondeg}}
        \label{fig:thmnondeg}
\end{figure}

The proof of Theorem \ref{thm:nondeg} proceeds as follows. By definition, the local tensors of $A^{pq}$ of an SRE state $| \rho \rrangle$ are normal. It is always possible to make them injective by blocking a finite number of sites \cite{2018blocking}. Hence, we assume $A^{pq}$ to be injective, meaning the matrices ${ A^{pq} }$ span the entire set of $D\times D$ matrices \cite{2017MPDO}, where $D$ is the entanglement bond dimension. A finite-depth channel $\mathcal{E}$ is non-degenerate if and only if each of its local gates is non-degenerate. Consider the first layer of the finite-depth channel, with each local gate labeled by $\mathcal{E}_1$. If $\mathcal{E}_1$ is non-degenerate, then the evolved local tensor $A_2^{pq}$ (refer to Fig \ref{firstlayer}) also spans the entire set of $D\times D$ matrices and thus is injective. Denote the Choi state generated by the local tensor $A_2$ as $| \phi \rrangle$.

Next, we examine the second layer of the quantum channel, where each local gate is denoted by $\mathcal{E}_2$, as illustrated in Fig. \ref{fig:secondlayer}. We begin by performing a Schmidt decomposition of $A_2$ into its left and right components, as shown in Fig. \ref{Schmidt}, and then recombine them in a different order to obtain a new local tensor, $B_2$, depicted in Fig. \ref{recombine}. We now demonstrate that the state generated by $B_2$, denoted by $| \psi \rrangle$, is SRE. A crucial observation is that $| \psi \rrangle$ is related to $| \phi \rrangle$ by translation by one unit cell:
\begin{equation}
    | \psi \rrangle = T | \phi \rrangle,
\end{equation}
where $T$ represents translation. By definition, $| \phi \rrangle$, as an SRE state, can be prepared from a translationally-invariant product state (denoted by $| 0 \rrangle$) using a finite-depth local unitary $U$ in the doubled space, \ie $| \psi \rrangle = U | 0 \rrangle$. Therefore, we have:
\begin{equation}
    | \psi \rrangle = T U | 0 \rrangle = T U T^{-1} | 0 \rrangle, 
\end{equation}
and $T U T^{-1}$ is also a finite-depth local unitary. Hence, $| \psi \rrangle$ is also SRE.

The local tensor $B_2$ generating the SRE state $| \psi \rrangle$ must be normal. Without loss of generality, we assume $B_2$ to be injective, as this can always be achieved by blocking a finite number of sites. Therefore, the evolved local tensor after the second layer remains injective if $\mathcal{E}_2$ is non-degenerate, as illustrated in Fig \ref{newevol}. One may iterate the same argument for any finite-depth local channel $\mathcal{E}$, and the resulting state $\mathcal{E}| \rho \rrangle$ is generated by an injective local tensor. We then conclude that $\mathcal{E}| \rho \rrangle$ is SRE.

\section{A proof of Theorem~\ref{thm:relationbetweenDs}}
\label{app:proofofD}

In this appendix, we provide a proof of Theorem~\ref{thm:relationbetweenDs}. For simplicity in notation, we assume the state possesses an exact $\Z_2$ symmetry, generated by $U = \prod_i X_i$. Additionally, we make the assumption that $\rho$ can be described by an MPDO with a finite bond dimension.

\begin{figure}
\begin{center}
  \includegraphics[width=.18\textwidth]{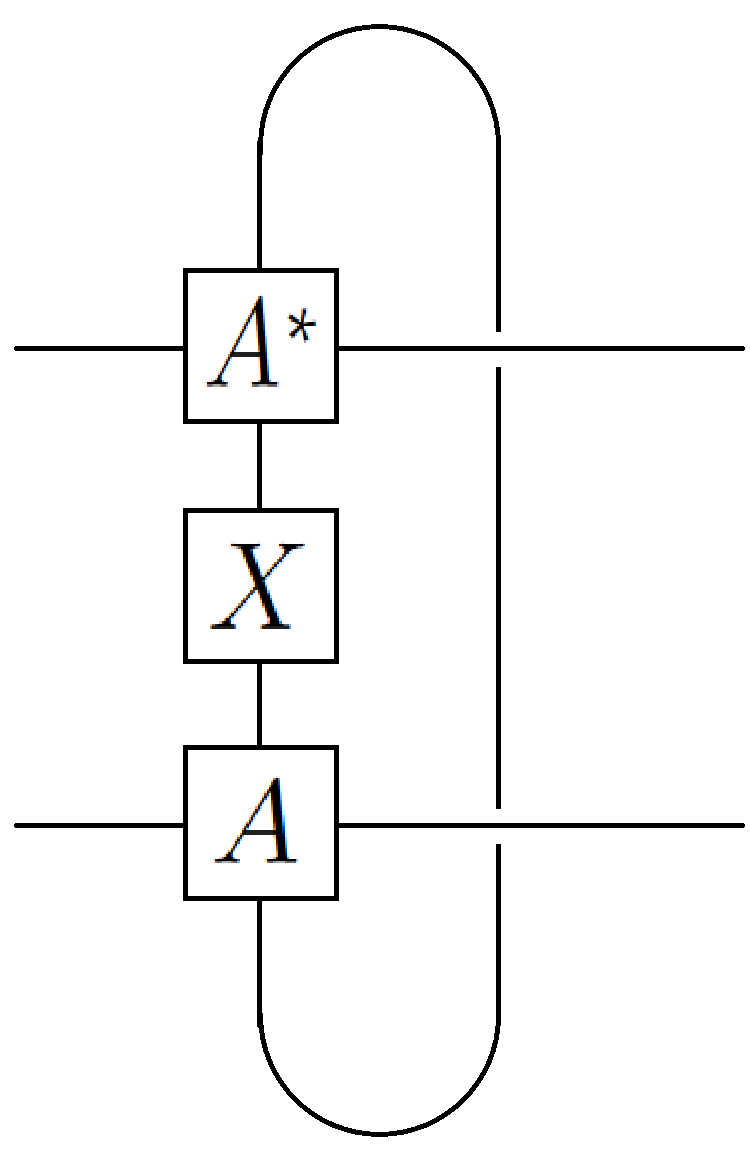} 
\end{center}
\caption{
A diagrammatic representation of the transfer matrix $E^2_X$ with an inserted $X$.
}
\label{Fig:E2x}
\end{figure}

MPDOs exhibiting a $\Z_2^+ \times \Z_2^-$ symmetry can be categorized into three classes:
\begin{enumerate}
    \item Symmetric injective MPDOs (possibly achieved by blocking several sites together).
    \item A direct sum of a finite number of symmetric injective MPDOs.
    \item Non-injective MPDOs in which the symmetry acts by exchanging injective blocks, each of which is not symmetric. This category includes the well-known GHZ state.
\end{enumerate}
The first two classes exhibit essentially identical behavior for our purposes, and so we will focus on the first class and the third class. Our objectives are to show (1) If the disorder parameter $D_1$ is non-zero for a suitable choice of endpoint operator, the MPDO must fall into class 1. (2) For states in class 1, there exists a choice of endpoint operator such that $D_3$ is also non-zero.

Let us examine an MPDO belonging to class 3, featuring an injective local tensor denoted as $M^{pq}$, with $p$ and $q$ taking values of 0 or 1. The tensor is \emph{not} symmetric under the exact $\Z_2$ symmetry: One cannot find a gauge transformation $V$, such that the ``Symmetries Push Through" \cite{2017jacobmps} condition is satisfied. However, for simplicity, we will assume that the injective tensor $M$ remains symmetric under the \emph{diagonal} $\Z_2$ symmetry. Namely, there exists a gauge transformation $V$, such that 
\begin{equation}
    M^{\Bar{p}\Bar{q}} = V M^{pq} V^\dagger,
\end{equation}
where $\Bar{0}=1$ and vice versa. In cases where the injective tensor is not symmetric even under the diagonal subgroup, the reasoning can be applied similarly. We denote the MPDO generated by the local tensor $M$ as $\rho_0$.

In order for the state $\rho$ to be symmetric under the exact $\Z_2$ symmetry, one has to symmetrize the state $\rho_0$. Namely, $\rho\propto \rho_0+ U\rho_0$. The local tensor $A^{pq}$ of the MPDO $\rho$ can then be represented as:
\begin{equation}
A^{pq}=
\begin{pmatrix}
M^{pq}&\\&M^{\Bar{p}q}
\end{pmatrix}.
\end{equation}
The transfer matrix $E^2$, and the transfer matrix with an $X$ insertion, denoted as $E^2_X$ (see Fig.~\ref{Fig:E2x}), can be written as:
\begin{equation}
    \begin{split}
        E^2 & = 
        \begin{pmatrix}
M^{pq}&\\&M^{\Bar{p}q}
\end{pmatrix}
\otimes
\begin{pmatrix}
M^{pq*}&\\&M^{\Bar{p}q*}
\end{pmatrix},
     \\
    E^2_X & = 
        \begin{pmatrix}
M^{\Bar{p}q}&\\&M^{pq}
\end{pmatrix}
\otimes
\begin{pmatrix}
M^{pq*}&\\&M^{\Bar{p}q*}
\end{pmatrix}.
\end{split}
\end{equation}
It can be demonstrated that for an injective tensor $M$ that is not symmetric under a single copy of a $\Z_2$ transformation, the largest eigenvalue of $E^2_X$ is strictly less than 1 \cite{2017jacobmps}. Consequently, $D_1$ cannot take a nonzero value in any state belonging to class 3. 

Therefore, if we observe a nonvanishing $D_1$, the MPDO must fall into class 1, where we have a symmetric injective MPDO. States in class 1 exhibit a nice property: the symmetry, acting in a finite but large region, affects the state nontrivially only near its boundary, as described by:
\begin{equation}
    \prod_{i\in A} X_i | \rho \rrangle = V_l V_r | \rho \rrangle,
\end{equation}
where $V_l$ and $V_r$ represent unitary operators that are nontrivial only near the left and right endpoint of $A$. By selecting $U_A = V_l^\dagger V_r^\dagger \prod_{i\in A} X_i$, we can ensure that
\begin{equation}
    D_3 = \llangle I | U_A | \rho \rrangle \sim O(1).
\end{equation}
This concludes the proof of Theorem~\ref{thm:relationbetweenDs}.

\end{document}